\documentclass{aa}  
\usepackage{graphicx}
\usepackage{lscape}
\usepackage{natbib}
\usepackage{txfonts}
\begin{document}
\newcommand{\kms}{km~s$^{-1}$}
\newcommand{\Msun}{M$_{\odot}$}
\newcommand\mycom[2]{\genfrac{}{}{0pt}{}{#1}{#2}}
\title {Are the orbital poles of binary stars in the solar neighbourhood anisotropically distributed? }

   \author{J-L. Agati
          \inst{1}
          \and D. Bonneau
          \inst{2}
          \and A. Jorissen
          \inst{3}
         	\and E. Souli\'e
          \inst{4} 
          \and S. Udry 
          \inst{5}
					\and P. Verhas
          \inst{6}
          \and J. Dommanget$\dagger$\thanks{Deceased October 1, 2014}
          \inst{7}
  }

	\institute{
					13, rue Beyle Stendhal, 38000 Grenoble, France
	\and
					Laboratoire Lagrange, UMR 7293, Univ. Nice Sophia-Antipolis, CNRS, Observatoire de la C\^ote d'Azur, 06300 Nice, France\\
					\email{daniel.bonneau@oca.eu}
	\and
					Institut d'Astronomie et d'Astrophysique, Universit\'e Libre de Bruxelles, CP. 226, Boulevard du Triomphe, 1050 Brussels, Belgium
	\and
					CEA/Saclay, DSM, 91191 Gif-sur-Yvette  Cedex  
	\and
					Observatoire de Gen\`eve, Chemin des Maillettes, CH-1290 Sauverny, Suisse
	\and
					251 Vieille rue du Moulin, 1180 Brussels, Belgium
	\and
					Observatoire Royal de Belgique, Avenue Circulaire 3, B-1180 Bruxelles, Belgique 
	}

	\date{Received 15/11/2013; accepted 30/10/2014}

  \abstract
   {We test whether or not the orbital poles of the systems in the solar neighbourhood are isotropically distributed on the
celestial sphere. The problem is plagued by the ambiguity on the position of the ascending node.
 
Of the 95 systems closer than 18 pc from the Sun with an orbit in the 6th Catalogue of Orbits of Visual Binaries, the pole ambiguity could be resolved for 51 systems using radial velocity collected in the literature and CORAVEL database or acquired with the HERMES/Mercator spectrograph. For several systems, we can correct the erroneous nodes in the 6th Catalogue of Orbits and obtain new combined spectroscopic/astrometric orbits for seven systems [WDS 01083+5455Aa,Ab; 01418+4237AB; 02278+0426AB (SB2); 09006+4147AB (SB2); 16413+3136AB; 17121+4540AB; 18070+3034AB].

We used of spherical statistics to test for possible anisotropy. After ordering the binary systems by increasing distance from the Sun, we computed the false-alarm probability for subsamples of increasing sizes, from $N = 1$ up to the full sample of 51 systems. Rayleigh-Watson and Beran tests deliver a false-alarm probability 
of 0.5\% for the 20 systems closer than 8.1~pc. To evaluate the robustness of this conclusion, we used a jackknife approach, for which we repeated this procedure after removing one system at a time from the full sample.
The false-alarm probability was then found to vary between 1.5\% and 0.1\%, depending on which system is removed. The reality of the deviation from isotropy can thus not be assessed with certainty at this stage, because only so few systems are available, despite our efforts to increase the sample. However, when considering the full sample of 51 systems, the concentration of poles toward the Galactic position $l = 46.0^\circ$, $b = 37^\circ$, as observed in the 8.1~pc sphere, totally vanishes (the Rayleigh-Watson false-alarm probability then rises to 18\%).
}

\maketitle
\titlerunning{Orbital poles of binary stars}

\section{Introduction}
\label{Sect.Intro}

In 1838, J.~H.~M\"adler \nocite{Madler1838} announced the probable existence of an anastonishing phenomenon related to the orbital plane of visual double
stars: they seem to be aligned with each other. This phenomenon has received quite some attention since then, but the most earliest studies that was based on
data secured with old observational techniques and equipments, did not bring significant progress. In the years 1950~-~1967, J.~Dommanget
(with the assistance of O.~Nys) re-investigated the question after collecting all available data needed to perform the first study based on
a sample of orbits for which the orbital poles were determined unambiguously. 
This research \citep{Dommanget1968}, updated in 1982~-~1987 \citep{Dommanget1988}, confirmed M\"adler's suspicion that the orbital poles of visual binaries in the solar neighbourhood do not seem to be oriented isotropically, but rather seem to cluster around the positions $(l,b) \sim (100^\circ, -15^\circ)$ and $(280^\circ,
+15^\circ)$. This claim was based on a limited number of objects (8 systems within 10~pc of the Sun and 70 systems up to 25~pc), 
and was not subject to any analysis of its statistical significance. It was later confirmed by \citet{Glebocki2000}, who found that the poles of  19 systems within
10~pc of the Sun are concentrated around positions near $(l,b) \sim	(110^\circ, -10^\circ)$  and $(-115^\circ, -5^\circ)$, although the global distribution of 252 systems is isotropic. 
An historical overview of all these studies may be found in \citet{Dommanget2014}.

In this paper, we present the results of a new research conducted since 2005 within the framework of a collaboration between professional astronomers and amateur astronomers that are members of the Double Stars Committee of the {\it Soci\'et\'e Astronomique de France} \footnote{http://saf.etoilesdoubles.free.fr/}.

Following the suggestion of  \citet{Dommanget2006}, we re-investigated the question by concentrating on a very local data set. Moreover, to the
best of our knowledge, it seems that the most recent  list of orbital poles dates back to \citet{Batten1967}. Because the more recent studies
did not publish the data sets on which they were based and because of the rapid increase of the number of available orbits,  we feel the
necessity to collect such data, and provide an updated list of orbital poles for binaries within 18~pc of the Sun.

As already noted by \citet{Finsen1933} and \citet{Dommanget1968,Dommanget1988,Dommanget2005}, a severe pitfall awaits the researcher on this topic: 
the ambiguity in the position of the ascending node, which, if not correctly assessed (Sect.~\ref{Sect:Ambiguity}), may ruin the distribution of the poles on
the sphere.

To avoid this problem and establish a valid statistical material, we have first a) assembled a list of known visual binary
orbits in the vicinity of the Sun for which a set of consistent and accurate radial-velocity measurements is available for solving the
ambiguity of the orbital ascending node and then b) computed the Galactic positions of the orbital poles of all systems for which the ambiguity of
the orbital ascending node has been solved.

At first sight, some large-scale coherence in the process of binary formation might imply a non-random orientation of the
poles on the celestial sphere. However, such a large-scale coherence is not expected to last after the coeval association has dissolved, since
at any given time in the history of the Galaxy, stars in a given volume emanate from very different birth places and times, and their presence
in a given volume is a momentary event. Therefore, given these potentially important (and problematic) consequences, any deviation from
isotropy in the pole distribution should be firmly established before embarking on any physical discussion on the above questions.

In this new research, emphasis has been place on
\begin{itemize}
\item increasing the sample size with regard to previous studies;
\item using dedicated tests to evaluate the statistical significance of any deviation from isotropy in the distribution of the orbital poles;
\item analysing the evolution, as a function of the volume sampled, of such possible deviations from isotropy;
\item discussing these results from the standpoint of Galactic kinematics.
\end{itemize}

This paper is structured as follows: In Sect.~\ref{Sect:Ambiguity}, we briefly recall the problem of unambiguously determining of the
direction of the orbital pole, all mathematical details being provided in Appendix~\ref{Sect:pole}. 
The master list of selected systems is presented in Sect.~\ref{Sect:Data}. For 51 systems up to 18 pc from the
Sun, the distribution of the orbital poles on the sky is presented in Sect.~\ref{Sect:Distribution}. This section describes the available
statistical tools used to analyse this distribution, and some indications about its degree of anisotropy are given. 
Possible selection effects in our sample is discussed in Sect.~\ref{Sect:Selection}. 
The problem of possible orbital coplanarity in multiple systems is briefly discussed in Sect.~\ref{Sect:Multiplicity}.
Section \ref{Sect:Galactic} discusses the question of the orientation of the orbital poles in the framework of Galactic kinematics. 
Conclusions and perspectives are given in Sect.~\ref{Sect:Conclusion}. 
In Appendix~\ref{Sect:notes}, notes are given on individuact.\ref{Sect:Galactic} discusses the issue of the orientation of the orbital poles in the framework of galactic kinematics. Conclusions and
perspectives are given in Sect.~\ref{Sect:Conclusion}. l systems, along with new radial-velocity measurements used in the present study  and, in
several cases, new combined astrometric-spectroscopic orbits that resulted. In Appendix~\ref{Sect:pole}, we recall the conventions
adopted to define the relative orbit of a binary star, and we describe the method used to determine the ascending node and the direction of the
orbital pole without any ambiguity. 

\section{Unambiguous determination of the orbital pole of a binary}
\label{Sect:Ambiguity}

The spatial orientation of the orbital planes of double stars is sutdied based on the knowledge of the orientation of their orbital poles. 

It is well known that an essential problem in this type of study is that the oriention of the orbital pole is ambiguous for visual and astrometric binaries because two true orbits are compatible with the apparent orbit calculated from the astrometric observations: although the orbital inclination $i$ is unambiguously determined, it is impossible to distinguish between the two possible values of the position angle $\Omega$ of the ascending node. 
The problem can only be solved if radial-velocity measurements are available for at least one component of the system. Then, two scenarios may occur:

i) For spectroscopic binaries resolved as visual or astrometric, the ascending node is immediately known unambiguously.

ii) For visual or astrometry binaries for which radial-velocity measurements are available, the correct ascending node may be selected by comparing the trend of the  measured radial velocities with the slope of the relative radial velocity curve computed from the orbit.

The conventions used in defining of the relative orbit of a binary as well as the details on the method used to solve the ambiguity of the ascending node and computating of the orientation of the orbital pole in Galactic coordinates are given in the Appendix~\ref{Sect:pole}.

\section{Data}
\label{Sect:Data}

The selection of the 95 systems in the master sample presented in Tables~\ref{Tab:starident} and \ref{Tab:starlist} was based on the
availability of a visual or an astrometric orbit in the \emph{6th Catalog of Orbit of Visual Binary Stars} at USNO \citep[6th COVBS,
http://ad.usno.navy.mil/wds/orb6.html,][]{hartkopf2001} and of a Hipparcos parallax \citep{ESA-1997} larger than 50 mas.\footnote{see
also \citet{Kirkpatrick2012}, who recently compiled an exhaustive list of stars nearer than 8~pc from the Sun, and \citet{Raghavan2010}, who
collected orbits for visual binaries with Hipparcos parallaxes larger than 40~mas and with primary spectral types F6-K3 (their Table~11).}
Table~\ref{Tab:starident} lists various identifiers for the systems, such as their designations in the WDS (J2000 coordinates and the
component designation taken in the 6th COVBS), CCDM, HIP or HIC, HD and SB9 (The 9th Catalogue of Spectroscopic Binary Orbits,
http://sb9.astro.ulb.ac.be/intro.html) catalogs, and the discoverer name. Table~\ref{Tab:starlist} lists various physical data for the
systems identified by their WDS designation (Col.~1). The parallax (Col.~2) is for all HIP entries from the revised  Hipparcos Catalogue
\citep{vanLeeuwen2007}, or from the paper listing the orbit otherwise.
The spectral type (Col.~3) is  generally taken from the Simbad database at the {\it Centre de Donn\'{e}es Astronomiques de Strasbourg} (CDS)
\citep{cds}. Columns 4 (orbital period), 5 (orbital semi-major axis), 6 (total mass of the system), 7 (mass ratio), 8 (Grade of the orbit) and 9
(reference of visual orbit) come from the 6th COVBS, or are directly derived from the catalogue data (the total mass of each system -- col.~6
-- is computed from the orbital elements and the parallax using the third Kepler law), with the exception of the mass ratio (Col.~7), which
is listed when available from SB2 or astrometry. Column~10 ('S.Orb') lists either the reference of the spectroscopic orbit used to lift the
ascending-node ambiguity, or 'New' if the spectroscopic orbit has been computed in the present paper (see Appendix~\ref{Sect:notes}), or 'VD' if only the radial-velocity drift was used to lift the ambiguity, or 'SD' if the sign of the velocity difference $V_{\mathrm{r}}$(B) - $V_{\mathrm{r}}$(A) was used to lift the ambiguity. Column~11, labelled 'A', contains a flag indicating whether the system has (or not) been considered in the final analysis: Y stands for yes, NO stands for 'no
reliable visual orbit', NV stands for 'no radial velocities', ND stands for 'no visible drift in the velocities', because either the period is
too long or the velocities too inaccurate, and IC stands for inconclusive analysis. For systems with a combined visual and spectroscopic solution computed 
by \citet{Pourbaix2000}, the orbital elements were adopted without an additional verification because the method naturally lifts the ambiguity on the ascending node. 
The last column of Table~\ref{Tab:starlist} identifies systems with an explanatory note ('n') in Appendix~\ref{Sect:notes} and triple or
higher-multiplicity systems noted ('t').

Radial-velocity measurements were searched for in the literature and among unpublished measurements obtained with the CORAVEL
spectrovelocimeter \citep{Baranne1979}. A few more radial velocities were obtained especially for this purpose with the HERMES spectrograph
installed on the Mercator 1.2m telescope \citep{Raskin2011}. In the end, 51 systems from the master list comprising 95 binaries
(Table~\ref{Tab:starlist}) have radial velocities available that enable lifting the ambiguity on the ascending node. Hence the
number of systems used in our analysis drops from 95 to 51, and the systems analysed all have $d < 18$~pc.

For each system, the ambiguity on the ascending node was solved using one of the methods described in Appendix~\ref{Thenode}. The
orbital pole orientation was been computed following the method of Appendix~\ref{Thepole}. In several cases, the radial velocities
collected from the HERMES spectrograph \citep{Raskin2011} or from the CORAVEL database (Table~\ref{Tab:VrHERMES}) enable computing
a new or updated spectroscopic binary orbit, as listed in Tables~\ref{Tab:New_SB} and \ref{Tab:SB2}. These solutions were computed
by combining the astrometric and spectroscopic data, following the method described by \citet{Pourbaix2000}. The orbital solutions are
drawn in Figs.~\ref{Fig:HIP5336} -- \ref{Fig:HIP88745}.

The positions of the orbital poles are listed in Table~\ref{Tab:poles}.
The column 'Notes' identifies the entries for which a comment is provided in Appendix~\ref{Sect:notes}. 
\medskip\\
\noindent References to visual and spectroscopic orbits listed in Table~\protect\ref{Tab:starlist}:
\medskip\\
{\small A1918b : \protect\citet{A1918b};
Abt2006: \protect\citet{Abt2006};
AST1999: \protect\citet{AST1999};
AST2001: \protect\citet{AST2001};
B1960c: \protect\citet{B1960c};
Bag2005: \protect\citet{Bag2005};
Baz1980b: \protect\citet{Baz1980b};
Chg1972: \protect\citet{Chg1972};
CIA2010: \protect\citet{CIA2010};
Cou1960b: \protect\citet{Cou1960b};
Doc1985c: \protect\citet{Doc1985c};
Doc2008d: \protect\citet{Doc2008d};
Doc2010d: \protect\citet{Doc2010d};
Dom1978: \protect\citet{Dom1978};
Dru1995: \protect\citet{Dru1995};
Duq1991b: \protect\citet{Duquennoy1991data};
Egg1956: \protect\citet{Egg1956};
Egn2008: \protect\citet{Egn2008};
Fek1983: \protect\citet{Fekel1983};
FMR2008b: \protect\citet{FMR2008b};
Frv1999: \protect\citet{Frv1999};
Gri1975: \protect\citet{Griffin1975};
Gri1998: \protect\citet{Griffin1998};
Gri2004: \protect\citet{Griffin2004};
Grr2000: \protect\citet{Grr2000};
HaI2002: \protect\citet{HaI2002};
Hei1974c: \protect\citet{Hei1974c};
Hei1984a: \protect\citet{Hei1984a};
Hei1986: \protect\citet{Hei1986};
Hei1987b: \protect\citet{Hei1987b};
Hei1988d: \protect\citet{Hei1988d};
Hei1990c: \protect\citet{Hei1990c};
Hei1990d: \protect\citet{Hei1990d};
Hei1994a: \protect\citet{Hei1994a};
Hei1994b: \protect\citet{Hei1994b};
Hei1996: \protect\citet{Heintz1996};
HIP1997d: \protect\citet{HIP1997d};
Hle1994: \protect\citet{Hle1994};
Hnk2011: \protect\citet{Hinkley2011};
Hop1958: \protect\citet{Hop1958};
Hop1973a: \protect\citet{Hop1973a};
Hrt1996a: \protect\citet{Hrt1996a};
Hrt2003: \protect\citet{Hrt2003};
Irw1996: \protect\citet{Irwin1996};
Jnc2005: \protect\citet{Jancart2005};
Kam1989: \protect\citet{Kamper1989};
Kiy2001: \protect\citet{Kiyaeva2001};
Knc2010: \protect\citet{Knc2010};
Lip1967: \protect\citet{Lip1967};
Lip1972: \protect\citet{Lip1972};
Llo2007: \protect\citet{Llo2007};
Maz2001:  \protect\citet{Mazeh2001};
Mnt2000a: \protect\citet{Mnt2000a};
Msn1995: \protect\citet{Mason1995};
Msn1999a: \protect\citet{Msn1999a};
Msn2011c: \protect\citet{Msn2011};
Mut2010b: \protect\citet{Mut2010b};
Mut2010c: \protect\citet{Mut2010c};
Nid2002: \protect\citet{Nidever2002};
Pbx2000b: \protect\citet{Pourbaix2000};
Pbx2002: \protect\citet{Pourbaix2002Cen};
PkO2006b: \protect\citet{Pk2006b};
Pop1996b: \protect\citet{Pop1996b};
Sca2002c: \protect\citet{Sca2002c};
Sca2007c: \protect\citet{Sca2007c};
Sgr2000: \protect\citet{Segransan2000};
Sgr2012: \protect\citet{Bay2012};
Sod1999: \protect\citet{Sod1999};
Str1969a: \protect\citet{Str1969a};
Str1977: \protect\citet{Str1977};
USN1988a: \protect\citet{USN1988a};
USN1988b: \protect\citet{USN1988b};
USN2002: \protect\citet{USN2002};
vAb1957: \protect\citet{vAb1957};
Wie1957: \protect\citet{Wie1957};
Zir2003: \protect\citet{Zir2003};
Zir2011: \protect\citet{Zir2011}.
}
\begin{table*}
\caption[]{
\label{Tab:starident}
Cross-identifications for the master list of 95 systems closer than 20
pc from the Sun and the are known to have an orbit in the 6th COVBS at USNO. The column labelled 'SB9'
lists the entry number in the 9th Catalogue of Spectroscopic Binary Orbits  \citep{Pourbaix2004}.
}
\begin{tabular}{llrrrl}
\hline\\
\multicolumn{1}{c}{WDS}& \multicolumn{1}{c}{CCDM}& \multicolumn{1}{c}{HIP/HIC}& \multicolumn{1}{c}{HD}& SB9& \multicolumn{1}{c}{Name}\\
         &        &   &    &     \\
\hline\\
00022+2705AB& 00022+2705AB&   171& 224930& 1468&      BU 733\\
00057+4549AB& 00057+4549AB&   473&     38&     &     STT 547\\
00184+4401AB& 00184+4401AB&  1475&   1326&     &      GRB 34\\
00321+6715Aa,Ab&  00325+6714A&    2552&      -&    -&  MCY 1\\
00321+6715AB& 00325+6714AB&  2552&      -&    -&       VYS 2\\
00373-2446AB& 00373-2446AB&  2941&   3443&    -&      BU 395\\
00491+5749AB& 00491+5749AB&  3821&   4614&    -&      STF 60\\
01032+2006AB& 01032+2006AB&  4927&      -&    -&     LDS 873\\
01083+5455Aa,Ab& 01080+5455A&  5336&   6582&   57&     WCK 1\\
01388-1758AB& 01390-1756AB&     -&      -&    -&     LDS 838\\
01398-5612AB& 01398-5612AB&  7751&  10360&    -&       DUN 5\\
01418+4237AB& 01418+4237AB&  7918&  10307& 2546&       MCY 2\\
01425+2016AB& 01425+2016AB&  7981&  10476&    -&     HJ 2071\\
02171+3413Aa,Ab& 02170+3414A& 10644&  13974&  117&       MKT\\
02278+0426AB& 02278+0426AB& 11452&  15285&    -&      A 2329\\
02361+0653Aa,P& 02361+0653A& 12114&  16160&    -&        PLQ 32\\
02442+4914AB& 02441+4913AB& 12777&  16895&    -&     STF 296\\
03095+4544AB& 03095+4544AB& 14669&      -&    -&       HDS 404\\
03121-2859AB& 03121-2859AB& 14879&  20010&    -&       HJ 3555\\
03575-0110AB& 03575-0110AB& 18512&  24916&    -&        BU 543\\
04153-0739BC& 04153-0739BC&     -&  26976&    -&     STF 518\\
04312+5858Aa,Ab& 04312+5858A& 21088&      -&    -& STN 2051\\
05025-2115AB& 05025-2115AB& 23452&  32450&    -&      DON 91\\
05074+1839AB& 05075+1839AB& 23835&  32923&    -&        A 3010\\
05167+4600Aa,Ab& 05168+4559AP& 24608&  34029&  306&    ANJ 1\\
05333+4449& 05333+4449A&     -&      -&    -&       GJ 1081\\
05544+2017& 05544+2017A& 27913&  39587& 1535&    $\chi^1$ Ori\\
06003-3102AC& 06003-3102AC& 28442&  40887&    -&     HJ 3823\\
06262+1845AB& 06262+1845AB& 30630&  45088&    -&       BU 1191\\
06293-0248AB& 06294-0249AB& 30920&      -&    -&      B 2601\\
06451-1643AB& 06451-1643AB& 32349&  48915&  416&      AGC 1\\
06579-4417AB& 06578-4417AB& 33499&      -&    -&      LPM 248\\
07100+3832&           -& 34603&      -&  434&       GJ 268\\
07175-4659AB& 07175-4659AB& 35296&  57095&    -&       I 7\\
07346+3153AB& 07346+3153AB& 36850&60178/9&    -&    STF 1110\\
07393+0514AB& 07393+0514AB& 37279&  61421&  467&       SHB 1\\
07518-1354AB& 07518-1354AB& 38382&  64096&  478&        BU 101\\
08592+4803A,BC& 08592+4803AB& 44127&  76644&    -&   HJ 2477\\
09006+4147AB& 09007+4147AB& 44248&  76943&  546&      KUI 37\\
09144+5241AB& 09144+5241AB& 45343&  79210&    -&    STF 1321\\
09313-1329AB& 09313-1329AB& 46706&      -&    -&        KUI 41\\
09357+3549AB& 09357+3549AB& 47080&  82885&    -&       HU 1128\\
10454+3831AB& 10454+3831AB& 52600&      -&    -&    HO 532\\
11182+3132AB& 11182+3132AB&     -&  98231&    -&    STF 1523\\
11182+3132Aa,Ab& 11182+3132A&     -&  98231&    -&    $\xi$~UMa\\
11247-6139AB& 11247-6139AB& 55691&  99279&    -&         BSO 5\\
12417-0127AB& 12417-0127AB& 61941& 110380&    -&    STF 1670\\
13100+1732AB& 13100+1732AB& 64241& 114378&    -&    STF 1728\\
13169+1701AB& 13169+1701AB& 64797& 115404&    -&      BU 800\\
13198+4747AB& 13198+4747AB& 65026& 115953&    -&    HU 644\\
13328+1649AB& 13328+1649AB& 66077&      -&    -&         VYS 6\\
13473+1727AB& 13473+1727AB& 67275& 120136&    -&       STT 270\\
13491+2659AB& 13491+2659AB& 67422& 120476&    -&    STF 1785\\
13547+1824& 13547+1824A& 67927& 121370&  794&       $\eta$~Boo\\
14035+1047& 14035+1047A& 68682& 122742&  799&        GJ 538\\
\hline\\
\end{tabular}
\end{table*}

\addtocounter{table}{-1}
\begin{table*}
\caption[]{
Continued.
}
\begin{tabular}{llrrrl}
\hline\\
\multicolumn{1}{c}{WDS}& \multicolumn{1}{c}{CCDM}& \multicolumn{1}{c}{HIP/HIC}& \multicolumn{1}{c}{HD}& SB9& \multicolumn{1}{c}{Name}\\
   &     &        &   &    &     \\
\hline\\
14396-6050AB& 14396-6050AB& 71683& 128620&  815&       RHD\\
14514+1906AB& 14513+1906AB& 72659& 131156&    -&    STF 1888\\
14540+2335AB& 14539+2333AB& 72896&      -&    -&         REU 2\\
14575-2125AB& 14574-2124AB& 73184& 131977&    -&      H N 28\\
14575-2125Ba,Bb& 14574-2124B& 73182& 131976& 1475&      H N 28B\\
15038+4739AB& 15038+4739AB& 73695& 133640&    -&      STF 1909\\
15232+3017AB& 15233+3018AB& 75312& 137107&  842&    STF 1937\\
15527+4227& 15527+4227A& 77760& 142373&    -&    $\chi$~Her\\
16240+4822Aa,Ab& 16240+4821A& 80346&      -& 1542&    HEN 1\\
16413+3136AB& 16413+3136AB&  81693& 150680&  915&      STF 2084\\
16555-0820AB& 16555-0820AB&  82817& 152751& 2683&      KUI 75\\
17121+4540AB& 17121+4540AB&  84140& 155876&    -&      KUI 79\\
17153-2636AB& 17155-2635AB&  84405& 155885/6&  -&     SHJ 243\\
17190-3459AB& 17190-3459AB&  84709& 156384&    -&       MLO 4\\
17191-4638AB& 17191-4638AB&  84720& 156274&    -&      BSO 13\\
17304-0104AB& 17304-0104AB&  85667& 158614&  969&      STF 2173\\
17349+1234& 17349+1234A&  86032& 159561&    -&         MCY 4\\
17350+6153AB& 17351+6152AB&  86036& 160269& 2557&      BU 962\\
17364+6820Aa,Ab& 17366+6822A&  86162&      -&    -&   CHR 62\\
17465+2743BC& 17465+2744BC&  86974& 161797&    -&        AC 7\\
18055+0230AB& 18055+0230AB&  88601& 165341& 1122&    STF 2272\\
18070+3034AB& 18071+3034AB&  88745& 165908&    -&       AC 15\\
18211+7244Aa,Ab& 18211+7245A&  89937& 170153& 1058&    LAB 5\\
18428+5938AB& 18428+5938A&  91768& 173739&    -&    STF 2398\\
18570+3254AB& 18570+3254AB&  93017& 176051& 2259&      BU 648\\
19074+3230Ca,Cb& 19075+3231C&      -&      -&    -&   KUI 90\\
19121+0254AB& 19122+0253A&  94349&      -&    -&         AST 1\\
19167-4553AB& 19167-4553AB&  94739& 179930&    -&      RST 4036\\
19255+0307Aa,Ab& 19254+0307A&  95501& 182640&    -&    BNU 6\\
20452-3120BC&           -& 102141& 196982&    -&     LDS 720\\
21000+4004AB& 21001+4004A& 103655&      -& 1280&       KUI 103\\
21069+3845AB& 21069+3844AB& 104214& 201091/2&  -&     STF 2758\\
21313-0947AB& 21313-0947AB& 106255&      -&    -&         BLA 9\\
21379+2743Aa,Ab& 21380+2743A& 106811&      -&    -& HDS 3080\\
22070+2521& 22070+2520A& 109176& 210027& 1354&     $\iota$ Peg\\
22234+3228AB& 22235+3228AB& 110526&      -&    -&        WOR 11\\
22280+5742AB& 22281+5741AB& 110893& 239960&    -&       KR 60\\
22385-1519AB&            -&      -&      -&    -&        BLA 10\\
23317+1956AB& 23319+1956AB& 116132&      -&    -&       WIR 1\\
23524+7533AB& 23526+7532AB& 117712& 223778&    -&      BU 996\\
\hline\\
\end{tabular}
\end{table*}

\begin{landscape}
\begin{table}
\caption[]{
\label{Tab:starlist}
Data and references for the 95 systems of the master list. See text for the description of the column contents.
}
\begin{tabular}{lrlrrrrrllll}
\hline\\
\multicolumn{1}{c}{WDS}&$\varpi$& Sp& $P$ & $a$& $M_{tot}$ & $\frac{M_{B}}{M_{A}}$& Gr& V.Orb.& S.Orb.& A& Notes\\
          & (mas) &     & (yr) &  ('') & ($M_\odot$) \\
\hline\\
00022+2705AB&     82.2&        G5 Vb&  26.27&   0.83&  1.49&    -& 2&  Sod1999&      Gri2004&      Y&  -\\
00057+4549AB&     88.4&         K6 V&  509.7&   6.21&  1.33&    -& 4&  Kiy2001&           SD&      Y& nt\\
00184+4401AB&    278.8&        M2.0V&   2600&  41.15&  0.48&    -& 5&  Lip1972&            -& NO, ND&  n\\
00321+6715Aa,Ab&  99.3&       M2.5Ve&  15.64&  0.511&  0.56&    -& 4& Doc2008d&            -&     NV&  t\\
00321+6715AB&     99.3&       M2.5Ve&  222.3&  3.322&  0.76&    -& 5& Doc2008d&            -&     NV& nt\\
00373-2446AB&     64.9&      G8V+G8V&  25.09&  0.667&  1.72&    -& 1& Pbx2000b&     Pbx2000b&      Y&  -\\
00491+5749AB&    168.0&          G3V&    480&  11.99&  1.58&    -& 3& Str1969a&           VD&      Y&  -\\
01032+2006AB&     61.6&           M2&  521.4&  3.366&  0.60&    -& 4& FMR2008b&            -&     NV&  -\\
01083+5455Aa,Ab& 132.4&         G5Vb&  21.75&  1.009&  0.94&    -& 4&  Dru1995&          New&      Y& nt\\
01388-1758AB&    373.7&  M5.5V+M6.0V&  26.52&   1.95&  0.20&    -& 3& USN1988b&            -&     NV&  -\\
01398-5612AB&    127.8&      K0V+K5V&  483.6&  7.817&  0.98&    -& 5&  vAb1957&            -& NO, ND&  n\\
01418+4237AB&     78.5&        G1.5V&   19.5&   0.58&  1.06&    -& 4&      New&          New&      Y&  n\\
01425+2016AB&    132.8&          K1V&  0.567&  0.005&     -&    -& 9& HIP1997d&            -&     ND&  n\\
02171+3413Aa,Ab&  92.7&        G0.5V&  0.027&  0.010&  1.57& 0.89& 1& Pbx2000b&     Pbx2000b&      Y&  t\\
02278+0426AB&     58.3&          M1V&  25.14&  0.543&  1.80& 0.89& 1&      New&          New&      Y&  n\\
02361+0653Aa,P&  139.3&          K3V&   61.0&  0.171&     -&    -& 9& Hei1994b&            -&     ND& nt\\
02442+4914AB&     89.9&          F7V&   2720&  22.29&  2.06&    -& 5&  Hop1958&            -& NO, ND&  n\\
03095+4544AB&     63.4&           M2&  28.31&  0.569&  0.90&    -& 4&  Bag2005&            -&     NV&  -\\
03121-2859AB&     70.2&          F6V&  269.0&   4.00&  2.55&    -& 4&  Sod1999&           VD&      Y&  -\\
03575-0110AB&     64.4&          K4V&   3200&   11.1&  0.50&    -& 5&  Dom1978&            -& NO, ND&  n\\
04153-0739BC&    198.2&           DA&  252.1&  6.943&  0.68&    -& 4& Hei1974c&            -& NO, ND& nt\\
04312+5858Aa,Ab& 179.3&           M4&   23.0&  0.070&     -&    -& 9&  Str1977&            -&     NV& nt\\
05025-2115AB&    116.6&          M0V&   44.0&   1.80&  1.90&    -& 4&  Sod1999&            -&     NV&  -\\
05074+1839AB&     64.8&          G4V&   1.19&  0.180&  15.1&    -& 3&  Egg1956&            -& NO, ND&  n\\
05167+4600Aa,Ab&  76.2& G5IIIe+G0III&  0.285&  0.056&  5.02& 0.95& 1& Pbx2000b&     Pbx2000b&      Y&  -\\
05333+4449&       65.2&           M2&  11.40&  0.041&     -&    -& 9& USN1988a&            -&     NV&  -\\
05544+2017&      115.4&          G0V&  14.12&  0.090&     -&    -& 9&  HaI2002&      Nid2002&      Y&  -\\
06003-3102AC&     61.0&        K6.5V&  390.6&   3.95&  1.78&    -& 5& Baz1980b&            -&     NV& nt\\
06262+1845AB&     67.9&         K3Vk&    600&   3.23&  0.30&    -& 5&  Hle1994&            -&     NO& nt\\
06293-0248AB&    242.3&        M4.5V&  16.12&   1.04&  0.30& 0.51& 4&  Sgr2000&      Sgr2000&      Y&  n\\
06451-1643AB&    379.2&          A1V&  50.09&   7.50&  3.08&    -& 2&   B1960c&       A1918b&      Y&  -\\
06579-4417AB&    124.8&         M0.3&  298.3&  4.396&  0.49&    -& 5&  Zir2003&            -&     NV&  n\\
07100+3832&      158.9&        M4.5V&  0.028&  0.011&  0.42& 0.85& 8&  Sgr2012&      Sgr2012&      Y&  n\\
07175-4659AB&     68.5&          K3V&   85.2&   1.00&  0.43&    -& 4& Mns2011c&            -&     NV&  n\\
07346+3153AB&     64.1&         A2Vm&  444.9&  6.593&  5.49&    -& 3& Doc1985c&           VD&      Y& nt\\
07393+0514AB&    284.5&       F5IV-V&  40.82&  4.271&  2.03&    -& 3&  Grr2000& Duq1991b, VD&      Y&  n\\
07518-1354AB&     60.6&          G0V&  22.70&  0.602&  1.90& 0.97& 2& Pbx2000b&     Pbx2000b&      Y&  -\\
08592+4803A,BC&   68.9&          A7V&  817.9&  9.092&  3.43&    -& 5& Hop1973a&            -& NO, ND& nt\\
\hline\\
\end{tabular}
\end{table}
\end{landscape}

\begin{landscape}
\addtocounter{table}{-1}
\begin{table}
\caption[]{
Continued.
}
\begin{tabular}{lrlrrrrrllll}
\hline\\
 \multicolumn{1}{c}{WDS}& $\varpi$& Sp& $P$ & $a$& $M_{tot}$ & $q$& Gr& V.Orb.& S.Orb.& A& Notes\\
          & (mas) &     & (yr) &  ('') & ($M_\odot$) \\
\hline\\
09006+4147AB&     62.2&    F4V + K0V&  21.77&  0.633&  3.42& 0.98& 1&      New&          New&      Y&  n\\
09144+5241AB&    172.1&        M0.0V&    975&  16.72&  0.97&    -& 4&  Chg1972&           VD&      Y&  n\\
09313-1329AB&     99.9&          M2V&  18.40&  0.630&  0.74& 0.94& 3&  Sod1999&            -&     ND&  n\\
09357+3549AB&     87.9&       G8IIIv&    201&  3.840&  2.06&    -& 5& Hei1988d&            -& NO, NV&  n\\
10454+3831AB&     71.9&          M2V&  160.7&  1.791&  0.60&    -& 4& Mnt2000a&            -&     NV& nt\\
11182+3132AB&    127.0&          G0V&  59.88&  2.536&  2.24& 1.15& 1&  Msn1995&      Gri1998&      Y&  nt\\
11182+3132Aa,Ab& 127.0&          G0V&  1.834&  0.054&  1.20& 0.44& 9& Hei1996 &      Gri1998&      Y&  nt\\
11247-6139AB&     72.9&     K5V+M0Ve&  399.4&  5.672&  2.95&    -& 4& Sca2002c&            -&     NV&  n\\
12417-0127AB&     85.6&      F0V+F0V& 169.10&  3.639&  2.69&    -& 2& Sca2007c&            -&     NV&  -\\
13100+1732AB&     56.1&      F5V+F5V&  25.97&  0.661&  2.43&    -& 1& Mut2010b&            -&     ND&  n\\
13169+1701AB&     90.3&          K2V&    770&   8.06&  1.20&    -& 4&  Hle1994&            -&     NO&  -\\
13198+4747AB&     93.4&          K0V&  48.91&  1.507&  1.76&    -& 2& Msn1999a&            -&     IC& nt\\
13328+1649AB&     78.4&        M2.5V&    430&   3.17&  0.36&    -& 5& Hei1990d&            -&     NV&  -\\
13473+1727AB&     64.3&         F6IV&   2000&  14.39&  2.80&    -& 5&  Hle1994&            -& NO, ND&  n\\
13491+2659AB&     74.6&      K4V+K6V&  155.7&  2.433&  1.43&    -& 2& Hei1988d&           SD&      Y&  n\\
13547+1824&       87.7&         G0IV&  1.353&  0.036&     -&    -& 9&  Jnc2005&      Jnc2005&      Y&  n\\
14035+1047&       58.9&          G8V&  9.897& 0.1053&     -&    -& 9& HIP1997d&      Nid2002&      Y&  -\\
14396-6050AB&    754.8&      G2V&  79.91&  17.57&  1.98& 0.85& 2&  Pbx2002&      Pbx2002&      Y&  -\\
14514+1906AB&    149.0&      G8V&  151.6&   4.94&  1.59&    -& 2&  Sod1999&           VD&      Y&  n\\
14540+2335AB&     98.4&      M3V&  34.10&   0.69&  0.30&    -& 4& Hei1990c&            -&     NV&  -\\
14575-2125AB&    171.2&      K4V&   2130&  32.34&  1.49&    -& 5&  Hle1994&            -& NO, ND& nt\\
14575-2125Ba,Bb& 168.8&    M1.5V&  0.846&  0.151&  1.00& 0.73& 3&  Frv1999&     Pbx2000b&      Y&  t\\
15038+4739AB&     79.9&    G0Vnv&   209.8&   3.666& 2.19&    -& 2&  Zir2011&           VD&      Y& nt\\
15232+3017AB&     56.0&      G2V&   41.63&   0.862& 2.11& 0.89& 1& Mut2010b&           VD&      Y&  -\\
15527+4227&       62.9&     F8Ve&   0.168&   0.001&    -&    -& 9& HIP1997d&            -& NO, ND&  n\\
16240+4822Aa,Ab& 124.1&    M0.3V&    3.74&   0.237& 0.50&    -& 2&  Llo2007&      Nid2002&      Y&  t\\
16413+3136AB&     93.3&     G0IV&   34.45&    1.33& 2.44&    -& 1&  Sod1999&             new&      Y& nt\\
16555-0820AB&    161.4&   M3.5Ve&   1.717&   0.226& 1.05& 1.61& 1&  Maz2001&         Maz2001&      Y& nt\\
17121+4540AB&    167.3&      K5V&   12.95&   0.762& 0.56&    -& 2& Hrt1996a&             New&      Y&  n\\
17153-2636AB&    168.5&      K2V&   568.9&    14.7&  1.5&    -& 4&  Irw1996&         Irw1996&      Y&  n\\
17190-3459AB&    146.3&  K3V+K5V&   42.15&    1.81& 1.07&    -& 2&  Sod1999&               -&     NV&  -\\
17191-4638AB&    113.6&      G8V&   693.2&   10.45& 1.62&    -& 5&  Wie1957&               -&     NV&  n\\
17304-0104AB&     61.2&   G9IV-V&    46.4&    0.98& 1.91& 0.92& 1& Hei1994a&        Pbx2000b&      Y&  n\\
17349+1234&       67.1& A5IV+K2V&   8.620&   0.427& 3.46& 0.36& 5&  Hnk2011&         Kam1989&      Y&  n\\
17350+6153AB&     70.5&     G0Va&    76.1&    1.53& 1.77&    -& 3&  Sod1999&    Duq1991b, VD&      Y&  -\\
17364+6820Aa,Ab& 220.8&    M3.5V&    24.5&   0.102&    -&    -& 9&  Lip1967&               -&     NV& nt\\
17465+2743BC&    120.3&    M3.5V&    43.2&   1.360& 0.77&    -& 2& Cou1960b&              VD&      Y& nt\\
\hline\\
\end{tabular}
\end{table}
\end{landscape}

\begin{landscape}
\addtocounter{table}{-1}
\begin{table}
\caption[]{
Continued.
}
\begin{tabular}{lrlrrrrrllll}
\hline\\
 \multicolumn{1}{c}{WDS}& $\varpi$& Sp& $P$ & $a$& $M_{tot}$ & $q$& Gr& V.Orb.& S.Orb.& A& Notes\\
          & (mas) &     & (yr) &  ('') & ($M_\odot$) \\
\hline\\
18055+0230AB&    196.7&      K0V&   88.37&   4.526& 1.56& 0.87& 1&  Egn2008&        Pbx2000b&      Y&  -\\
18070+3034AB&     63.9&      F7V&    56.4&    1.00& 1.20&    -& 2&  Sod1999&             New&      Y& nt\\
18211+7244Aa,Ab& 124.1&      F7V&   0.768&   0.124& 1.71& 0.78& 1&  CIA2010&        Pbx2000b&      Y&  t\\
18428+5938AB&    280.2&    M3.0V&     408&   13.88& 0.73&    -& 4& Hei1987b&               -&     ND&  n\\
18570+3254AB&     67.2&  G0V+K1V&   61.41&   1.276& 1.81&    -& 2& Mut2010e&Duq1991b,Abt2006&      Y&  n\\
19074+3230Ca,Cb& 120.2&      M3V&    5.77&   0.288&  0.4& 0.93& 4&  Sgr2000&         Sgr2000&      Y&  t\\
19121+0254AB&     97.8&    M3.5V&   2.466&   0.148& 0.57& 0.51& 2&  AST1999&         AST2001&      Y&  n\\
19167-4553AB&     62.1&     M0Vk&    7.70&   0.260& 1.24&    -& 3&  Sod1999&               -&     NV&  -\\
19255+0307Aa,Ab&  64.4&     F0IV&   3.426&   0.056&    -&    -& 9& HIP1997d&         Kam1989&      Y& nt\\
20452-3120BC&     93.5&     M4Ve&     209&    3.18& 0.90&    -& 5&  Hrt2003&               -&     ND&  t\\
21000+4004AB&     65.4&    M3.0V&   29.16&   0.725& 1.60& 0.32& 4& Doc2010d&        Pbx2000b&      Y& nt\\
21069+3845AB&    286.8&      K5V&     678&   24.27& 1.32& 0.62& 4& PkO2006b&              SD&      Y&  n\\
21313-0947AB&    117.5&   M4.5Ve&    1.92&   0.141& 0.45& 0.56& 4&  Sgr2000&         Sgr2000&      Y&  -\\
21379+2743Aa,Ab&  75.1&      M0V&    68.0&    1.50&    -&    -& 9&  Sod1999&               -&     NV&  t\\
22070+2521&       85.3&      F5V&   0.028&   0.010& 2.27& 0.62& 8&  Knc2010&         Fek1983&      Y&  -\\
22234+3228AB&     64.5&     M0Ve&    90.0&   1.401& 1.27&    -& 5&  USN2002&               -&     NV&  n\\
22280+5742AB&    249.9&  M3V+M4V&   44.67&   2.383& 0.45& 0.65& 2&  Hei1986&             SD &      Y&  n\\
22385-1519AB&    293.6&      M6V&    2.25&   0.347& 0.33& 0.54& 2&  Sgr2000&         Sgr2000&      Y&  -\\
23317+1956AB&    161.8&    M3.5V&     359&    6.87& 0.59&     & 5& Hei1984a&               -&  NO,NV&  n\\
23524+7533AB&     91.8&      K3V&     290&    4.14& 1.09&    -& 5&  Hle1994&               -&  NO,NV& nt\\
\hline\\
\end{tabular}
\end{table}
\end{landscape}

\begin{landscape}
\begin{table}
\caption[]{
\label{Tab:poles}
Position of the 51 orbital poles in Galactic coordinates for systems closer than 18 pc. $\omega$ is the position of the longitude of periastron of the visual orbit (B with respect to A; it differs by 180$^\circ$ from the value for the spectroscopic orbit of A with respect to the centre of mass). 
$N$ ranks the systems in order of increasing distance.
}
\begin{tabular}{lrrrrrrrrrrrrrr}
\hline\\
 & \multicolumn{3}{c}{Geometrical elements} &   \multicolumn{2}{c}{System coordinates} & Parallax & \multicolumn{5}{c}{Pole} & Distance & $N$  & Note\\
\cline{5-6}\cline{8-12}
\multicolumn{1}{c}{WDS} & $i$ & $\Omega$ & $\omega$ & $\alpha$ &  $\delta$ & &     $l$ & $b$  & $x$ & $y$ & $z$ &  \\
& ($^{\circ}$) & ($^{\circ}$) & ($^{\circ}$) & ($^{\circ}$) & ($^{\circ}$) & (mas) &  ($^{\circ}$) & ($^{\circ}$) & & & & (pc)\\
\hline\\
00022+2705AB     &  49.0& 290.0&  96.0&   0.542&  27.082&  82.2& 223.5&  63.0& -0.32904& -0.31178&  0.89136& 12.2 & 32 & -\\
00057+4549AB     &  54.9&  13.5& 267.2&   1.418&  45.812&  88.4& 245.5&  -9.2& -0.40949& -0.89835& -0.15902& 11.3 & 29 & nt\\
00373-2446AB     &  78.6& 292.0& 318.1&   9.332& -24.767&  64.9& 144.2&  12.4& -0.79180&  0.57162&  0.21519& 15.4 & 42 & - \\
00491+5749AB     &  34.8& 278.4& 268.6&  12.271&  57.817& 168.0& 296.0&  39.3&  0.33970& -0.69529&  0.63339&  6.0 & 12 & -\\
01083+5455Aa,Ab  & 104.7& 224.6& 330.0&  17.068&  54.920& 132.4&  60.2&  36.3&  0.40124&  0.69932&  0.59157&  7.6 & 17 & nt\\
01418+4237AB     & 105.0&  33.0&  22.0&  25.444&  42.614&  78.5& 214.6& -24.5& -0.74952& -0.51624& -0.41438& 12.7 & 34 & n\\
02171+3413Aa,Ab  & 163.0&  37.0& 171.0&  34.260&  34.225&  92.7& 161.2& -28.8& -0.82900&  0.28284& -0.48246& 10.8 & 28 & t\\
02278+0426AB     &  73.0& 290.1&  49.1&  36.941&   4.432&  58.3& 138.1&  54.2& -0.43560&  0.39148&  0.81055& 17.1 & 50 & n\\
03121-2859AB     &  81.0& 117.0&  43.0&  48.018& -28.989&  70.9& 347.2&  -7.8&  0.96589& -0.22017& -0.13634& 14.1 & 37 & t\\
05167+4600Aa,Ab  & 137.2& 220.9& 269.0&  79.172&  45.999&  76.2& 121.1&  -6.0& -0.51300&  0.85197& -0.10480& 13.1 & 35 & -\\
05544+2017       &  95.9& 126.4& 291.5&  88.596&  20.276& 115.4& 269.8& -66.5& -0.00114& -0.39882& -0.91703&  8.7 & 25 & -\\
06293-0248AB     &  51.8& 210.7&  43.0&  97.348&  -2.814& 242.3&  78.1& -20.5&  0.19297&  0.91631& -0.35091&  4.1 &  7 & n\\
06451-1643AB     & 136.5&  44.6& 147.3& 101.289& -16.713& 379.2& 268.0&   6.7& -0.03496& -0.99260&  0.11628&  2.6 &  2 & -\\
07100+3832       & 100.4& 270.0&  32.0& 107.508&  38.529& 159.0&  92.6&  21.6& -0.04169&  0.92898&  0.36777&  6.3 & 15 & n\\
07346+3153AB     & 114.6&  41.5& 253.3& 113.650&  31.889&  64.1& 262.4&  35.1& -0.10871& -0.81102&  0.57482& 15.6 & 44 & nt\\
07393+0514AB     &  31.1& 277.3& 272.2& 114.827&   5.228& 284.6&  59.2&   5.1&  0.51046&  0.85536&  0.08824&  3.5 &  4 & n\\
07518-1354AB     &  80.4& 102.9&  73.1& 117.943& -13.897&  60.6& 329.2& -43.4&  0.62419& -0.37154&  -0.68727& 16.5 & 49 & -\\
09006+4147AB     & 133.2&  23.8& 212.7& 135.161&  41.783&  62.2& 255.3&  71.0& -0.08263& -0.31472&  0.94558& 16.1 & 46 & n\\
09144+5241AB     &  21.0& 173.8&  44.0& 138.595&  52.687& 172.1& 332.0& -62.2&  0.41137& -0.21885& -0.88480&  5.8 & 9 & n\\
11182+3132A      &  91.0& 318.0& 144.0& 169.546&  31.529& 127.0&  65.6&  14.6&  0.40030&  0.88102&  0.25212&  7.9 & 18 & nt\\
11182+3132AB     & 122.1& 101.9& 127.9& 169.546&  31.529& 127.0& 267.6&  27.9& -0.03655& -0.88288&  0.46818&  7.9 & 19 & nt\\
13491+2659AB     &  47.4& 335.4&  20.0& 207.268&  26.980&  74.6&  71.8& -53.5&  0.18603&  0.56546& -0.80352& 13.4 & 36 & n\\
13547+1824       & 115.7&  75.2& 146.3& 208.671&  18.399&  87.8& 327.7&  12.6&  0.82479& -0.52194&  0.21748& 11.4 & 30 & n\\
14035+1047       &  93.5& 252.3&   9.0& 210.885&  10.788&  58.9& 143.9&  24.5& -0.73486&  0.53655&  0.41484& 17.0 & 49 & -\\
14396-6050AB     &  79.2& 204.9& 231.7& 219.920& -60.835& 754.8& 210.3&  47.8& -0.57995& -0.33945&  0.74056&  1.3 &  1 & -\\
14514+1906AB     & 139.0& 347.0& 203.0& 222.847&  19.101& 149.0&  47.3&  23.9&  0.62020&  0.67219&  0.40436&  6.7 & 16 & n\\
14575-2125Ba,Bb  & 107.6&  15.9& 307.6& 224.358& -21.407& 168.8&  22.1& -26.5&  0.82935&  0.33632& -0.44616&  5.9 & 11 & t\\
15038+4739AB     &  83.6&  57.1&  39.9& 225.949&  47.654&  80.0& 352.0&  -8.8&  0.97878& -0.13669& -0.15263& 12.5 & 33 & nt\\
15232+3017AB     &  58.1&  22.8& 219.9& 230.801&  30.288&  56.0&  12.1& -62.0&  0.45854&  0.09852& -0.88320& 17.9 & 51 & -\\
16240+4822Aa,Ab  & 154.0& 278.5&  68.7& 246.035&  48.354& 124.1& 105.1&  32.5& -0.22040&  0.81412&  0.53725&  8.1 & 20 & t\\
16413+3136AB     & 132.0& 235.2& 297.0& 250.323&  31.602&  93.3& 124.7&  56.6& -0.31360&  0.45246&  0.83483& 10.7 & 27 & nt\\
16555-0820AB     & 163.1& 163.2& 115.6& 253.872&  -8.334& 161.4& 355.9&  31.2&  0.85285& -0.06111&  0.51857&  6.2 & 14 & nt\\
\hline\\
\end{tabular}
\end{table}
\end{landscape}

\begin{landscape}
\addtocounter{table}{-1}
\begin{table}
\caption[]{
Continued.
}
\begin{tabular}{lrrrrrrrrrrrrrr}
\hline\\
 & \multicolumn{3}{c}{Geometrical elements} &   \multicolumn{2}{c}{System coordinates} & Parallax & \multicolumn{5}{c}{Pole} & Distance & $N$  & Note\\
\cline{5-6}\cline{8-12}
\multicolumn{1}{c}{WDS} & $i$ & $\Omega$ & $\omega$ & $\alpha$ &  $\delta$ & &     $l$ & $b$  & $x$ & $y$ & $z$ &  \\
& ($^{\circ}$) & ($^{\circ}$) & ($^{\circ}$) & ($^{\circ}$) & ($^{\circ}$) & (mas) &  ($^{\circ}$) & ($^{\circ}$) & & & & (pc)\\
\hline\\
 17121+4540AB     & 149.1& 160.0&  99.0& 258.032&  45.670& 167.3&  45.2&  62.2&  0.32882&  0.33098&  0.88449&  6.0 & 13 & n\\
 17153-2636AB     &  99.6& 255.1& 276.4& 258.839& -26.600& 167.1&  81.6&  50.3&  0.09338&  0.63199&  0.76933&  6.0 & 10 & n\\
 17304-0104AB     &  99.1& 332.3& 148.1& 262.599&  -1.063&  61.2&  91.6& -28.1& -0.02432&  0.88198& -0.47066& 16.3 & 47 & n\\
 17349+1234       & 125.0& 232.0& 162.0& 263.733&  12.561&  67.1&  92.4&  62.9& -0.01906&  0.45448&  0.89055& 14.9 & 39 & n\\
 17350+6153AB     & 104.0& 151.0& 307.0& 263.747&  61.876&  70.5& 329.1&  59.8&  0.43199& -0.25880&  0.86395& 14.2 & 38 & -\\
 17465+2743BC     &  66.2& 240.7& 354.0& 266.605&  27.717& 120.3& 189.4&  26.0& -0.88700& -0.14733&  0.43764&  8.3 & 22 & nt\\
 18055+0230AB     & 121.1& 301.7&  13.4& 271.363&   2.502& 196.7&  88.6&   2.0&  0.02510&  0.99905&  0.03564&  5.1 &  8 & -\\
 18070+3034AB     &  39.8& 227.2& 290.7& 271.757&  30.562&  64.0& 219.0&  13.4& -0.75569& -0.61253&  0.23180& 15.6 & 45 & nt\\
 18211+7244Aa,Ab  &  74.4&  50.3& 299.3& 275.260&  72.734& 124.1&  18.3& -43.3&  0.69070&  0.22838& -0.68613&  8.1 & 21 & t\\
 18570+3254AB     & 114.2&  48.8& 281.7& 284.256&  32.902&  67.2&  30.2& -45.4&  0.60754&  0.35323& -0.71142& 14.9 & 40 & n\\
 19074+3230Ca,Cb  &  77.3& 265.1& 150.9& 286.928&  32.544& 120.2& 174.1&  26.0& -0.89409&  0.09184&  0.43837&  8.3 & 23 & t\\
 19121+0254AB     & 131.5& 358.8& 206.5& 288.056&   2.888&  97.8&  67.8& -44.0&  0.27214&  0.66594& -0.69460& 10.2 & 26 & n\\
 19255+0307Aa,Ab  & 150.0& 337.0& 191.0& 291.374&   3.115&  64.4&  64.7& -24.0&  0.39046&  0.82565& -0.40725& 15.5 & 43 & nt\\
 21000+4004AB     &  53.0&  96.8& 156.1& 315.020&  40.071&  65.4& 308.7& -23.9&  0.57205& -0.71280& -0.40579& 15.3 & 41 & nt\\
 21069+3845AB     &  51.0& 178.0& 149.0& 316.725&  38.749& 286.8& 305.8&  38.3&  0.45917& -0.63651&  0.61969&  3.5 &  5 & n\\
 21313-0947AB     &  46.0& 144.4& 188.7& 322.827&  -9.791& 120.5& 286.5&  44.2&  0.20316& -0.68798&  0.69671&  8.3 & 24 & -\\
 22070+2521       &  95.8& 356.3&  92.8& 331.752&  25.345&  85.3& 182.8& -35.3& -0.81561& -0.03923& -0.57727& 11.7 & 31 & -\\
 22280+5742AB     & 167.2& 334.5&  31.0& 336.998&  57.696& 249.9& 117.4&  -1.3& -0.46041&  0.88741& -0.02292&  4.0 &  6 & n\\
 22385-1519AB     & 112.6& 161.2& 338.7& 339.64 & -15.299& 294.3&   6.0&   1.8&  0.99401&  0.10469&  0.03149&  3.4 &  3 & -\\
\hline\\
\end{tabular}
\end{table}
\end{landscape}

\section{Distribution of the orbital poles on the sky}
\label{Sect:Distribution}

\begin{figure} \centering 
\includegraphics[width=9cm]{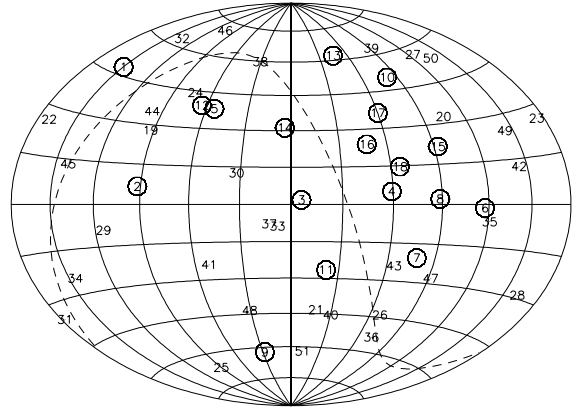}
\caption{Positions of the orbital poles on the sky, with the grid representing Galactic coordinates. The Galactic centre is at the centre of the grid. 
The systems are labelled by their numbers of Table~\ref{Tab:poles}.
The circled labels identify the systems up to $N = 18$ whose distribution is the most significantly different from isotropic according to Fig.~\protect\ref{Fig:statistics}. The dashed line displays the celestial equator. For comparison, the ecliptic pole is located  at $l=96.4^\circ, b = 30^\circ$. }
\label{Fig:aitoff} 
\end{figure}

Figure~\ref{Fig:aitoff} displays the distribution on the sky in galactic coordinates of the orbital poles listed in Table~\ref{Tab:poles}, 
for systems up to 18~pc.
After estimating in Sect.~\ref{Sect:error} the typical errors affecting the orbital-pole positions, we describe in Sect.~\ref{Sect:statistics} the available statistical tools for studying the possible deviation from isotropy of the distribution of orbital poles on the sky.

\subsection{Estimating the error on the orbital-pole positions} 
\label{Sect:error}

\begin{figure} \centering 
\includegraphics[width=9cm]{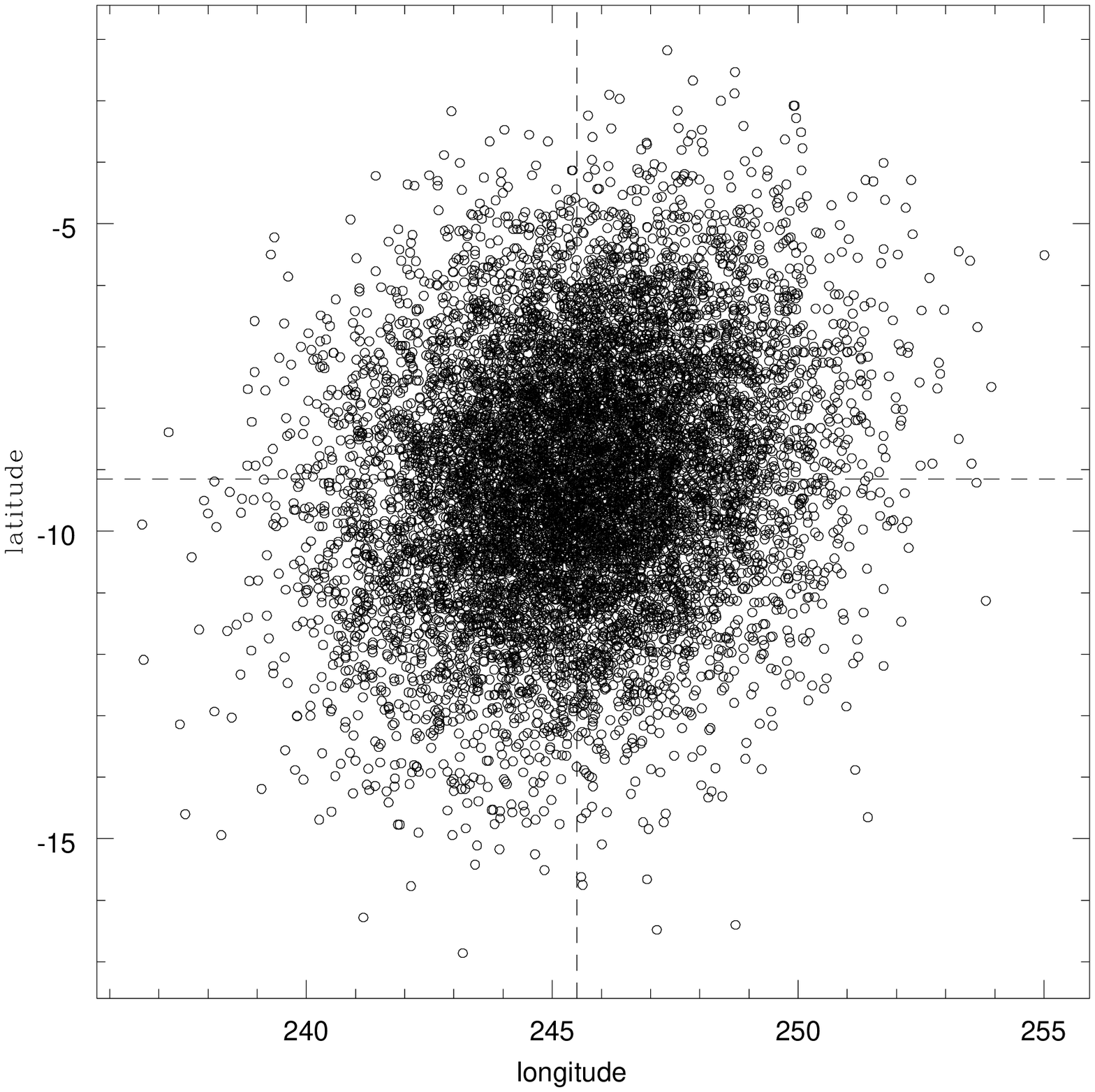}
\includegraphics[width=9cm]{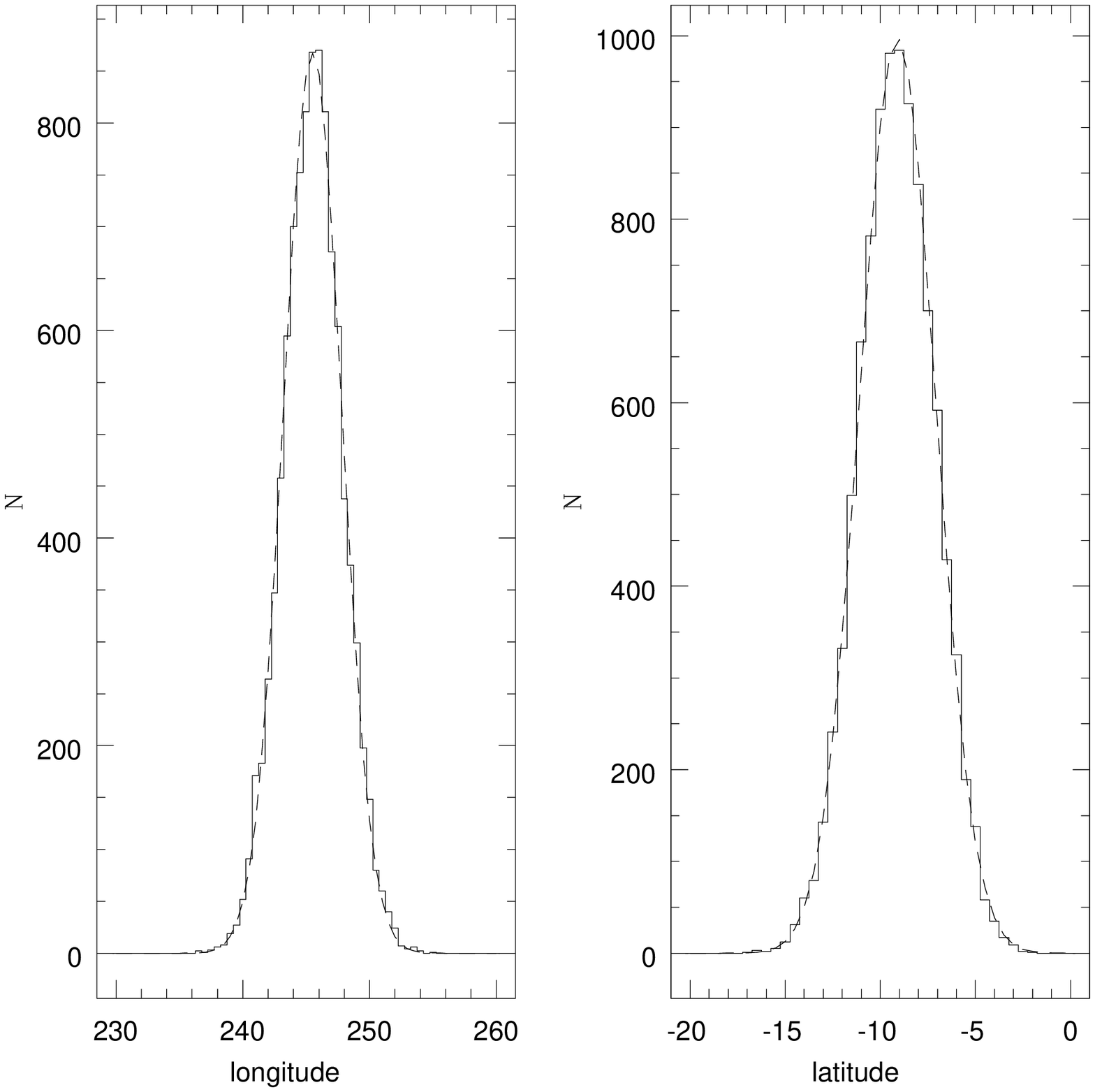}
\caption{Top panel: The distribution of the orbital-pole positions on the sky, drawn from the normal distributions for $i = 54.9^\circ\pm2.4^\circ$ and $\Omega = 13.5^\circ\pm2.3^\circ$ corresponding to the benchmark system WDS~00057+4549. The intersection of the dashed lines marks the nominal pole position. As expected, the distribution is centred on this nominal position. Bottom panel: The histograms corresponding to the longitude and latitude distributions. The dashed lines correspond to Gaussian distributions centred on the nominal pole position, with standard deviations of $2.3^\circ$ and $2.0^\circ$ in Galactic longitude and latitude, respectively.}
\label{Fig:pole_Monte_Carlo} 
\end{figure}

To evaluate the typical error affecting the orbital-pole positions, we used a Monte Carlo approach because of the complexity of the algebraic relations that link the orbital elements and the system position to the orbital-pole position (Eqs.~\ref{EquCoord} to \ref{Galcoord2}). We used the system WDS~00057+4549 (HD~38) as a test case  because it is one of the systems with the longest-period for which we lifted the ascending-node ambiguity. 
Even though the error on the orbital period is large ($P = 509.6\pm100.0$~yr), only the uncertainties on the inclination $i$ and on the longitude of the ascending node $\Omega$ affect the uncertainty on the pole position. For  WDS 00057+4549, we therefore adopted:
$\alpha = 1.4209^\circ$, $\delta = 45.81^\circ$, $i = 54.9^\circ\pm2.4^\circ$, and $\Omega = 13.5^\circ\pm2.3^\circ$, according to the orbit computed by \citet{Kiyaeva2001}. We drew $10^4$ ($i, \Omega$) pairs from this average and standard-deviation values, assuming that they follow a normal distribution; we then derived the orbital pole position for each of these cases from Eqs.~\ref{EquCoord} to \ref{Galcoord2}. 
The results are displayed in Fig.~\ref{Fig:pole_Monte_Carlo}. The Gaussian fits to the resulting longitude and latitude distributions (bottom panel), show -- as expected -- that the uncertainties on $i$ and $\Omega$ directly propagate onto the polar position ($2.3^\circ$ and $2.0^\circ$ in Galactic longitude and latitude, respectively, for WDS~00057+4549). If the uncertainties on $i$ and $\Omega$ are unequal, the resulting distribution of the orbital poles will simply be tilted with respect to the axis frame. In the analysis of the pole distribution presented below, one should therefore keep in mind that these positions are subject to an uncertainty of the same order as that on $i$ and $\Omega$.

\subsection{Available statistical tools} 
\label{Sect:statistics}

The current section is mainly based on \citet{Gillett1988}, \citet{Upton1989}, and \citet{Briggs1993}. In this section, we discuss how to test the null
hypothesis that the pole distribution is isotropic. The simplest test is the Rayleigh test \citep{Fisher1953,Briggs1993}. For the purpose of Rayleigh statistics,
the $i=1,...N$ poles are represented by unit vectors $(x_i, y_i, z_i)$ (with $x_i^2 + y_i^2 +z_i^2 = 1$), and one computes 
\begin{equation} 
\label{Eq:dipole}
\mathcal{R}^2 =
\left(\Sigma_{i=1}x_i\right)^2 + \left(\Sigma_{i=1}y_i\right)^2 +
\left(\Sigma_{i=1}z_i\right)^2, 
\end{equation} 
where $\mathcal{R}$ representing the module of the vectorial sum of the  $N$ unit vectors $(x_i, y_i, z_i)$. 
This vectorial sum gives the orientation  of the dipole of the distribution. If the poles are isotropically distributed, the unit vectors representing their
orientation will tend to cancel each other out and the resulting $\mathcal{R}$ will be close to zero, while if the poles have a preferential orientation, $\mathcal{R}$ will tend to be larger than expected for  the isotropic distribution. However, there could be very specific anisotropic situations leading to low $\mathcal{R}$
values, for instance one with an even number of poles distributed in a bimodal distribution with the modes aligned in exactly opposite directions, and equal
numbers of poles populating both modes. This underlines the fact that a test such as the Rayleigh test should always be carried out against an
alternative model, and this model must be chosen on physical grounds \citep{Gillett1988}. The most natural alternative to be used with Rayleigh's test is the Fisher
distribution \citep{Fisher1953,Watson1956},  which is unimodal and assumes symmetry about the mean. Obviously, this bimodal situation does not
follow a Fisher distribution, and Rayleigh's test has therefore no power to distinguish it from isotropy (i.e., the second-kind risk of accepting the null
hypothesis of isotropy while it is false is high). The combined statistic of \citet{Gine1975}, which are described below (Eq.~\ref{Eq:Gine}),  is the
most powerful against all alternatives \citep{Gillett1988}.

The Rayleigh statistics may be used as a test of isotropy against a unimodal distribution by calculating the probability $\alpha_{\mathcal{R}}$  of obtaining
a value of $\mathcal{R}$ greater than or equal to the observed value $\mathcal{R_{\rm obs}}$ assuming that the data are drawn from the isotropic distribution 
\citep{Watson1956,Watson1983}: 

\begin{equation} 
\label{Eq:R}
\alpha_{\mathcal{R}} = \mathrm{P}(\mathcal{R} \geq \mathcal{R_{\rm obs}}) =
\int_{\mathcal{R_{\rm obs}}}^{N}f_\mathcal{R}(R)\;\mathrm{d}R, 
\end{equation}

where $f_\mathcal{R}$ is the probability density function for $\mathcal{R}$ (i.e., the probability of finding $\mathcal{R}$ between $R$ and $R+\mathrm{d}R$
is $f_\mathcal{R}\mathrm{d}R$). Its detailed expression may be found in \citet{Fisher1953} and \citet{Briggs1993}. If $\alpha_{\mathcal{R}}$ is small,
then the poles are very unlikely to have been drawn from an isotropic parent population. Under such circumstances, it is said that the observed pole
distribution is inconsistent with isotropy at the $1-\alpha_{\mathcal{R}}$ confidence level. It means that there is a (first-kind) risk
$\alpha_{\mathcal{R}}$ of rejecting the null hypothesis of isotropy while it is actually true.  \citet{Stephens1964} provides a table of critical values for
${\mathcal{R}}$ for $\alpha_{\mathcal{R}} = 1, 2, 5$ and 10\%.

While the Rayleigh statistics is valid for any value of $N$, there exists a more convenient form of it at large $N$ ($\ge 50$), known as the Rayleigh-Watson
statistics $\mathcal{W}$ \citep{Watson1956,Watson1983}: 
\begin{equation}
\mathcal{W} = \frac{3}{N}\mathcal{R}^2. 
\end{equation} \citet{Upton1989} used the table of \citet{Stephens1964} to devise an improved approximation, which works well
even for low values of $N$: 
\begin{equation} \label{Eq:W*} 
\mathcal{W^*} =
\frac{3}{N-c(\alpha)}\mathcal{R}^2.
\end{equation} 
$\mathcal{W^*}$ follows a $\chi^2$ distribution with three degrees of freedom. For convenience, the values of the correction factor $c(\alpha)$ for the Rayleigh test are reproduced from \citet{Upton1989} in Table~\ref{Tab:calpha}. A value $\mathcal{W}^*  > 11.34$ indicates that there is a significant concentration of poles toward one direction (with a confidence level of 99\%).

\begin{table} \caption[]{\label{Tab:calpha} Values of the correction factor
$c(\alpha)$ for the Rayleigh test, from \citet{Upton1989}. }
\begin{tabular}{rrrr} 
\hline 
Significance level $\alpha$ & 10\% & 5\% & 1\% \\
\hline\\ 
$c(\alpha)$ & 0.121 & 0.278 & 0.630 \\ 
Reference value
$\chi^2_3(\alpha)$ & 6.25 & 7.81 & 11.34 
\smallskip\\ 
\hline\\ 
\end{tabular}
\end{table}

Instead of considering $\mathcal{R}$ as the length of the resulting dipole vector, it may be useful to consider it as the distance from the origin achieved
by a random walk of $N$ unit steps. A well-known result from the randow-walk theory then states that $\left< \mathcal{R}^2 \right> = N$, from which it
follows that  $\left< \mathcal{W} \right> = 3$. It is therefore not surprising that the statistic $\mathcal{W}$ is asymptotically distributed as a $\chi^2$
distribution with three degrees of freedom \citep{Watson1956,Watson1983}, so
that 

\begin{equation} 
\alpha_{\mathcal{W^*}} = \mathrm{P}(\mathcal{W^*} \geq
\mathcal{W^*_{\rm obs}}) = \int_{\mathcal{W^*_{\rm
obs}}}^{\infty}f_{\chi_3^2}(x)\;\mathrm{d}x = F_{\chi_3^2}(\mathcal{W^ *} \geq
\mathcal{W^*_{\rm obs}}), 
\end{equation} 

where $f_{\chi_3^2}$ is the probability density of $\chi_3^2$ and $F_{\chi_3^2}$ is the upper cumulative probability distribution function of $\chi_3^2$. Obviously, the latter function is much easier to evaluate than the corresponding one for the Rayleigh statistics (Eq.~\ref{Eq:R}). An additional advantage of this statistics is that it is asymptotically independent from $N$.

Again, the isotropy of the data may be tested by calculating the observed statistics $\mathcal{W_{\rm obs}}^ *$ and the corresponding probability $\alpha_{\mathcal{W}}$, which has a similar interpretation as the one given above for $\alpha_{\mathcal{R}}$. For large $N$, $\alpha_{\mathcal{R}} = \alpha_{\mathcal{W}}$, and the two tests are equivalent.

We stress here an important result from the hypothesis-testing theory: although the {\it first-kind risk} (of rejecting the null hypothesis of isotropy while it is
true, also known as the false-alarm probability) is controlled by the adopted confidence level $1-\alpha_{\mathcal{R}}$, the {\it second-kind risk} (of accepting the null hypothesis of isotropy while it is false, also known as the power of the test) is not controlled as easily, since this second-kind risk depends on the exact nature of the true underlying distribution. This was well illustrated above with the Rayleigh test which is said to have no power against a bimodal distribution. Because different statistical tests have different powers it is important not to use only one if one does not know the actual underlying distribution. 

Since the Rayleigh test is not appropriate when the alternative is a bimodal axial distribution, for testing uniformity when the alternative is such, we must turn to tests based upon the properties of statistics involving the eigenvalues of the quadrupolar
matrix $\mathbf{M}_N$ \citep{Watson1966,Watson1983,Briggs1993}: 

\begin{equation} 
\label{Eq:array} 
\mathbf{M}_N =
\frac{1}{N} \Sigma_{i=1}^N
\left[ 
\begin{array}{lll} 
x_i x_i & x_i y_i & x_i z_i \\ 
y_i x_i & y_i y_i & y_i z_i  \\ 
z_i x_i & z_i y_i & z_i z_i \\ 
\end{array}
\right] . 
\end{equation} 

Since $\mathbf{M}_N$ is real and symmetric, it has three eigenvalues $\lambda_k$, and since it has unit trace, the sum of its eigenvalues $\lambda_k$ is one\footnote{The eigenvalues are ordered so that $\lambda_3 \ge \lambda_2 \ge \lambda_1$.}. Moreover, the diagonal elements are sums of squares. In the coordinate system in which $\mathbf{M}_N$ is diagonal, the diagonal elements are the eigenvalues, and therefore the eigenvalues are not negative. If the poles were drawn from the isotropic distribution, then, excepting statistical fluctuations, the eigenvalues $\lambda_k$ should be equal, and since $\mathbf{M}_N$ has unit trace, $\lambda_k = 1/3$.

If the alternative to isotropy is a {\it rotationally asymmetric} axial distribution\footnote{In other words, the data points are non-uniformly distributed along a ring.}, the Bingham statistics $\mathcal{B}$ is appropriate: it measures the deviation of the eigenvalues  $\lambda_k$ of the quadrupolar matrix $\mathbf{M}_N$ from the value 1/3 expected for isotropy \citep{Bingham1974,Watson1983}: 

\begin{equation} 
\label{Eq:Bingham} 
\mathcal{B}
= \frac{15 N}{2}\;\Sigma_{k=1}^3\left(\lambda_k - \frac{1}{3}\right)^2.
\end{equation} 

The statistics $\mathcal{B}$ is asymptotically distributed as a $\chi^2$ distribution with five degrees of freedom, so that 

\begin{equation}
\label{Eq:alphaB}
\alpha_{\mathcal{B}} = \mathrm{P}(\mathcal{B} \geq \mathcal{B_{\rm obs}}) =
\int_{\mathcal{B_{\rm obs}}}^{\infty}f_{\chi_5^2}(x)\;\mathrm{d}x =
F_{\chi_5^2}(\mathcal{B} \geq \mathcal{B_{\rm obs}}), 
\end{equation} 
where $f_{\chi_5^2}$ is the probability density function of $\chi_5^2$ and $F_{\chi_5^2}$ is the upper cumulative probability distribution function of
$\chi_5^2$. Again, if $\alpha_{\mathcal{B}}$ is very small, the hypothesis of isotropy is contradicted and rejected with a first-kind risk of $\alpha_{\mathcal{B}}$. \citet{Briggs1993} showed that $\alpha_{\mathcal{B}}$ as derived from  Eq.~\ref{Eq:alphaB} overestimates the true first-kind risk when $N < 20$ 
for $\alpha_{\mathcal{B}} = 0.05$, and when $N < 40$ for $\alpha_{\mathcal{B}} = 0.01$. 
In other words, under these circumstances, $1 -\alpha_{\mathcal{B}}$ will underestimate the statistical confidence with which the hypothesis of isotropy can be rejected.

If the alternative to isotropy is a rotationally {\it symmetric} axial distribution \citep[either a bipolar distribution along a given direction, or a girdle with rotational symmetry, both known as the Dimroth-Watson distribution;][]{Upton1989}, the Bingham test is still applicable, but two of the three eigenvalues will then be close to each other \citep[][and below]{Bingham1974}. 
Assuming that the pole of the distribution is located at $\phi_0 = 0^{\circ}$ and $\theta_0 = 0^{\circ}$ [$(\phi, \theta)$ being a polar coordinate system on the
sphere, with $\phi$ the azimutal angle, and $\theta$ the polar angle], \citet{Mardia2000} showed that $(\hat{\phi}_0, \hat{\theta}_0)$, the maximum likelihood estimate of the polar direction, is related to the eigenvectors of the $\mathbf{M}_N$ matrix; a girdle distribution will have its symmetry axis along $\mathbf{e}_1$: 

\begin{equation}
\label{Eq:polegirdle} 
\mathbf{e}_1 = (\sin\hat{\theta}_0 \cos\hat{\phi}_0,
\sin\hat{\theta}_0 \sin\hat{\phi}_0, \cos\hat{\theta}_0 ) \;\; \mathrm{and} \;\; \lambda_1 < \lambda_2 \sim \lambda_3,
\end{equation}
whereas an axial distribution will have $\mathbf{e}_3$ as symmetry axis\footnote{we recall that the eigenvalues are ordered so that 
$\lambda_3 \ge \lambda_2 \ge \lambda_1$.}: 

\begin{equation} 
\label{Eq:polebipolar} 
\mathbf{e}_3
= (\sin\hat{\theta}_0 \cos\hat{\phi}_0, \sin\hat{\theta}_0 \sin\hat{\phi}_0,
\cos\hat{\theta}_0 ) \;\; \mathrm{and} \;\; \lambda_1 \sim \lambda_2 < \lambda_3. 
\end{equation} 
In the general case of a {\it rotationally asymmetric} axial distribution, one has 
$\lambda_2 - \lambda_1$ and $\lambda_3 - \lambda_1$ large and non-equal (girdle), or 
$\lambda_3 - \lambda_1$ and $\lambda_3 - \lambda_2$ large and non-equal (polar).

\citet{Anderson1972} showed that $\lambda_1$ and $\lambda_3$ themselves are the most appropriate test statistics for the comparison of the hypothesis of
uniformity against that of a Dimroth-Watson distribution. \citet{Upton1989} have used the tables computed by \citet{Anderson1972} to devise critical values of $\lambda_1$ and $\lambda_3$ at confidence level $\alpha$. A significant departure from isotropy towards the girdle form of the Dimroth-Watson distribution is judged to have occurred if$\lambda_1 < \lambda_1(\alpha)$, where 

\begin{equation} 
\label{Eq:lambda1}
\lambda_1 (\alpha) = \frac{1}{3} +  \frac{a_1(\alpha)}{\sqrt{N}} +
\frac{b_1(\alpha)}{N} +  \frac{c_1(\alpha)}{N\sqrt{N}}, 
\end{equation} 

whereas a significant departure towards the bipolar form of the distribution is judged to have occurred if $\lambda_3 > \lambda_3(\alpha)$, where 

\begin{equation}
\label{Eq:lambda3} 
\lambda_3 (\alpha) =  \frac{1}{3} +
\frac{a_3(\alpha)}{\sqrt{N}} + \frac{b_3(\alpha)}{N}, 
\end{equation} 

where the coefficients $a_{1,3}(\alpha)$, $b_{1,3}(\alpha)$, and $c_{1}(\alpha)$ have values that depend on the significance level, $\alpha$.  These values may be found in Table~10.15 of \citet{Upton1989}. The values  $\lambda_1(\alpha)$ and $\lambda_3(\alpha)$ have been plotted as dashed lines in the lower panel of Fig.~\ref{Fig:statistics} for $\alpha = 1$\% and 10\%.

In summary, by combining the Rayleigh-Watson statistics with the Bingham statistic, the isotropy of the distribution may be distinguished among several alternatives, since a value $\mathcal{W} > 3$ reveals that there is a concentration of poles toward one direction, but values close to 3 indicate that the distribution is either isotropic or bipolar. The bipolarity of the distribution is in turn assessed from $\lambda_3 > \lambda_3(\alpha)$. 
To summarize [see also Table~1 of Briggs (1993) and Table~10.7 of  Upton \& Figgleton (1989)]: 

\begin{itemize} 
\item[$\bullet$] unipolar distribution around one end of the eigenvector $\mathbf{e}_3$:
\end{itemize} 
\begin{equation}
\label{Eq:unipolar}
\mathcal{W} > 3:\;\; \lambda_{1,2} < 1/3 < \lambda_3, 
\end{equation}
\begin{itemize} 
\item[$\bullet$]  
bipolar distribution around both ends of the eigenvector $\mathbf{e}_3$: 
\end{itemize} 
\begin{equation} 
\mathcal{W} \sim 3:\;\;  \lambda_{1,2} < 1/3 < \lambda_3 , 
\end{equation} 
\begin{itemize} 
\item[$\bullet$]  
girdle orthogonal to the eigenvector $\mathbf{e}_1$: 
\end{itemize} 
\begin{equation} 
\label{Eq:girdle}
\mathcal{W} \sim 3:\;\; \lambda_1 < 1/3 < \lambda_{2,3}. 
\end{equation}

Finally, one should mention the non-parametric statistical tests that do not presuppose any alternative model. They are thus more general than tests against a specific alternative, but for that reason are also less powerful \citep[i.e., have a higher risk of accepting the null hypothesis of isotropy while it is false;][]{Gillett1988,Mardia2000}. To test against alternatives that are asymmetric with respect to the centre of the sphere, Beran's $B$ test \citep{Beran1968} is the most powerful: 

\begin{equation} 
\label{Eq:Beran} 
B = N - \frac{4}{N\pi} \Sigma_{i<j}^{N} \Psi_{ij}  = N - \frac{2}{N\pi} \Sigma_{i}^{N}\Sigma_{j}^{N} \Psi_{ij}. 
\end{equation} 

To test against symmetric or axial alternatives, Gin\'e's test $G$ \citep{Gine1975} is the most powerful: 

\begin{equation} 
\label{Eq:Gine} 
G = \frac{N}{2} - \frac{4}{N\pi} \Sigma_{i<j}^{N} \sin \Psi_{ij} =  \frac{N}{2} - \frac{2}{N\pi} \Sigma_{i}^{N}\Sigma_{j}^{N} \sin \Psi_{ij}. 
\end{equation}

The combined Gin\'e statistic $F = B + G$ tests against all alternatives. In Eqs.~\ref{Eq:Beran} and \ref{Eq:Gine}, $N$ is the number of data points, and
$\Psi_{ij} = \mathrm{cos}^{-1}(\mathbf{x_i} \cdot \mathbf{x_j})$ is the smaller of the angles between poles $\mathbf{x_i}$ and $\mathbf{x_j}$. 
High values of these statistics suggest departure from isotropy. Critical values for various percentage levels of significance for $B,G$ and $F$ have been calculated by \citet{Keilson1983}. 
In the large-sample limiting distribution of $F$, the 10\%, 5\%, and 1\% confidence levels are 2.355, 2.748, and 3.633, respectively.

\subsection{Isotropic or anisotropic?} \label{Sect:analysis}

\begin{figure}[t] \centering
\includegraphics[width=9cm]{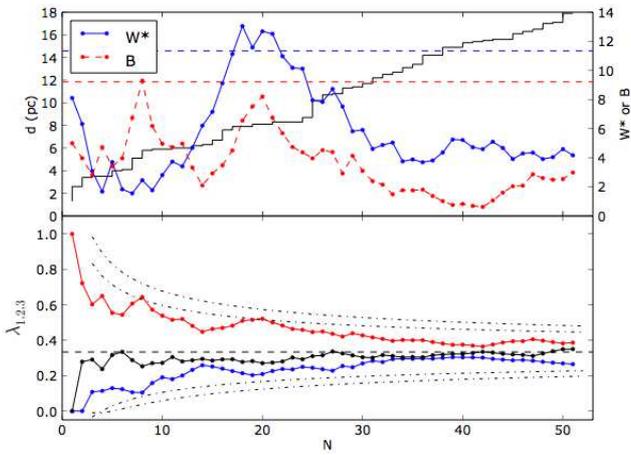} 
\caption[]{ Upper panel: The evolution of the Rayleigh-Watson $\mathcal{W}^*$ (blue dashed line, and right-hand scale; the correction to $\mathcal{W}$ -- Eq.~\protect\ref{Eq:W*} -- has been made for a confidence level $\alpha$ of 1\%) and Bingham $\mathcal{B}$ (red dot-dashed line, and right-hand scale) statistics as a function of the number of systems. The horizontal blue dashed curve corresponds to a first-kind risk of 1\% of rejecting the null hypothesis of isotropy while it is true, based on the $\mathcal{W}^*$ statistics. The horizontal red dashed line is the 10\% first-kind risk for the $\mathcal{B}$ statistics. 
The number of systems as a function of distance is given by the black solid line and left-hand scale. 
Isotropy corresponds to $\mathcal{B} \rightarrow 0$ and $\mathcal{W} \rightarrow 3$.
Lower panel:  The 3 eigenvalues of the orientation matrix (Eq.~\protect\ref{Eq:array}). Isotropy corresponds to  $\lambda_{1,2,3} = 1/3$, as seen at large distances. 
The critical values for $\lambda_{1,3}$ corresponding to $\alpha = 1$\% and 10\% as  given by Eqs.~\ref{Eq:lambda1} and \ref{Eq:lambda3} are displayed by the dash-dotted lines.}
\label{Fig:statistics} 
\end{figure}

In this section, we apply the methods outlined in Sect.~\ref{Sect:statistics} to the sample of orbital poles listed in Table~\ref{Tab:poles}.
The top panel of Fig.~\ref{Fig:statistics} displays the evolution of the Rayleigh-Watson $\mathcal{W}^*$ and Bingham $\mathcal{B}$ statistics as
a function of distance, thus for an increasing number of systems.

The various statistical indicators described in Sect.~\ref{Sect:statistics} are listed in Table~\ref{Tab:stat} separately for the 20 systems up to 8.1~pc, 
and for all 51 systems closer than 18~pc. For estimating $\alpha_{\mathcal{W*}}$ and $\alpha_{\mathcal{B}}$, the asymptotically equivalent $\chi_3^2$ and  $\chi_5^2$ distributions were used, respectively.

For the 20 systems with $d \le 8.1$~pc, the Rayleigh-Watson test $\mathcal{W*}$  signals a very significant  deviation from isotropy
\footnote{As apparent from Fig.~\ref{Fig:statistics}, restricting the sample to 18 systems would yield an even more significant result, but this would imply considering only one orbital pair from the triple system WDS~11182+3132 (Table~\ref{Tab:poles}), which is not physically sound. 
A discussion  of the problem of triple systems is presented  in Sect.~\ref{Sect:Multiplicity}.}, with a (first-kind) risk of $\alpha_{\mathcal{W*}} = 0.5$\% of  rejecting the null hypothesis of isotropy while it is actually true (Table~\ref{Tab:stat}). The Beran test (for isotropy against alternatives that are asymmetric
with respect to the centre of the sphere) provides an identically significant result \citep[$B = 3.32$ or $\alpha \sim 0.7\%$;][]{Keilson1983}. 

As seen from Table~\ref{Tab:stat}, the Bingham and Gin\'e tests are less conclusive, but this is easily understood. The Gin\'e test, being 'universal' (it tests against all alternatives), is less efficient than Rayleigh-Watson's in the sense that it requires a stronger deviation from isotropy to yield similarly low $\alpha$ values. 
Bingham test being devised against symmetric alternatives, which is not the situation encountered in our data (Fig.~\ref{Fig:aitoff}), it is not efficient either in the situation under study, and should not be given much consideration.

Another way to understand why the Rayleigh-Watson and Bingham statistics yield vastly different significance levels is remembering that the
Rayleigh-Watson and Bingham statistics test for the presence of a dipole or of a quadrupole, respectively,  in the distribution of poles on the
sphere \citep[][and Sect.~ \ref{Sect:statistics}]{Hartmann1989,Briggs1993}. As already discussed in relation with Eq.~\ref{Eq:dipole}, the vector $\vec{\mathcal{R}}$ is
the dipole moment of the distribution,  whereas the eigenvectors of the quadrupole matrix $\bf{M}_N$ are related to the orientation of the quadrupolar component of the distribution (see the discussion in relation with Eqs.~\ref{Eq:unipolar} -- \ref{Eq:girdle}). Therefore, the dipolar vector and the eigenvector $\vec{e}_3$ associated with the highest eigenvalue of $\bf{M}_N$ ought be different. For the considered orbital-pole distribution, the largest dipole moment is obtained for the $N = 20$ closest systems and points in the direction $l = 46.0^\circ, b = 36.7^\circ$, with a clear deficit of poles in the opposite direction. On the other
hand, the eigenvector $\vec{e}_3$ associated with the highest eigenvalue
$\lambda_3 = 0.52$ for $N = 20$  points towards $l = 77.4^\circ, b = 8.4^\circ$. 
These directions are similar, albeit not identical, and obviously point in the direction of a higher concentration of poles (Fig.~\ref{Fig:aitoff}).

The Rayleigh-Watson (dipolar) test yields a deviation from isotropy that is more significant than the Bingham (quadrupolar) test. 
Figure~\ref{Fig:statistics} indeed signals that the eigenvalues $\lambda_{1,2,3}$ and $\mathcal{W*}$ fulfill Eq.~\ref{Eq:unipolar}
that corresponds to a unipolar distribution. 

\begin{figure}[t] \centering
\includegraphics[width=9cm]{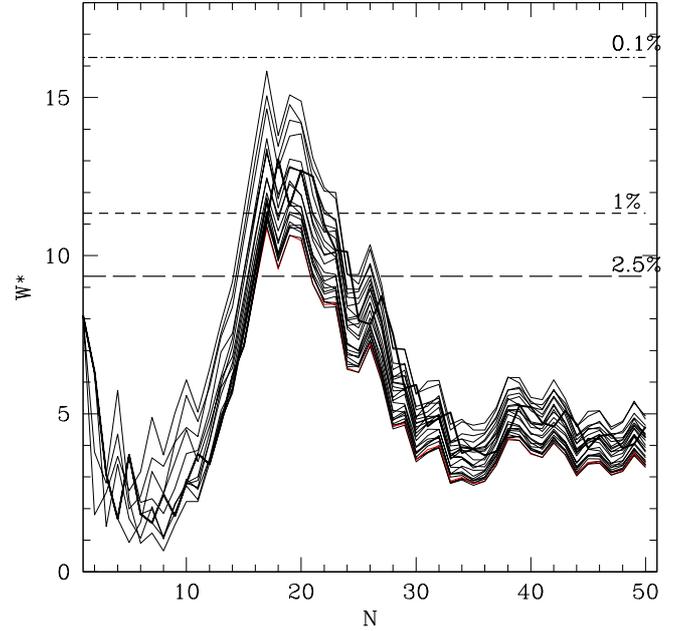} 
\caption[]{
Evolution of the Rayleigh-Watson $\mathcal{W}^*$ statistics as a function of the number of systems in a jackknife approach. 
Each curve thus corresponds to the full sample with one system removed. The thick line corresponds to the full sample 
(same as the blue dashed line in the upper panel of Fig.~ \ref{Fig:statistics}). 
The lowest thin (red) line corresponds to the sample with system No.~16 removed (WDS~14514+1906).
False-alarm probabilities of 2.5, 1, and 0.1\% are depicted by the dashed horizontal lines.
}
\label{Fig:jackknife} 
\end{figure}

To evaluate the robustness of the low first-kind risk obtained for the Rayleigh-Watson statistics, a jackknife approach was used, that is, we repeated the above procedure of computing $\mathcal{W*}$ as a function of distance (or sample size, with systems ordered by increasing distance from the Sun) after removing one system at a time from the full sample. 
Since $\mathcal{W*}$ peaks around $N = 20$, it was sufficient to do so for the 25 systems closest to the Sun, removing one system at a time. 
This way, 25 curves of $\mathcal{W*}$ as a function of $N$ were generated, one for each of the 25 samples thus generated. 
These curves are presented in Fig.~\ref{Fig:jackknife} and are similar to the curve presented in the upper panel of Fig.~\ref{Fig:statistics}. 
The  thick line in Fig.~\ref{Fig:jackknife} corresponds to all systems being considered (thus identical to the curve in the upper panel of Fig.~\ref{Fig:statistics}). The  lower (red) line is the one corresponding to the sample with  system No.~16 removed (since this system is the one weighting heavily in the original $\mathcal{W*}$ peak; the number refers to Table~\ref{Tab:poles} and Fig.~\ref{Fig:aitoff}). 
As expected, this curve has the lowest significance, peaking at only 1.5\% instead of 0.5\% in the full sample.
The other thin curves are for all the other samples, as explained above. 
Figure~\ref{Fig:jackknife} thus provides a fair evaluation of the robustness of the alarm signal: indeed, after removing system No.~16, the false-alarm probability just misses 1\% and is thus not especially significant; but removing closer systems (such as Nos.~1 or 2, i.e., systems far away from the pole cluster seen in Fig.~\ref{Fig:aitoff}) makes the curve rocket to 0.1\% significance!
In summary, the fluctuations caused by the small sample size causes the highest $\mathcal{W*}$ significance level to vary between 1.5 and 0.1\%! 
The reality of the deviation from isotropy can thus not be assessed with certainty at this stage, given the small number of systems available, 
and despite our efforts to increase it.

The fact that this deviation from isotropy does not appear for subsamples with  cutoff distances shorter than 8~pc moreover decreases its {\it physical} significance. One could have envisioned for instance a situation where the $\mathcal{W*}$ index maintains statistically significant values all the way to 8~pc, and then fades to insignificant values at larger distances. But the first part of that statement is not fulfilled: the closest 15 systems (up to 6~pc) have $\alpha(\mathcal{W*}) > 5$\% (Fig.~\ref{Fig:statistics}). At distances larger than 8~pc, the significance of the deviation from isotropy fades away again, at least for the Bingham statistics: $\mathcal{B} \rightarrow 0$, $\lambda_{1,2,3} \rightarrow 1/3$.
The rise of $\mathcal{B}$ above 2.5 for $N > 47$ is not at all significant since it corresponds to $\alpha_{\mathcal{B}} = 0.79$. 
In contrast, the Rayleigh-Watson statistics $\mathcal{W*}$ does not drop to its isotropic value of 3, but rather stays around 4, corresponding to a first-risk probability  of about 18\% of rejecting the null hypothesis of isotropy despite being true. Indeed, the addition of systems farther away than 8~pc does not fully destroy the anisotropy that appeared in the 8~pc sphere, since a deficit of poles remain around $l = 270^\circ$, $b = -30^\circ$ (Fig.~\ref{Fig:aitoff}).

We thus conclude that {\it the poles of the binary systems within 8~pc of the Sun exhibit a weak tendency to cluster 
around $l = 46^{\circ}, b = 37^{\circ}$   (which, incidentally, is not far from the pole of the ecliptic). The systems numbered 15--18 and 20 (see Table~\ref{Tab:poles}) are responsible for that clustering, which fades away if shorter or larger sampling distances are considered.}

\begin{table*} 
\caption[]{\label{Tab:stat} 
Various isotropy indicators, listed separately for the 20 systems up to 8.1~pc, and for all 51 systems up to 18~pc. 
See Sect.~\ref{Sect:statistics} for the definition of the various indicators. $\bar{x}$, $\bar{y}$ and  $\bar{z}$ are the average of $(x_i, y_i, z_i)$. The Watson statistics has first been corrected by Eq.~\ref{Eq:W*} for $\alpha = 1$\%, before being approximated by the $\chi_3^2$ statistic so as to
compute $\alpha_{\mathcal{W^*}}$. 
The columns labelled $B$ and $G$ refer to Beran and Gin\'e statistics, respectively (Eqs.~\ref{Eq:Beran} and \ref{Eq:Gine}). } 
\begin{tabular}{rrrrrrrrrrrrrrrrr} 
\hline\\ 
$N$ & $\bar{x}$ & $\bar{y}$ & $\bar{z}$ & $\mathcal{R}$ & $\mathcal{W}$ &
$\mathcal{W^*}$ & $\alpha_{\mathcal{W}}$ & $\lambda_3$ & $\lambda_2$ &
$\lambda_1$ & $\mathcal{B}$ & $\alpha_{\mathcal{B}}$ & $B$ & $\alpha_{B}$ & $G$ & $\alpha_{G}$ \\ 
\hline\\
\noalign{Systems closer than 8.1 pc}\\ 
20 & 0.254 & 	0.262 &	0.268  & 9.05 & 12.28 & 12.68 & 0.005 &
0.52 & 0.27 & 0.21 & 8.20 & 0.14& 3.32 & 0.007 & 0.65 & 0.19 \\ 
\medskip\\ 
 \noalign{All systems up to 18 pc}\\
51 & 0.071 &	0.109 &	0.100& 8.36 & 	4.11  & 4.17 & 0.18 & 0.39 & 0.35 & 0.26  &		 
2.98 & 0.70 & 1.22 & 0.20 & 0.32 & 0.80 \\ 
\hline\\ 
\end{tabular} 
\end{table*}

\section{Impact of selection effects}
\label{Sect:Selection}

\begin{figure}
\includegraphics[width=8cm]{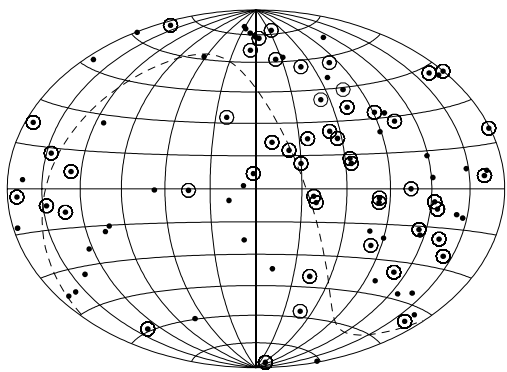}
\caption[]{\label{Fig:allsystems}
Location of binary systems from Table~\protect\ref{Tab:starlist} in Galactic coordinates, with the galactic centre in the middle of the map. The circles identify the systems with an unambiguously defined pole , as plotted in Fig.~\protect\ref{Fig:aitoff}. The dashed line is the celestial equator.
} 
\end{figure}

As A. Blaauw  appropriately commented after Batten's talk on the topic of the present paper during the IAU symposium 30 \citep{Batten1967}, the possibility that observational selection effects could actually be causing  an anisotropy in the pole distribution should be considered.
Figure~\ref{Fig:allsystems} indeed reveals that the studied systems are concentrated in the northern equatorial hemisphere, since lifting the pole ambiguity requires radial velocities that are easier to obtain for northern hemisphere targets (combined with the fact that the astronomy of visual double stars relies on a longer time-base in the northern hemisphere; hence long periods are more commonly derived for northern targets). Could this anisotropic initial distribution be the cause for the pole anisotropy observed in the 8~pc sphere around the Sun?

\begin{figure}
\includegraphics[width=8cm]{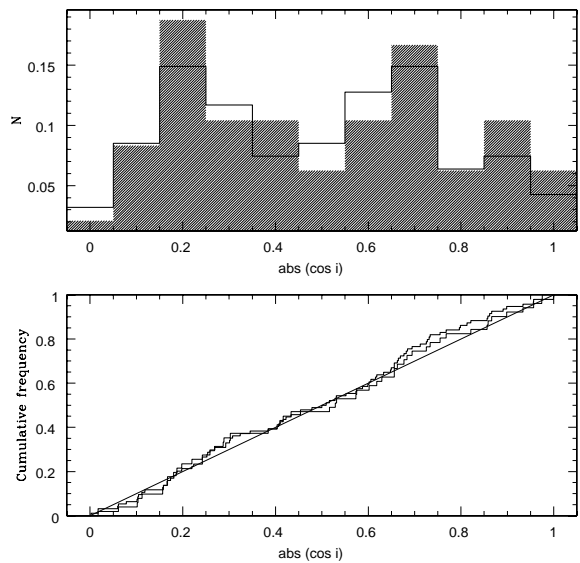}
\caption[]{\label{Fig:cosi_allsystems}
Distribution of  $\cos i$ for all binaries from Table~\ref{Tab:starlist} (thick lines)
and for those with the poles analysed (thin line in the bottom panel and shaded histogram in the upper panel).
}
\end{figure}

For this to be the case, there should be a correlation between the position of the binary on the sky and the position of its orbital pole. 
A non-uniform celestial distribution of the systems studied (as observed in Fig.~\ref{Fig:allsystems}) would then propagate onto their orbital-pole distribution. A correlation between system position and pole position would manifest itself as a non-uniform distribution of the scalar product between the unit vector defining the system position (denoted $\mathbf{1}_*$ hereafter) and the unit vector defining the pole position (denoted $\mathbf{1}_p$ hereafter).
From Eq.~\ref{EquCoord}, it is easy to verify that this scalar product $\mathbf{1}_* \cdot \mathbf{1}_p$ equals $-\cos i$. 
Thus, an observational bias acting on our ability to derive the poles for systems with a given range of inclinations $i$ would introduce a correlation between  $\mathbf{1}_*$ and $\mathbf{1}_p$. 
To check whether this is the case, we plot in Fig.~\ref{Fig:cosi_allsystems} the distribution of $\cos i$ values for all systems of Table~\ref{Tab:starlist} and  for those with the poles analysed (thin line and shaded histogram). 
There is no significant difference between these distributions, meaning  that the requirement of having radial velocities available for lifting the pole  ambiguity does not introduce a selection bias on the orbital inclination $i$. 
On the other hand, the full-sample and pole-known distributions of $\cos i$ are close to being uniform\footnote{The maximum absolute difference 
between the uniform distribution and the distribution for pole-known systems is 0.11, which is not very significant (about 20\% probability that such a difference be due to chance.)} except for two marginal peaks at 0.2 and 0.7. 
However, given the small amplitude of these peaks, it seems very unlikely that they could be at the origin of the segregation observed in the pole distribution.
Furthermore, there is no obvious observational selection effect that we could think of that could cause peaks at $\cos i \sim 0.2$ and 0.7.

\section{Coplanarity in multiple systems}
\label{Sect:Multiplicity}

When searching for some anisotropy in the space orientation of orbital poles, which could either be signatures of the process of binary
formation or of tidal effects from the Galactic disc, one could be misled by evolutionary processes acting within the stellar system
itself that modify the orbital inclination. Within a triple system, weak perturbations from the outer body can have strong long-term
effects on the inner binary. The simplest of these is precession of the orbital plane, which occurs if the orbital planes of the inner and outer
binary are not aligned \citep{Fabrycky2007}. If the inner and outer binary orbits are circular, this precession is analogous to the
precession of two rigid rings with the same mass, radius, and angular momentum as the binary orbits; both the mutual inclination and the
scalar angular momenta of the rings remain fixed, while the two angular momentum vectors precess around the direction defined by the total
angular momentum vector of the triple system.

An unexpected aspect of this behaviour was discovered by \citet{Kozai1962} \citep[see also][]{Kiseleva1998,Eggleton2001,Fabrycky2007,Muterspaugh2008}. 
Suppose the inner binary orbit is initially circular, with the initial mutual inclination between inner and outer binaries equal to $i_{\rm initial}$.
Kozai found that there is a critical angle $i_c$ such that if $i_{\rm initial}$ is between $i_c$ and $180^\circ - i_c$, then the orbit of the
inner binary cannot remain circular as it precesses; both the eccentricity of the inner binary $e_{\rm in}$ and the mutual inclination $i$ execute periodic oscillations known as Kozai cycles. The amplitude of the eccentricity and inclination oscillations is independent of the strength of the perturbation from the outer body (which depends on the mass of that outer body, and on $a_{\rm out}$ and $e_{\rm out}$), but the oscillation amplitude does depend on $i_{\rm initial}$; for initial
circular orbits with $i_{\rm initial} = i_c$ or $180^\circ - i_c$, the eccentricity is 0, but if $i_{\rm initial} = 90^\circ$,  the highest eccentricity is unity, that is, the two inner bodies collide.

In the framework of the present study, it is important to stress that Kozai cycles occur independent of distances or component masses, their only requirement is that the mutual inclination is between $i_c = 39.2^\circ$ and $180^\circ - 39.2^\circ = 140.8^\circ$ \citep{Fabrycky2007,Muterspaugh2008}.

Hence, for triple systems, the current orbital pole may not at all be representative of its initial orientation if the Kozai mechanism is operating. Therefore, a discussion of the pole orientation should properly identify the triple systems, and possibly exclude them from the analysis. The triple and higher-multiplicity systems in our sample are identified by flag 't' in Tables~\ref{Tab:starlist} and \ref{Tab:poles}, the notes in the Appendix provide more details in most cases.

\begin{figure} \centering 
\includegraphics[width=9cm]{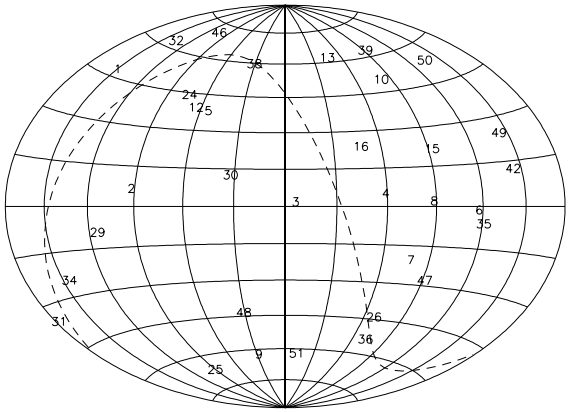}
\caption{Same as Fig.~\ref{Fig:aitoff}, but for the binary stars without 
indications for higher multiplicity. }
\label{Fig:aitoff_triple} 
\end{figure}

\begin{figure}[t] \centering
\includegraphics[width=9cm]{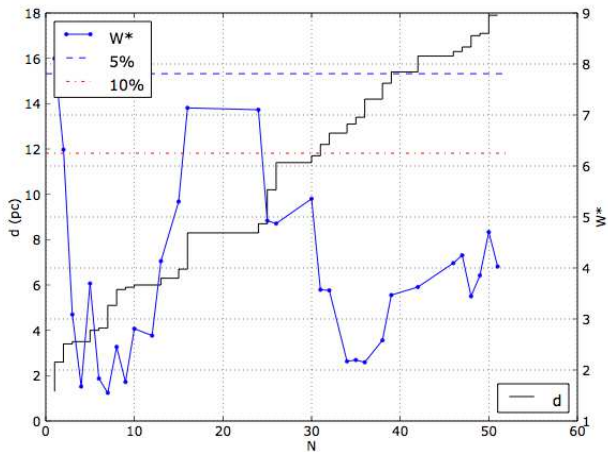} 
\caption[]{ 
Same as Fig.~\protect\ref{Fig:statistics}, but restricted to binary systems without
evidence for higher multiplicity. The horizontal blue dashed and red dash-dotted curves correspond to a first-kind
risk of 5\% and 10\%, respectively, of rejecting the null hypothesis of isotropy while it is true,
based on the $\mathcal{W}^*$ statistics. 
The systems have the same number labels as in Table~\ref{Tab:poles} and Fig.~\ref{Fig:statistics}.
}
\label{Fig:W_triple} 
\end{figure}

Because the Kozai mechanism does alter the orbital inclinations of non-coplanar triple systems, we now investigate the pole-anisotropy problem separately for  the sample devoid of triple systems (Figs.~\ref{Fig:aitoff_triple} and \ref{Fig:W_triple}). These figures reveal that the Rayleigh-Watson test still predicts some deviation from
isotropy in the sample restricted to 8~pc, but the first-kind error of the Rayleigh-Watson statistics this time is not better than 5\%. We thus conclude that triple systems in the analysed sample do not degrade a possible anisotropy, quite the contrary, since the first-kind error of rejecting the null hypothesis of isotropy even increases when including the triple systems.

Nevertheless, the fact that several triple systems are now known not to be coplanar is certainly not in favour of a concentration of orbital poles on the celestial sphere. 
The quadruple system WDS~11182+3132 belongs to the sample of systems closer than 8~pc to the Sun; the AB and Aa,Ab poles are known (Nos.~18 and 19 on Fig.~ \ref{Fig:aitoff}) and are separated by about 100$^\circ$ on the sky.
Although such a complete knowledge is not very frequent\footnote{\citet{Sterzik2002} revealed that there are only 22 
triple systems with both orbits determined visually. Eight of these have questionable elements and only three have had their ascending nodes identified using radial velocity measurements.}, non-coplanarity is also observed in $\eta$~Vir \citep[HIP~60129: the orbital planes of the two systems are 
inclined by about 30$^\circ$ with respect to each other;][]{Hummel2003}, and similarly for the triple system V819~Her \citep[HD 157482;][]{OBrien2011}. 
In addition to V819~Her and $\eta$~Vir, unambiguous mutual inclinations have been calculated for several other systems: $\kappa$~Peg, $\epsilon$~Hya, $\zeta$~UMa, and Algol \citep[see Table~5 of][and references therein]{OBrien2011}, and $\mu$~Ori, $\xi$~UMa, $\epsilon$~Hya, and 88~Tau \citep{Muterspaugh2008}. 
These studies, along with the earlier ones by \citet{Fekel1981} and \citet{Sterzik2002}, strongly argued against coplanarity, despite predictions from older theories of formation of multiple stars \citep{Bodenheimer1978}.
Modern theories of stellar formation \citep[][and references therein]{Sterzik2002,Fabrycky2007} indeed predicted that triple systems will form non-coplanarily. Clearly, this result jeopardizes the possibility of observing coplanarity on a larger scale such as that of the solar neighbourhood.
Therefore, the weak evidence for pole alignment observed in a sphere of 8~pc around the Sun does not receive support from our analysis of the properties of triple systems. In the next section, we show that neither do theories of binary-star formation coupled with the dispersive effect caused by Galactic rotation provide supportive arguments for a possible alignment of the orbital poles in the solar neighbourhood. 
 
\section{Pole alignment in the context of binary-star formation theories and Galactic kinematics}
\label{Sect:Galactic}

This section addresses the following two questions: 
(i) Do current theories of binary-star formation predict alignment of orbital poles for all binaries resulting from the same event of star formation, 
and (ii) assuming that the answer to the first question be positive, is this coherence in the pole orientation of young binary systems preserved by the Galactic kinematics in a volume-limited sample (the solar neighbourhood) consisting of much older systems?  

Theories for wide and close binary-star formation have been reviewed by \citet{Bodenheimer2001} and \citet{Bonnell2001}, respectively. 
The favoured mechanism for producing most binary stellar systems is the fragmentation of a molecular cloud core during its gravitational collapse. 
Fragmentation can be divided into two main classes: direct fragmentation and rotational fragmentation \citep[see references in][]{Bate2000}. 
Direct fragmentation depends critically on the initial density structure within the molecular cloud core (e.g., non-spherical shape or density perturbations are needed to trigger the collapse), whereas rotational fragmentation is relatively independent of the initial density structure of the cloud because the fragmentation occurs due to non-axisymmetric instabilities in a massive, rotationally supported disc or ring. Not many studies of binary-star formation however include a discussion about the possible spatial alignment of the orbital poles of binary systems resulting from the same fragmentation event. 
This lack of information results from the difficulty of conducting the numerical simulations till the end of the binary formation, since a phase of accretion follows the initial fragmentation of the molecular cloud. 
To obtain the final parameters of a stellar system, the calculation must thus be performed until all of the original cloud material has been accumulated by one of the protostars or by their discs. Following simultaneously several among such events is an even more challenging task thus.  
Although \citet{Bate2000} specifically investigated the properties of binary systems after this phase of accretion, there is no discussion whatsoever of the question of orbital-pole alignment. This was addressed in a very qualitative way by \citet{Boss1988}. 
A short discussion about the question of pole alignment has been found in the context of a variant of the fragmentation scenario, namely the shock-induced gravitational fragmentation followed by capture (SGF+C). That mechanism was initially proposed by \citet{Pringle1989}. 
Collisions between stable molecular cloud clumps produce dense shocked layers, which cool radiatively and fragment gravitationally. The resulting fragments then condense to form protostellar discs, which at the same time collapse and, as a result of tidal and viscous interactions, capture one another to form binary systems. When the initial clumps are sufficiently massive, a large number ($> 10$) of protostellar discs is produced.  
A prediction of the SGF+C mechanism \citep{Turner1995} is that {\it the resulting binary systems are likely to have their orbital and spin angular momenta aligned -- the preferred direction of alignment being that of the global angular momentum of the original clump/clump pair. However, with more realistic, that is less highly organized, initial conditions, such alignments should still arise locally but are likely to be less well coordinated across large distances.} 

Even though there is currently no strong evidence from theories of binary formation for pole alignment in binary systems resulting from the same molecular-cloud collapsing event, we assumed for now such a coherence and followed what happened as a result of Galactic kinematics dispersing these coeval binary systems across the Galaxy.  
After several Galactic rotations, this coeval association will (fully or partly) evaporate and form a tube called supercluster. \citet{Woolley1961} has presented an interesting argument that makes it possible to tag stars that formed at the same place and time, and that are still found together today in the solar neighbourhood. In general, stars found today in the solar neighbourhood emanate from various birth locations in the Galaxy because the diversity of stellar orbits in the Galaxy. Woolley's argument goes as follows. Stars in the supercluster mentioned above still share common $V$ velocities 
[with the usual notation $(U,V,W)$ for the Galactic velocity components directed towards the Galactic centre, along the direction of Galactic rotation, and towards the north Galactic pole, respectively] when located in the same region of the tube (for example in the solar neighbourhood) for the following reason, as put forward by \citet{Woolley1961}: if the present Galactocentric radius  of a star on a quasi-circular epicyclic orbit equals that of the Sun (denoted $r_\odot$), and if such a star is observed with a peculiar velocity $v = V + V_\odot$, then its guiding-centre radius $r_g$ writes 

\begin{equation}
\label{Eq:rg}
r_g = 
r_\odot - x_g = 
r_\odot + \frac{v}{2B},
\end{equation}

where $x_g$ is the position of its guiding-centre in the Cartesian reference frame associated with the local standard of rest (LSR) (in the solar neighbourhood approximation, the impact of $y_g$ is negligible), and $B$ is the second Oort constant. \citet{Woolley1961} pointed out that disc stars (most of which move on quasi-circular epicyclic orbits) that formed at the same place and time, and that because they stayed together in the Galaxy after a few galactic rotations (since they are all currently observed in the solar neighbourhood) must necessarily have the same period of revolution around the Galactic centre, and thus the same guiding-centre $r_g$, and thus the same velocity $V = v - V_\odot$ according to Eq.~\ref{Eq:rg}. 

Therefore, according to this argument, binary systems that formed in the same cluster or association and that stayed together after a few Galactic revolutions are observed today in the solar neighbourhood as sharing the same $V$ velocity. In other words, if there are any orbital-pole concentrations on the celestial sphere that can be ascribed to angular momentum conservation for binary systems originating from the same cluster or association, they should be identified from similar $V$
velocities.

\begin{figure}[t] \centering
\includegraphics[width=9cm]{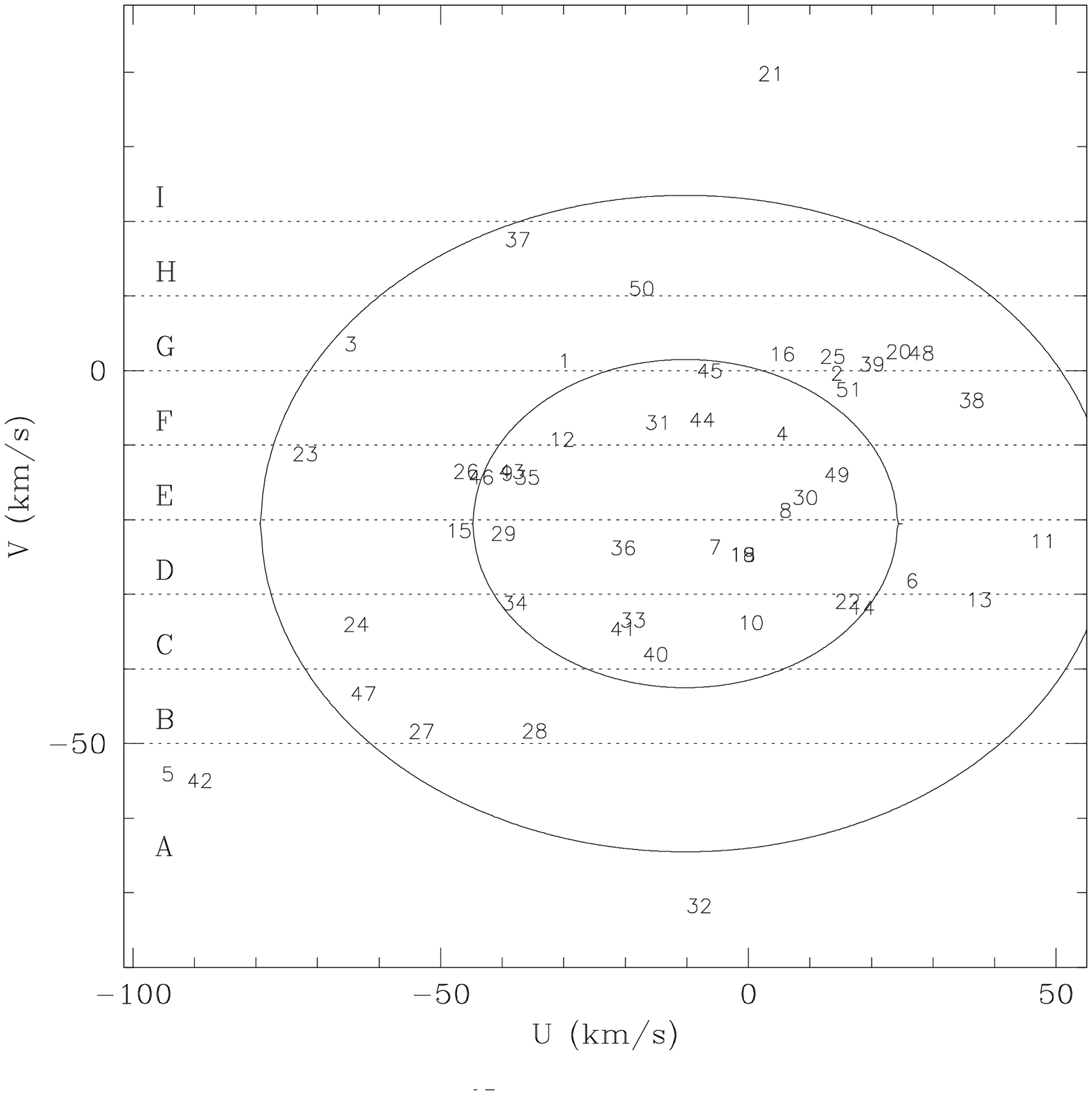} 
\caption[]{ Distribution of the systems in the $(U,V)$ plane, with the same labels as in Fig.~\ref{Fig:aitoff}.
System~17 (01083+5455) falls outside the boundaries, at $U = -43$~\kms\ and $V = -157$~\kms.
To fix the ideas, two velocity ellipso\"\i ds are shown, both centred on the reflex solar motion in the LSR: $-U_\odot = -10.2$~\kms,
$-V_\odot = -20.5$~\kms, and with $\sigma_U = 34.5$~\kms, $\sigma_V = 22.5$~\kms,  
as derived by \citet{Famaey2005} from their most precise sample. Horizontal bands and the 
corresponding labels are used in Fig.~\protect\ref{Fig:aitoff_galactic}.
 }
\label{Fig:galactic} 
\end{figure}

We first computed the $(U,V,W)$ velocity components from the Hipparcos proper motion and from the systemic velocity, according to standard prescriptions for converting equatorial to Galactic coordinates. The resulting distribution in the $(U,V)$ plane is shown in Fig.~\ref{Fig:galactic}.

\begin{figure}[t] \centering
\includegraphics[width=9cm]{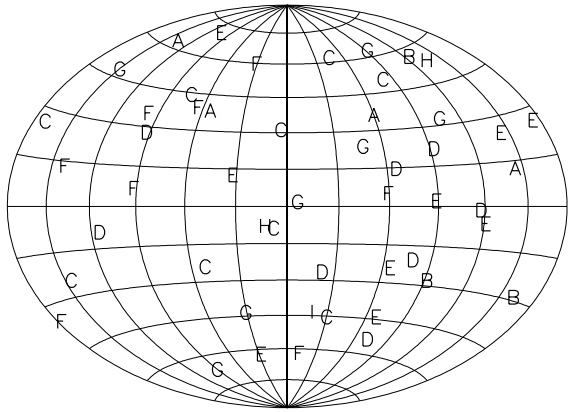} 
\caption[]{ Same as Fig.~\ref{Fig:aitoff} but with a letter coding the value of the galactic $V$ velocity 
(see Fig.~\protect\ref{Fig:galactic}). A necessary condition for binary systems to originate
from the same cluster/association at birth is to have similar $V$ velocities (see text), 
hence they should be labelled by the same letter.
}
\label{Fig:aitoff_galactic} 
\end{figure}

As expected, the sample is distributed according to the general velocity ellipso\"\i d, with overdensities observed at the location of the Sirius 
$<\!U\!> \sim 6.5$~km~s$^{-1}$, $<\!V\!> \sim 3.9$~km~s$^{-1}$) and Hyades ($<\!U\!> \sim -40$~km~s$^{-1}$, $<\!V\!> \sim -20$~km~s$^{-1}$) streams 
\citep[e.g.,][]{Dehnen1998,Famaey2005}. 
Because of Woolley's argument expressed above, the $(U,V)$ diagram was divided into horizontal bands of width 10~\kms, as indicated in Fig.~\ref{Fig:galactic}, with a letter assigned to each of these $V$-bands.  
These letters were used to label the position of each pole on a Galactic-coordinate map (Fig.~\ref{Fig:aitoff_galactic}). 

The absence of clustering of similar letters in restricted regions of the celestial sphere provides definite proof that any possible pole concentration on the sphere (see Sect.~\ref{Sect:analysis}) cannot be traced back to a similar birth location for those systems. It must therefore either be ascribed to statistical fluctuations in our small-sized sample, or to some as yet unidentified physical process that removes orbital poles from the region $90^\circ \le l \le 300^\circ$, $b \le 0^\circ$, on a time scale of the order of that characterizing the convergence of systems with different Galactic orbits in the solar neighbourhood. Our Galaxy being in general symmetric with respect to its equatorial plane, it is however difficult to imagine a physical process that would break such a symmetry (the fact that the Sun lies somewhat above the Galactic plane? see below), as it seems to be required to account for the observed asymmetry.

Conversely, the poles of the binary systems belonging to the Sirius or Hyades streams do not concentrate either on the sphere, which is a further confirmation that these streams are not evaporated clusters, but rather have a dynamical origin \citep{Famaey2005,Famaey2007,Famaey2008,Pompeia2011}.

Finally, one should mention, for the sake of completeness, that the location of the Sun slightly above the Galactic plane \citep[$\sim 20$~pc;][]{Humphreys1995} could be responsible for the (Galactic) North-South asymmetry observed in the orbital-pole distribution, if some effect related to Galactic tides were the cause of the pole anisotropy.
That seems unlikely, however, given the low intensity of  Galactic tides with respect to the two-body interaction. \citet{Heisler1986} have expressed Galactic tides in terms of the Oort constants. Based on the size of the last closed Hill  surface\footnote{The Hill (or zero-velocity) surface is defined as the surface on which the Jacobi integral of the restricted three-body problem is constant.}, they conclude that in the solar neighbourhood, the Galactic tides have a non-negligible impact only on systems with semi-major axes of several $10^4$~AU (such as Oort-cloud comets around the Sun). The binary systems defining the pole statistics are very much tighter than this, so that there seems to be no way for the Galactic tides to imprint their signature on the orbital-pole distribution.


\section{Conclusion} \label{Sect:Conclusion} 

Among the 95 systems with an orbit in the 6th Catalogue of Orbits of Visual Binaries that are closer than 18~pc from the Sun, we were able to lift the pole ambiguity for 51 systems, thanks to radial-velocity data either collected from the literature, obtained from the CORAVEL database, or acquired with the HERMES/Mercator spectrograph. 
Of these 51 systems, several had an erroneous node choice listed in the 6th USNO orbit catalogue and had thus to be corrected.
An interesting side product of the present study is  a number of new spectroscopic orbits obtained from CORAVEL and HERMES radial velocities, as well as new combined spectroscopic/astrometric orbits for seven systems  [WDS 01083+5455aa,Ab; 01418+4237AB; 02278+0426AB (SB2); 09006+4147AB (SB2); 16413+3136AB; 17121+4540AB; 18070+3034AB].

This new sample of 51 orbital poles was subjected to several methods of spherical statistics (Rayleigh-Watson, Bingham, Beran and Gin\'e) to test possible deviations from isotropy. After ordering the binary systems by increasing distance from the Sun, the false-alarm probability (of rejecting the null hypothesis of  isotropy while it is true) was computed for subsamples of increasing sizes, from $N = 1$ up to the full sample of 51 systems. Two tests (Rayleigh-Watson's and Beran's) delivered a false-alarm probability of 0.5\% when considering the subsample of 21 systems closer than 8.1~pc to the Sun. The poles tend to cluster toward Galactic position $l = 46^\circ$, $b = 37^\circ$. This direction is not so distant from that of the ecliptic pole ($l = 96^\circ$, $b = 30^\circ$).
To evaluate the robustness of this clustering, a jackknife approach was used by repeating the above procedure after removing one system at a time from the full sample.
The above false-alarm probability was then found to vary between 1.5\% and 0.1\%, depending on which system is removed from the sample. The reality of the deviation from isotropy can thus not be assessed with certainty at this stage, given the small number of systems available, despite our efforts to increase it. 
Our so-far uncertain conclusion should be seen as an incentive to foster further studies on this problem, especially in the Gaia era. 

For the full sample extending up to 18~pc, the clustering totally vanishes (the Rayleigh-Watson first-kind risk then rises to 18\%). If present, coplanarity of orbits must thus be restricted to small spheres with a radius of the order of 10~pc, but several dynamical arguments were presented to show that the physical origin of any such coplanarity, should it exist, will be difficult to identify. In this context, the recent result of \citet{Rees2013} on the non-random orientation of {\it{ bipolar}} planetary nebulae in the Galactic Bulge, using a statistical method much like the ones used in our present study, came as a surprise.
Rees \& Zijlstra found evidence for an anisotropy in the orientation of the bipolar planetary nebulae, whose specific geometry is usually attributed to the binarity of the central star.
Assuming that the symmetry axis of the bipolar nebula coincides with the polar axis of the orbit, the result of Rees \& Zijlstra is therefore indirect evidence of a deviation from anisotropy for the poles of a subsample of binaries in the Galactic Bulge. 
The physical explanation proposed for the Galactic Bulge (related to the impact of the Galactic magnetic field on the binary formation process) is, however, unlikely to hold also for the solar neighbourhood, where the field is of lower intensity.

\begin{acknowledgements}{
\\
We dedicate this paper to the memory of J. Dommanget, who initiated this study and could unfortunately not see it accomplished. 
He would certainly have preferred a more significant result, but he had faith in statistics.
\\
We thank the anonymous referee for his constructive remarks and attitude.
We thank A. Lobel, P. Neyskens and C. Siopis, who acted as observers on the HERMES spectrograph and secured the necessary spectra to solve the pole ambiguity for several binary systems. Based on observations made with the Mercator Telescope, operated on the island of La Palma by the Flemish Community, at the Spanish Observatorio del Roque de los Muchachos of the Instituto de Astrofisica de Canarias. Based on observations obtained with the HERMES spectrograph, which is supported by the Fund for Scientific Research of Flanders (FWO), Belgium, the Research Council of K.U. Leuven, Belgium, the Fonds National de la Recherche Scientifique (FNRS), Belgium, the Royal Observatory of Belgium, the Observatoire de Gen\`eve, Switzerland and the Th\"uringer Landessternwarte Tautenburg, Germany.
We gratefully thank D. Pourbaix ({\it Institut d'Astronomie et d'Astrophysique de l'Universit\'e Libre de Bruxelles}) for his help in the calculation of the new orbits, 
as well as G. Marcy for sharing his velocity measurements of WDS~00057+4549AB.
This research has made use of the Simbad database, operated at the {\it Centre de Donn\'ees Astronomiques de Strasbourg} (CDS), France.  \footnote{Available at http://cdsweb.u-strasbg.fr/}
This research has made use of the Washington Double Star Catalog maintained at the U.S. Naval Observatory. \footnote{Available at http://ad.usno.navy.mil/wds/}
}
\end{acknowledgements}
\bibliographystyle{aa}
\bibliography{agati}

\appendix
\section{Notes on individual systems and new orbits}
\label{Sect:notes}
\begin{table*}
\caption[]{\label{Tab:VrHERMES}
HERMES and CORAVEL radial velocities for components of some visual binaries lacking velocities so far. 
The quoted error corresponds to the fitting error, but does not account for the pressure-driven zero-point shift, which can amount up to 200~m~s$^{-1}$ for HERMES velocities \citep{Raskin2011}.}
\begin{tabular}{lcccc}
\hline\\
JD$ - 2400000$ & $Vr(A)$ & $Vr(B)$ & HER/COR & Note \\ 
\hline\\
\noalign{WDS 09006+4147AB}\\
55520.795 & $29.06\pm 0.09$ & $ 22.06\pm 0.08$ & H & a\\ 
55521.640 & $29.03\pm 0.10$ & $ 21.93\pm 0.09$ & H & -\\ 
55579.783 & $29.44\pm 0.09$ & $ 21.83\pm 0.08$ & H & -\\
55589.620 & $29.53\pm 0.09$ & $ 21.68\pm 0.08$ &  H &-\\ 
55647.599 & $29.64\pm 0.08$ & $ 21.45\pm 0.07$ &  H &-\\ 
56046.402 & $31.11\pm 0.09$ & $ 19.92\pm 0.05$ & H & -\\ 
56244.717 & $31.54\pm 0.13$ & $ 19.48\pm 0.07$ & H & -\\
\\
\noalign{WDS 09144+5241A}\\
56039.412 & $10.620\pm0.002$ & -&  H &-\\
\\
\noalign{WDS 09144+5241B}\\
56039.418 & -& $11.848\pm0.002$&  H &-\\
\\
\noalign{WDS 10454+3831AB}\\
56040.468 & $-3.380\pm0.003$& -&  H &-\\
56040.485 & -& $-3.175\pm0.005$ &  H & b\\
\\
\noalign{WDS 13491+2659AB}\\
 45441.525  &$-20.32\pm 0.32$ & - & C & -\\
 45475.401  &$-19.29\pm 0.45$ &$-22.70\pm 0.42$  & C &-\\
 45789.537  &$-20.27\pm 0.33$ & $-23.18\pm 0.37$  & C &-\\
 46214.395  &$-20.06\pm 0.32$ & $-23.95\pm 0.35$  & C &-\\
 46562.422  &$-20.48\pm 0.31$ & $-23.80\pm 0.37$  & C &-\\ 
 51018.433  &$-20.27\pm 0.44$ & $-23.01\pm 0.46$  & C &-\\
\\
\noalign{WDS 17153-2636A}\\
56038.716 & $0.532\pm0.002$& -&  H &-\\
\\
\noalign{WDS 17153-2636B}\\
56038.719 & -& $0.246\pm0.001$&  H &-\\
\\
\noalign{WDS 17364+6820Aa,Ab}\\
556038.682 & $-30.0\pm0.3$ & $-27.0\pm0.5$  &H & c\\
\\
\noalign{WDS 18428+5938AB}\\
55655.719 & $-1.061\pm0.006$& -& H &-\\
55660.748 & $-1.048\pm0.005$& -&  H &- \\
56037.736 & $-1.074\pm0.005$& -&  H &- \\
56038.695 & $-1.088\pm0.005$& -& H & - \\
56038.699 &   -& $0.779\pm0.006$&  H &-\\
\\
\noalign{WDS 19255+0307Aa,Ab}\\
55657.742 & $-28.68\pm0.26$& $-37.70\pm0.09$&  H & \\
\hline\\
\end{tabular}
a: Component A ($V = 4.2$), of spectral type F4V, is a fast rotator. 
It has the largest peak on the HERMES cross-correlation function (Fig.~\ref{Fig:CCF76943}). 
For details see note in Appendix~\ref{Sect:notes}.\\
b: barely separable.\\
c: "blind" assignment based on the depth of the cross-correlation  peak: component A is assumed to have the deeper peak.\\
\end{table*}

\begin{landscape}
\begin{table}
\caption[]{
\label{Tab:New_SB}
New combined astrometric-spectroscopic orbits computed from CORAVEL and/or HERMES velocities. The value of $\omega$ given in  the present table corresponds to that of the visual orbit of B around A (it thus differs by 180$^\circ$ from that of A around the centre of mass of the system, obtained for purely spectroscopic binaries). 
}
\begin{tabular}{lllllllllllllllllllllllll}
\hline\\
\multicolumn{1}{c}{WDS} & \multicolumn{1}{c}{$P$} & \multicolumn{1}{c}{$a$} &\multicolumn{1}{c}{$i$} &   \multicolumn{1}{c}{$e$}  & \multicolumn{1}{c}{$\omega_B$}  &\multicolumn{1}{c}{$\Omega$} &\multicolumn{1}{c}{$T_0$}  & \multicolumn{1}{c}{$V_0$} & \multicolumn{1}{c}{$K_A$} \\
& \multicolumn{1}{c}{(yr)} & \multicolumn{1}{c}{$('')$} & \multicolumn{1}{c}{($^\circ$)} & & \multicolumn{1}{c}{($^\circ$)}  & \multicolumn{1}{c}{($^\circ$)} &\multicolumn{1}{c}{($yr$)} &\multicolumn{1}{c}{(km/s)} & \multicolumn{1}{c}{(km/s)} \\
\hline\\
01083+5455Aa,Ab & 21.40 $\pm$ 0.11 & 1.07  $\pm$ 0.04 & 104.7 $\pm$  1.9 & 0.53 $\pm$  0.02 & 330.0  $\pm$ 3.4 & 224.6  $\pm$  2.0 &1975.9  $\pm$ 0.1 & -97.35  $\pm$  0.04  & 2.13  $\pm$  0.11\\ 
01418+4237AB & 19.73 $\pm$ 0.09 & 0.61  $\pm$ 0.03 & 99.4  $\pm$  1.5 & 0.43 $\pm$  0.02 & 223.4  $\pm$ 2.3 & 30.2 $\pm$ 1.7    &1997.0  $\pm$ 0.1 & 3.31    $\pm$  0.03  & 2.93  $\pm$  0.07\\
16413+3136AB & 34.447  $\pm$ 0.004 & 1.333  $\pm$ 0.004 & 132.0  $\pm$  0.2 & 0.452  $\pm$  0.001 & 297.0  $\pm$ 0.2 & 235.2  $\pm$  0.2 &1933.26  $\pm$ 0.009 & -70.441  $\pm$  0.008  & 4.01  $\pm$  0.01\\ 
17121+4540AB & 12.85$\pm$0.05 & $0.88\pm0.02$ & 134.5$\pm$1.9 & 0.80$\pm$0.01 & 99.9$\pm$1.8 & 165.6$\pm$2.5 & 1990.84$\pm$0.03 & -30.45$\pm$0.05 & 1.8$\pm$0.1\\
18070+3034AB & 56.37  $\pm$ 0.06 & 1.07  $\pm$ 0.02 & 39.8  $\pm$ 1.6 &  0.772  $\pm$ 0.004 & 290.7  $\pm$ 1.0 & 227.2  $\pm$ 1.3 &  1997.72 $\pm$  0.03 & 0.267  $\pm$ 0.027 &  3.23  $\pm$ 0.03 \\
\hline\\
\end{tabular}
\end{table}
\end{landscape}

\begin{table}
\caption[]{
\label{Tab:SB2}
New combined astrometric-spectroscopic orbits (2 observable spectra) computed from CORAVEL or HERMES velocities.  The value of $\omega$ given in  the present table corresponds to that of the visual orbit of B around A (it thus differs by 180$^\circ$ from that of A around the centre of mass of the system, obtained for purely spectroscopic binaries).  $\varpi_{\rm dyn}$ is the orbital parallax, which is obtained from the ratio $a$ ('')/ $a$ (AU).
}
\begin{tabular}{llll}
\hline\\
WDS& 02278+0426AB& 09006+4147AB\\
\hline\\
$P$ (yr) & 25.14 $\pm$  0.04 & 21.78 $\pm$  0.02\\
$a$ ('')   &  0.5429  $\pm$ 5.2e-03 & 0.633 $\pm$ 0.004\\
$a$ (AU)      &      10.390  $\pm$ 0.086 & 11.76 $\pm$ 0.04\\
$i$ ($^\circ$)    &       73.0  $\pm$ 0.7 & 133.2 $\pm$ 0.6\\
$\omega_{\rm 2\; around\; 1}$ ($^\circ$)   &      49.1  $\pm$ 2.6 & 212.7 $\pm$ 1.9\\
$\Omega$ ($^\circ$)  &       290.1 $\pm$ 1.0 & 23.8 $\pm$ 0.8\\
$e$               &    0.21  $\pm$ 0.01 & 0.154 $\pm$ 0.004\\
$T_0$ (Besselian year) & 1962.8 $\pm$  0.2 & 1972.0 $\pm$ 0.1\\
$V_0$ (km/s)    &      6.80  $\pm$ 0.15 & 25.6 $\pm$ 0.1\\
$\varpi_{\rm dyn} ('')$     &     0.052  $\pm$ 0.002 & 0.054 $\pm$ 0.001\\
$M_1$ (M$_\odot$) \tablefootmark{a} &  0.95  $\pm$ 0.10 & 1.73:\\
$M_2$ (M$_\odot$)&  0.85 $\pm$  0.10 & 1.69:\\
$M_2/(M_1+M_2)$ & 0.47 $\pm$ 0.02 & 0.5\\
$K_1$ (km/s)   &       5.6  $\pm$ 0.3  $^a$ & 5.9 $\pm$ 0.2 \\
$K_2$ (km/s)     &      6.4  $\pm$ 0.3 & 6.0 $\pm$ 0.1 \\
\hline\\
\end{tabular}
\tablefoot{
\tablefootmark{a} {Most likely, spectroscopic component 1 is WDS 02278+0426 A (see text for details).}
}
\end{table}

\begin{table}
\caption[]{
\label{Tab:VrHIP473}
CORAVEL velocities (labelled C in the last column), \citet{Tokovinin2002} velocities (labelled T), and \citet{Chubak2012} velocities (labelled M) for the AB components of WDS 00057+4549AB = HIP 473. 
}
\begin{tabular}{lrrrrrr}
\hline\\
 \multicolumn{1}{c}{JD - 2400000} &  \multicolumn{2}{c}{$Vr$ A} &  \multicolumn{2}{c}{$Vr$ B}& $Vr$ B-A\\
           &   \multicolumn{2}{c}{(km/s)} &   \multicolumn{2}{c}{(km/s)}&  (km/s)\\
\hline\\
43791.437  &   1.83 & 0.43 & -1.99 &0.46  & -3.82  & C \\
43791.450  &   1.16 & 0.48 & -1.95 &0.49  & -3.11  & C \\
43795.447  &   1.45 & 0.43 & -1.59 &0.47  & -3.04  & C \\
43806.403  &   1.16 & 0.43 & -1.92 &0.53  & -3.08  & C \\
43808.388  &   1.32 & 0.43 & -1.74 &0.53  & -3.06  & C \\
43813.391  &   0.19 & 0.51 & -1.41 &0.52  & -1.60  & C \\
44211.319  &   0.67 & 0.47 & -0.35 &0.61  & -1.02  & C \\
44849.517  &   1.48 & 0.45 & -2.26 &0.45  & -3.74  & C \\
45264.406  &   1.06 & 0.37 & -1.91 &0.36  & -2.97  & C \\
45615.458  &   1.63 & 0.48 & -2.57 &0.46  & -4.20  & C \\
46008.457  &   1.86 & 0.48 & -1.99 &0.53  & -3.87  & C \\
46016.404  &   1.12 & 0.56 & -2.81 &0.54  & -3.93  & C \\
46372.418  &   0.85 & 0.33 & -2.53 &0.50  & -3.38  & C \\
46373.385  &   0.87 & 0.34 & -2.69 &0.44  & -3.56  & C \\
46413.273  &   0.96 & 0.43 & -2.84 &0.48  & -3.80  & C \\
46725.421  &   0.87 & 0.52 & -2.55 &0.58  & -3.42  & C \\
47035.584  &   1.01 & 0.50 & -2.18 &0.51  & -3.19  & C \\
47556.262  &  -0.01 & 0.39 & -2.14 &0.40  & -2.13  & C \\
47733.622  &   0.12 & 0.54 & -1.13 &0.58  & -1.25  & C \\
47853.376  &   1.17 & 0.48 & -1.54 &0.56  & -2.71  & C \\
48119.635  &   0.66 & 0.38 & -2.44 &0.39  & -3.10  & C \\
48237.300  &   0.72 & 0.39 & -1.91 &0.37  & -2.63  & C \\
48495.535  &   0.81 & 0.38 & -2.39 &0.38  & -3.20  & C \\
48571.438  &   0.87 & 0.31 & -2.54 &0.40  & -3.41  & C \\
48860.589  &   0.66 & 0.50 & -2.74 &0.51  & -3.40  & C \\
49036.326  &   0.58 & 1.07 &   -   & -    &        & C \\
49592.549  &   0.37 & 0.37 & -2.43 &0.35  & -2.80  & C \\
49598.459  &   1.43 & 0.40 & -1.96 &0.39  & -3.39  & C \\
49950.579  &   0.20 & 0.57 & -4.64 &0.59  & -4.84  & C \\
49975.449  &   0.34 & 0.24 & -2.30 &0.25  & -2.64  & T \\
50324.457  &   0.06 & 0.34 &       &      &        & T \\
50341.514  &   1.73 & 0.46 & -3.97 &0.51  & -5.70  & C \\
50650.553  &   1.57 & 0.27 & -3.06 &0.34  & -4.63  & T \\
50658.553  &   0.63 & 0.42 & -2.46 &0.56  & -3.09  & T \\
50667.570  &   2.76 & 0.73 & -1.12 &0.63  & -3.88  & T \\
50758.298  &   0.21 & 0.54 & -0.87 &0.52  & -1.08  & T \\
51065.517  &   2.00 & 0.32 & -1.92 &0.26  & -3.92  & T \\
54172.5    &   1.654 & 0.14 & -1.700 &0.10  & -3.35  &M\\
\hline\\
\end{tabular}
\end{table}

\begin{figure}
\includegraphics[width=9cm]{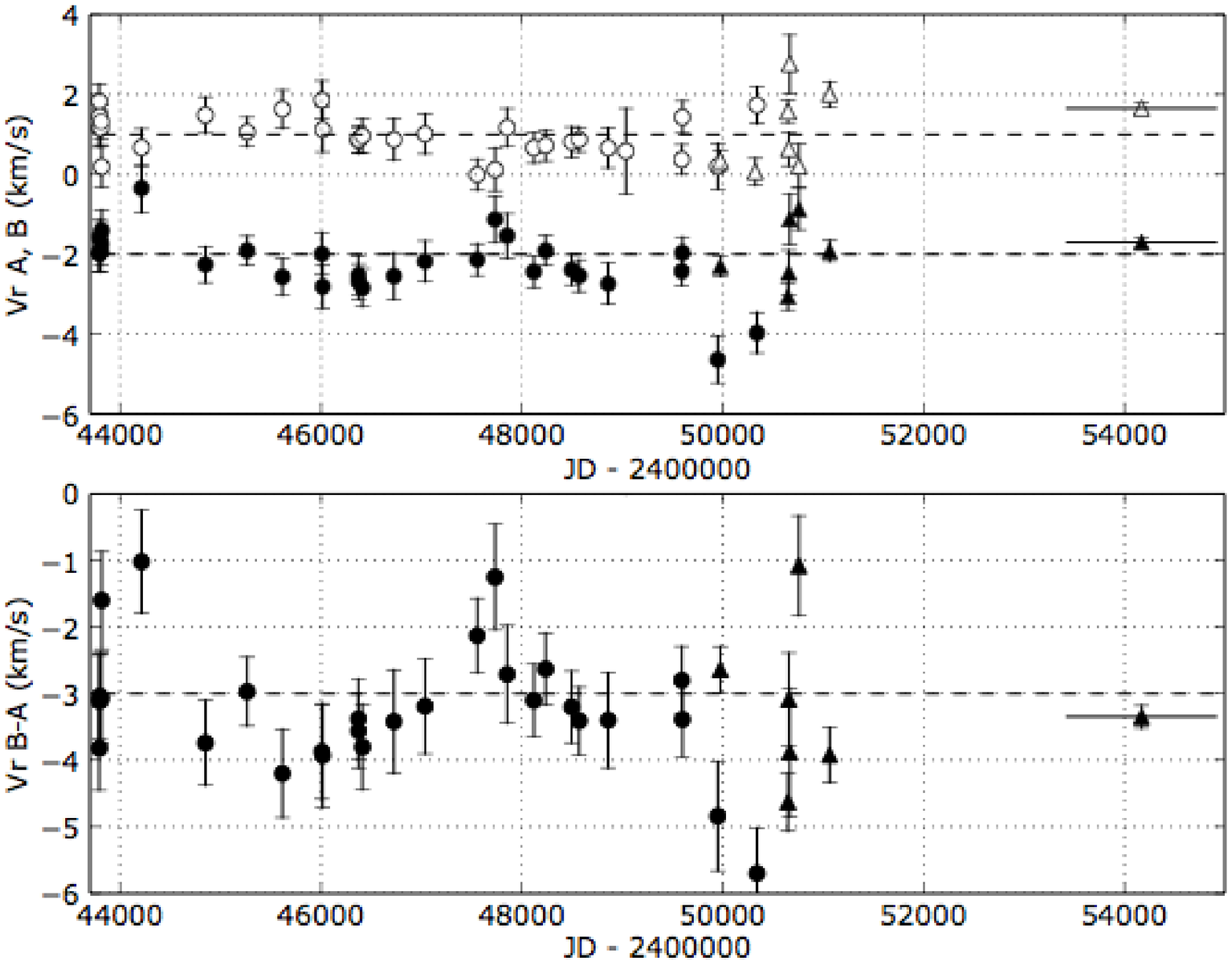}
\caption[]{\label{Fig:HIP473}
Top panel: Radial velocities for the two components of the pair HIP 473 AB = WDS 00057+4549AB. Dots refer to CORAVEL velocities and triangles to velocities from \citet{Tokovinin2002} or \citet{Chubak2012} for the last point, with the time spanned by the measurements represented by the horizontal bar. 
Open symbols refer to component A and filled symbols to component B. The parallel dashed lines are just a guide to the eye. Bottom panel: Radial-velocity difference B-A, with their errors taken as the root-mean-square of the errors on A and B. 
}
\end{figure}

\noindent{\bf WDS 00057+4549AB.} The Sixth Catalog of Orbits of Visual Binary Stars lists two orbits for the AB pair of that system
(separated by about $6\arcsec$): one with a period of 1550 yr \citep{Pop1996b} and another with a period of 509~yr \citep{Kiyaeva2001}, indicating that the solution is not yet well constrained. 
Radial velocities obtained by CORAVEL and by \citet{Tokovinin2002}, as listed in Table~\ref{Tab:VrHIP473} and displayed in Fig.~\ref{Fig:HIP473}, correspond to a radial velocity difference $V_{\mathrm{B}}~-~V_{\mathrm{A}}$ slightly decreasing from $-2.7$~km/s in 1980 to $-3.8$~km/s in 1996. The value $-3.0$~km/s has been used by \citet{Kiyaeva2001} to determine unambiguously the ascending node.
This result is confirmed thanks to the precise radial-velocity measurements of Marcy (2013, private communication) that show a linear variation for $V_{\mathrm{B}}~-~V_{\mathrm{A}}$ amounting to $-6.4$~m~s$^{-1}$~y$^{-1}$ from 2000 to 2010, while a value of $-11.5$~m~s$^{-1}$~y$^{-1}$ is calculated from the visual orbit.
Thus we have computed the position of the orbital pole but it is not expected to be precise.

\citet{Kiyaeva2001} also suggested that ADS 48 ABF is a hierarchical triple system whose orbits (AB) and (AB-F) are not coplanar. This hypothesis was contradicted by \citet{Cvetkovic2012}, who concluded that the F component has a common proper motion with the AB pair, but that it is not bound by gravity.

\noindent{\bf WDS 00184+4401AB.} Indeterminate visual orbit.

\noindent{\bf WDS 00321+6715Aa,Ab and AB.} The preliminary visual orbits of this triple hierarchical system were computed by \citet{Doc2008d}. No radial velocity measurements are available to resolve the ambiguity of the ascending node. 

\begin{figure}
\includegraphics[width=8cm]{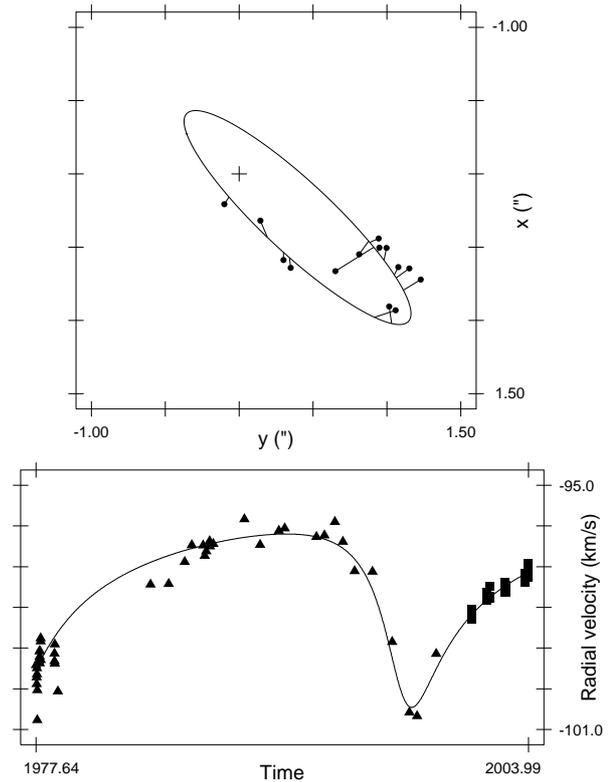}
\caption[]{\label{Fig:HIP5336}
Combined astrometric (upper panel) and spectroscopic (lower panel) orbits for WDS 01083+5455Aa,Ab = HIP 5336 = HD 6582 = $\mu$~Cas. The solid line in the lower panel corresponds to the predicted velocity curve for A. Triangles are CORAVEL measurements, squares are measurements from \citet{Abt2006}. Individual measurements for the visual orbit were taken from the WDS database.
}
\end{figure}
\begin{table}
\caption[]{
\label{Tab:VrHIP5336}
CORAVEL velocities for the components of WDS 01083+5455Aa,Ab = $\mu$~Cas.  
}
\begin{tabular}{rrrrrr}
\hline\\
Date & JD - 2400000 &  \multicolumn{2}{c}{$Vr$} \\
        &          &   \multicolumn{2}{c}{(km/s)} \\
\hline\\
 210877 &43377.577&  -99.42& 0.56\\
 050977 &43392.532&  -99.72& 0.41\\
 050977 &43392.535&  -99.88& 0.49\\
 070977 &43394.540&  -99.64& 0.45\\
 070977 &43394.542&  -99.49& 0.41\\
 170977 &43404.505& -100.03& 0.51\\
 170977 &43404.509& -100.77& 0.52\\
 171077 &43434.481&  -99.30& 0.47\\
 171077 &43434.483&  -99.32& 0.45\\
 011177 &43449.431&  -99.21& 0.62\\
 011177 &43449.435&  -99.07& 0.48\\
 011177 &43449.435&  -99.07& 0.48\\
 011177 &43449.438&  -99.37& 0.51\\
 191177 &43467.378&  -98.83& 0.39\\
 191177 &43467.381&  -98.76& 0.39\\
 191177 &43467.383&  -99.29& 0.44\\
 180878 &43739.617&  -99.13& 0.37\\
 190878 &43740.621&  -99.32& 0.34\\
 240878 &43745.611&  -98.91& 0.33\\
 240878 &43745.615&  -99.38& 0.33\\
 211078 &43803.444& -100.06& 0.35\\
 081083 &45616.487&  -97.44& 0.35\\
 240984 &45968.530&  -97.42& 0.34\\
 070885 &46285.624&  -96.88& 0.35\\
 201285 &46420.266&  -96.47& 0.37\\
 020886 &46645.615&  -96.48& 0.32\\
 300886 &46673.600&  -96.73& 0.36\\
 021086 &46706.588&  -96.62& 0.36\\
 071286 &46772.252&  -96.37& 0.37\\
 111286 &46776.302&  -96.50& 0.36\\
 170287 &46844.400&  -96.44& 0.33\\
 121088 &47447.622&  -95.83& 0.37\\
 170889 &47756.617&  -96.46& 0.36\\
 170890 &48121.622&  -96.12& 0.36\\
 141290 &48240.385&  -96.06& 0.36\\
 200892 &48855.626&  -96.27& 0.35\\
 250193 &49013.283&  -96.23& 0.36\\
 180893 &49218.623&  -95.90& 0.33\\
 200194 &49373.279&  -96.39& 0.35\\
 080994 &49604.572&  -97.11& 0.35\\
 200895 &49950.602&  -97.12& 0.44\\
 130996 &50340.559&  -98.85& 0.36\\
 120897 &50673.624& -100.58& 0.39\\
 090198 &50823.228& -100.67& 0.32\\
 200199 &51199.245&  -99.14& 0.34\\
\hline\\
\end{tabular}
\end{table}

\noindent{\bf WDS 01083+5455Aa,Ab.}  \citet{Dru1995} obtained the first relative astrometric orbit for that system from their direct detection of the faint B component. Combining that orbit with earlier photocentric positions  \citep{Lippincott1981,Russell1984}, masses could be derived: $M_A = 0.74\pm0.06$~M$_\odot$ and $M_B = 0.17\pm0.01$~M$_\odot$.

\citet{Duquennoy1991} derived a preliminary spectroscopic orbit,  and \citet{Abt2006} obtained thereafter more radial velocities; however, \citet{Abt2006} did not attempt to compute a combined orbit, and merged the orbital elements of\citet{Duquennoy1991} and  \citet{Dru1995}. Because this is not fully satisfactory and is prone to confusion (as further discussed below), especially regarding the ambiguity on $\omega$, we decided to compute a new combined orbit (Fig.~\ref{Fig:HIP5336}), using  data of \citet{Duquennoy1991} as well as more recent unpublished CORAVEL data (Table~\ref{Tab:VrHIP5336}), plus velocities reported by \citet{Abt2006}. The astrometric positions of B relative to A taken from the WDS database at USNO were used as well.  As shown in Table~\ref{Tab:New_SB}, the value of $\omega$  for the astrometric orbit of B around A is 329.4$^\circ\pm3.8^\circ$ (thus a change of 180$^\circ$ from the relative orbit of\citet{Dru1995}; their Table~9), and we stress that the value (332.7$^\circ\pm3.1^\circ$) listed by \citet{Abt2006} for the spectroscopic orbit (of A around the centre of mass of the system) is incorrect, as it should differ by 180$^\circ$ from the astrometric value, confirming the suspected confusion. It is in fact just the value reported by \citet{Dru1995}, which should not have been copied without change! Instead, the value ($147.9^\circ\pm4.9^\circ$) of\citet{Duquennoy1991} for the spectroscopic orbit is correct.

\noindent{\bf WDS 01398-5612AB.} Indeterminate visual orbit. In the absence of recent radial-velocity measurements it is not possible to lift the ambiguity on the position of the ascending node.

\noindent{\bf WDS 01418+4237AB.} Spectroscopic orbits were published by \citet{Duquennoy1991} and \citet{Abt2006}.
The present orbit (Table~\ref{Tab:New_SB}) improves upon those, as it includes several unpublished CORAVEL velocities given in Table~\ref{Tab:VrHD10307}.
\citet{Lippincott1983} and \citet{Sod1999} published astrometric orbits. 
Here, we obtained a combined astrometric/spectroscopic solution by considering the astrometric measurements from the Fourth Catalogue of Interferometric Measurements of Binary Stars\footnote
{Available from http://www.usno.navy.mil/USNO/astrometry/optical-IR-prod/wds/int4}. This solution agree well with the existing solutions quoted above.

\begin{table}
\caption[]{
\label{Tab:VrHD10307}
New CORAVEL velocities for the components of WDS 01418+4237AB in complement of velocities published by \citet{Duquennoy1991}.  
}
\begin{tabular}{rrrrrr}
\hline\\
Date & JD - 2400000 &  \multicolumn{2}{c}{$Vr$} \\
        &          &   \multicolumn{2}{c}{(km/s)} \\
\hline\\
 160890& 48120.647&    4.12& 0.33\\
 070990& 48142.581&    4.31& 0.33\\
 050891& 48474.627&    3.32& 0.32\\
 240891& 48493.570&    3.44& 0.30\\
 180892& 48853.646&    2.64& 0.32\\
 250193& 49013.276&    2.91& 0.32\\
 180893& 49218.649&    2.14& 0.32\\
 200194& 49373.273&    1.86& 0.31\\
 080994& 49604.602&    0.87& 0.32\\
 190895& 49949.643&   -0.27& 0.38\\
 130996& 50340.573&   -0.64& 0.33\\
 090198& 50823.239&    1.06& 0.32\\
 200199& 51199.263&    3.04& 0.32\\
\hline\\
\end{tabular}
\end{table}

\noindent{\bf WDS 01425+2016AB.} This system was announced as an astrometric binary with a period of 0.567~yr by the {\it Double and Multiple Star Annex} (DMSA) of the Hipparcos Catalogue \citep{ESA-1997}. However, reprocessing the Hipparcos data along the method outlined by \citet{Pourbaix2000} for a combined solution including the CORAVEL velocities does not allow us to confirm the Hipparcos DMSA/O solution. This system was therefore dropped from our final list.

\begin{figure}
\includegraphics[width=8cm]{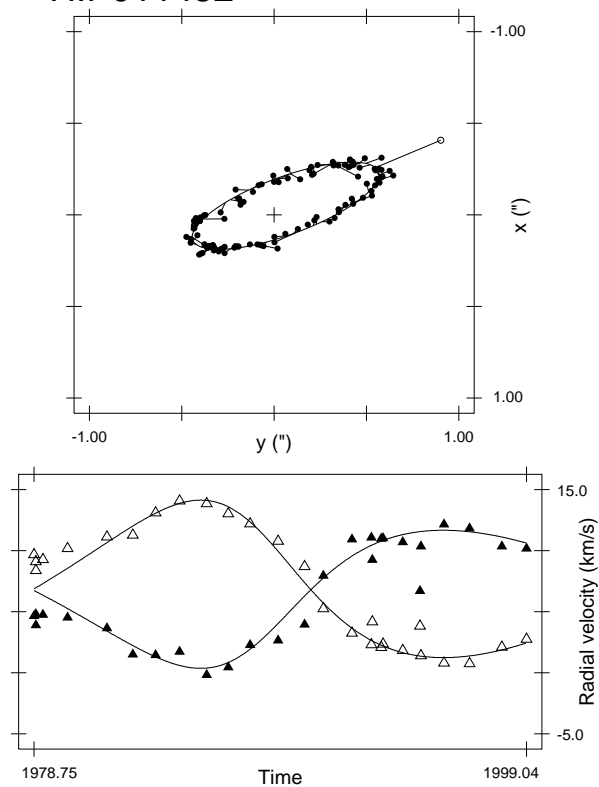}
\caption[]{\label{Fig:HIP11452}
Combined astrometric (upper panel) and spectroscopic (lower panel) orbits for  WDS 02278+0426AB. In the lower panel, filled and open triangles correspond to CORAVEL velocities for components 1 and  2, respectively.  Individual measurements for the visual orbit were taken from the WDS database at USNO.
}
\end{figure}
\begin{table}
\caption[]{
\label{Tab:VrHIP11452}
CORAVEL velocities for the components of WDS 02278+0426AB.  
}
\begin{tabular}{rrrrrr}
\hline\\
Date & JD - 2400000 &  \multicolumn{2}{c}{Vr (2)\tablefootmark{a} $\pm$} &   \multicolumn{2}{c}{Vr (1) $\pm$}\\
        &          &   \multicolumn{2}{c}{(km/s)} &  \multicolumn{2}{c}{(km/s)}\\
\hline\\
011078& 43783.569 &  9.67& 0.67 &  4.65& 0.73   \\
281078& 43810.476 &  9.11& 0.86 &  4.81& 0.96    \\
301078&43812.523  &  8.38& 1.04 &  3.89& 1.17    \\
120279& 43917.269 &  9.28& 0.73 &  4.76& 0.81   \\
180280&44288.279  & 10.17& 0.64 &  4.52& 0.70   \\
051081&44883.540  & 11.13& 0.66 &  3.65& 0.72   \\
281082&45271.562  & 11.28& 0.57 &  1.49& 0.62   \\
061083& 45614.559 & 13.10& 0.58 &  1.45& 0.63   \\
011084& 45975.561 & 14.07& 0.63 &  1.72& 0.69   \\
111185& 46381.452 & 13.86& 0.68 & -0.18& 0.75   \\
300986& 46704.567 & 13.03& 0.62 &  0.46& 0.67   \\
270887& 47035.618 & 12.19& 0.61 &  2.28& 0.67   \\
201088& 47455.516 & 10.78& 0.61 &  2.66& 0.67   \\
231189& 47854.441 &  8.69& 0.50 &  3.96& 0.53   \\
040990& 48139.643 &  5.24& 0.83 &  7.94& 0.75   \\
061191& 48567.490 &  3.26& 0.66 & 10.93& 0.72   \\
240892& 48859.605 &  2.30& 0.62 & 11.08& 0.67   \\
050992& 48871.837 &  4.17& 0.60 &  9.25& 0.65   \\
210193& 49009.264 &  2.07& 0.51 & 10.97& 0.54   \\
180293& 49037.264 &  2.36& 1.18 & 11.02& 1.33   \\
081293& 49330.432 &  1.82& 0.54 & 10.70& 0.58   \\
280894& 49593.912 &  3.82& 0.58 &  6.69& 0.63   \\
080994& 49604.626 &  1.39& 0.54 & 10.35& 0.57   \\
240895& 49954.651 &  0.79& 0.71 & 12.15& 0.78   \\
120996& 50339.632 &  0.75& 0.68 & 11.82& 0.75   \\
090198& 50823.312 &  2.10& 0.56 & 10.35& 0.60   \\
150199& 51194.289 &  2.74& 0.57 & 10.18& 0.62  \\ 
\hline\\
\end{tabular}
\tablefoot{
\tablefootmark{a} {Most likely, spectroscopic component 2 is WDS 02278+0426B (see text for details).}
}
\end{table}

\noindent{\bf WDS 02278+0426AB.} The most recent and thorough analysis of that system was presented by \citet{Andrade2007}, including a revised visual orbit. CORAVEL data (Table~\ref{Tab:VrHIP11452}) make it possible now to compute a combined SB2/astrometric orbit, which thus gives access for the first time to the masses of the system components. The combined orbital elements and other physical parameters of the system (such as orbital parallax and component masses) are listed in Table~\ref{Tab:SB2}, which identifies the components as 1 and 2 rather than A and B. It must be stressed that there is no easy way to  identify for sure the spectroscopic component 1 with either visual component A or B, since both components fall in the spectrograph slit and are of almost equal brightness: Hipparcos gives $m_{Hp, A} = 9.45$ and $m_{Hp, B} = 9.63$.
An indirect way to identify components A and B is by noting that in Table~\ref{Tab:SB2}, the velocity semi-amplitude $K_2$ is higher than $K_1$, hence $M_2$ is smaller than $M_1$. Since both components lie on the main sequence, component 2 should thus be less luminous than component 1, hence component 2  should be identified with B. It is possible to check quantitatively this assumption by using the mass-luminosity relationship provided by \citet{Kroupa1993}, namely  $M_{V,1} =  5.4$ for $M_1 = 0.95$~M$_\odot$ and $M_{V,2} =  6.2$ for $M_2 = 0.85$~M$_\odot$, or $M_{V,2} - M_{V,1} = 0.8$, somewhat larger than the observed $m_{Hp, B} - m_{Hp, A} = 0.18$, but certainly consistent with the error bars on $M_{1,2}$. Although \citet{Andrade2007} classified HD~15285~A as K7V based on several spectral features, among which the presence of weak TiO bands, its inferred mass of 0.95~M$_\odot$ is more typical of a G5V type, while the mass of component B would flag it as K0V \citep{Allen2000}. 

These masses were derived from the orbital parallax of $52\pm2$~mas, to be compared with the Hipparcos value of $60.2\pm1.7$~mas \citep{ESA-1997} or $58.3\pm1.1$~mas \citep{vanLeeuwen2007}, which yield total masses  of 1.16 and 1.28~M$_\odot$, respectively, to be compared with 1.8~M$_\odot$ for the orbital parallax.
From the fractional mass $M_2/(M_1+M_2) = 0.47$ (Table~\ref{Tab:SB2}), one obtains, $(M_1, M_2) = (0.62, 0.54)$ and (0.68, 0.60)~M$_\odot$. These values agree much better with a K7V spectral type for component A.

\noindent{\bf WDS 02361+0653A.} The 22 CORAVEL measurements available for component A (spanning 6837~d) yield a standard deviation of  0.29~\kms, identical to the  average error on a single measurement. The constancy of the radial velocity was confirmed by \citet{Chubak2012}, who obtained 62 measurements spanning 1299~d, with a standard deviation of 0.142~\kms. 
Hence, radial velocities cannot be used to lift the ambiguity on the orbital orientation.

\noindent{\bf WDS 02442+4914AB.} Indeterminate visual orbit and roughly constant radial velocity between 1900 and 1990.

\noindent{\bf WDS 03575-0110AB.} Indeterminate visual orbit and almost constant radial velocity between 1984 and 1993.

\noindent{\bf WDS 04153-0739BC.} In the triple system STF 518 A-BC, B is the white dwarf omi Eri B and C the flare star DY Eri. 

\noindent{\bf WDS 04312+5858Aa,Ab.} Triple system, with the A component being an astrometric binary in the visual binary STI~2051~AB. 

\noindent{\bf WDS 05074+1839AB.} Visual orbit highly conjectural and radial velocity from \citet{Duquennoy1991} is constant.

\noindent{\bf  WDS 06003-3102AC.}  {\it{A quadruple system}} \citep{Tokovinin2005}: HU 1399 AB ($P = 67.7$~y, $a = 0.912''$), HJ 3823 AC ($P = 390.6$~y, $a = 3.95''$ by Baize 1980, 
not confirmed by Tokovinin et al. 2005), TOK 9 C,CE ($P = 23.7$~y, $a = 0.12''$). \nocite{Baz1980b} A,B  and C,CE are coplanar \citep{Tokovinin2005}.

\noindent{\bf  WDS 06262+1845AB.} Triple system, with A known as SB2 with $P = 6.99$~d \citep{Griffin1975}.

\noindent{\bf WDS 06293-0248AB.} Solution for the SB2 orbit includes masses and orbital parallax \citep{Segransan2000}. 
Although not clearly stated, the orbital elements listed by \citet{Segransan2000}
correspond to the spectroscopic orbit (hence $\omega$ and $\Omega$ listed in our Table~\ref{Tab:poles} differ by 180$^\circ$ from theirs).

\noindent{\bf WDS 06579-4417AB.} Indeterminate visual orbit.

\noindent{\bf WDS 07100+3832.} The spectro-visual orbit of this system was only recently computed by \citet{Bay2012}.

\noindent{\bf WDS 07175-4659AB.} The visual orbit is still very uncertain. The total mass of the system is abnormally low given the spectral type.

\noindent{\bf WDS 07346+3153AB.} A difficult system since both A and B are spectroscopic binaries \citep{Hansen1940}, with $P = 9.21$~d and $P =2.93$~d, 
respectively. \citet{Batten1967} has lifted the ambiguity on the ascending node for that system, and we used the orbital elements of the 6th Catalogue of Visual Binary Orbits to compute the position of the pole, agreeing with that of \citet{Batten1967}, allowing for the fact that our poles follow
the right-hand rule, opposite to that of Batten.

\noindent{\bf WDS07393+0514AB.} The pole ambiguity was lifted from the radial-velocity drift present in the \citet{Duquennoy1991data} data.

\noindent{\bf WDS 08592+4803A,BC.} Indeterminate visual orbit.  It might be a quadruple system, since Aa seems to  be  a SB1 system  with a period $P \sim 11$~y \citep{Abt1965} although the spectroscopic orbit is not of an excellent quality.

\begin{figure}
\includegraphics[width=8cm]{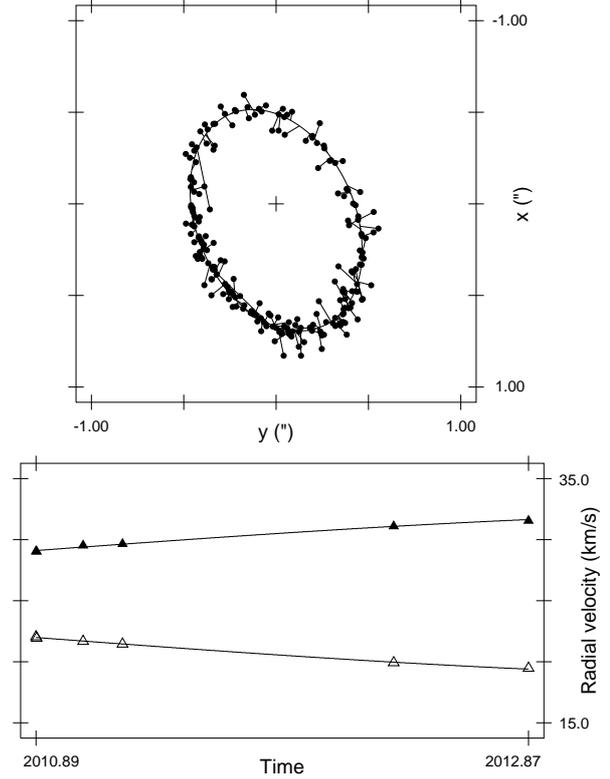}
\caption[]{\label{Fig:HD76943}
Combined astrometric (upper panel) and spectroscopic (lower panel) orbits for WDS 09006+4147AB. 
Filled triangles in the lower panel refer to component A. 
Individual measurements for the visual orbit were taken from the WDS database.
}
\end{figure}
\begin{figure}
\includegraphics[width=9cm]{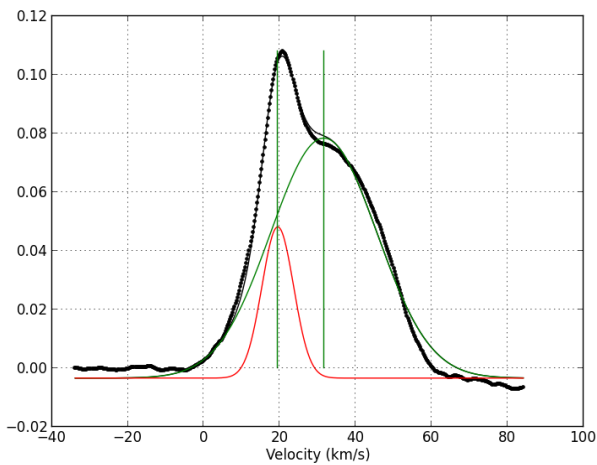}
\caption[]{\label{Fig:CCF76943}
Cross-correlation function of the spectrum of WDS 09006+4147AB (= HD~76943) obtained 
with the HERMES/Mercator spectrograph \citep{Raskin2011} on HJD~2456244.717 (November 13, 2012),
correlated with a F0V template. Both components (separated by 0.5'' on the sky) enter the fiber 
(2.5'' on the sky). Component A corresponds to the wide peak centred at 31.5~\kms.
}
\end{figure}

\noindent{\bf WDS 09006+4147AB.}
The period of 21.05~y for that visual binary is well established \citep{Mut2010b}. Based on an older (albeit similar) value of the visual period from \citet{Heintz1967}, \citet{Abt1976} proposed a spectroscopic orbit based on Heintz' visual elements plus approximate values for the systemic velocity and the semi-amplitude, which could account for Abt \& Levy's own velocity measurements and older ones from \citet{Underhill1963}. Although it is rather uncertain, this orbit is still the one quoted by the SB9 catalogue \citep{Pourbaix2004}. The spectroscopic orbit is difficult to derive because (i) the components are not separated widely enough on the sky to record their spectra
separately, (ii) the velocity semi-amplitude is small, and (iii) component A appears to rotate relatively fast. Therefore, the available unpublished CORAVEL measurements appear useless, since the two components appear hopelessly blended  in the cross-correlation function. More recent HERMES/Mercator spectra \citep{Raskin2011} have a  resolution high enough to allow a comfortable double-Gaussian fit of the cross-correlation function (CCF; Fig.~\ref{Fig:CCF76943}). Although such a fit has not necessarily a unique solution, we found it re-assuring that the CCFs for all seven available dates (Table~\ref{Tab:VrHERMES}) are well fitted with Gaussians having the same properties: height $\sim 0.08$ and $\sigma =  14$~\kms\ for the F4V component A ($V = 4.2$), and height $\sim 0.04$ and $\sigma =  4$~\kms\ for the K0V
component B ($V = 6.5$). It thus appears that component A  is a relatively fast rotator. The assignment of either of the two CCF peaks to component A thus relies on the assumption that component A, being of spectral type F4V, should correspond to the deepest peak  when computed with a F0V template.

The validity of our double-Gaussian fits, and of the resulting orbital elements listed in Table~\ref{Tab:SB2}, may be assessed a posteriori by the comparison of the orbital parallax ($53.8\pm1.0$~mas) with the Hipparcos parallax  \citep[$60.9\pm1.3$~mas;][]{ESA-1997}. The discrepancy probably comes from the fact that the limited phase coverage for the radial velocities does not yet impose strong enough constraints on the velocity semi-amplitudes (Fig.~\ref{Fig:HD76943}). The astrometrically derived masses \citep[$M_A = 1.37$~M$_{\sun}$, $M_B = 1.04$~M$_{\sun}$, in relatively good agreement with the spectral types;][]{Martin1998,Sod1999} are different from the types derived from our combined orbit  ($M_A = 1.73$~M$_{\sun}$, $M_B = 1.69$~M$_{\sun}$), which also hints at inaccurate velocity semi-amplitudes.

\noindent{\bf WDS 09144+5241AB.} For this long-period binary, the determination of the orbital pole is based on the agreement between the variation of the radial velocities $V_{\mathrm{A}}$ and $V_{\mathrm{B}}$ measured between 1910 and 1997 and the value predicted using the elements of the visual orbit. This result is confirmed by the value of the measurement of ($V_{\mathrm{B}}~-~V_{\mathrm{A}}$) performed with the HERMES spectrograph in 2012 (see Table~\ref{Tab:VrHERMES}).

\noindent{\bf WDS 09313-1329AB.} Mass-ratio derived from the astrometry \citep{Sod1999}.  Two HERMES radial-velocity measurements obtained one year apart neither showed any significant drift ($Vr = 7.93$~\kms\ on JD = 2455657.5 and $Vr = 7.89$~\kms\ on JD = 2456038.4), nor any line doubling, as would be expected for a system consisting of two components only 0.6 arcsec apart. 

\noindent{\bf WDS 09357+3549AB.} Indeterminate visual orbit.

\noindent{\bf WDS 10454+3831AB.} {\it{Triple system}}, Aa being the speckle binary CHR 191. A value of ($V_{\mathrm{B}}~-~V_{\mathrm{A}}$) obtained with the HERMES spectrograph in 2012 is listed in Table~\ref{Tab:VrHERMES}.

\noindent{\bf WDS 11182+3132AB.} Quadruple system, A and B are both SB1 with $P = 670.24$~d and $P = 3.98$~d  respectively \citep{Griffin1998}.

\noindent{\bf WDS 11247-6139AB.} Abnormally high total mass, inconsistent with the spectral types

\noindent{\bf WDS 13100+1732AB.} Spectral types and masses of components from \citet{tenBrummelaar2000}.

\noindent{\bf WDS 13198+4747AB}. {\it{Triple system}} \citep[Aa = CHR 193;][]{Beuzit2004}. CORAVEL sees this star as SB3, but no satisfactory attribution of the various peaks to the corresponding components could be made, making it very difficult to isolate the trend of $Vr$(A). 

\noindent{\bf WDS 13473+1727AB.} Indeterminate orbit. The component $\tau$~BooAb is a star with planet.

\noindent{\bf WDS 13491+2659AB.}  The pole ambiguity was lifted from unpublished CORAVEL velocities yielding the sign of $Vr$(B)$ - Vr$(A) ($= -3.3$~\kms) during the observation span 1983 -- 1988 (Table~\ref{Tab:VrHERMES}).

\noindent{\bf WDS 13547+1824.} \citet{Jancart2005} obtained a combined spectroscopic/astrometric orbit based on Hipparcos data and the spectroscopic orbit of \citet{Bertiau1957}. 
The value of $\omega$ from \citet{Jancart2005} has been adopted in Table~\ref{Tab:poles}. 

\noindent{\bf WDS 14514+1906AB.} The visual orbital period is 151~yr \citep{Sod1999}, so that no strong trend must be expected from the radial velocities over three decades. CORAVEL observations \citep{Duquennoy1991data} exhibit a weak upward trend between 1977 (1~\kms) and 1991 (1.4~\kms), which is however not significant based on the average measurement uncertainty of 0.34~\kms. Later measurements by \citet{Abt2006} seem to confirm that trend, but the trend is made fragile by a possible  zero-point offset between the two data sets. Nevertheless, the pole quoted in Table~\ref{Tab:poles} is based on the assumption that the radial-velocity trend is real, which then implies that the node given in the 6th USNO Catalogue is the correct one.

\noindent{\bf WDS 14575-2125AB.} Indeterminate visual orbit.

\noindent{\bf WDS 15038+4739AB.} Triple star: Bb = eclipsing SB2 with $P = 0.27$~d \citep{Lu2001}.

\noindent{\bf WDS 15527+4227.}  The Hipparcos Double and Multiple Star Annex  \citep[DMSA;][]{ESA-1997} claims HIP~77760 to be an astrometric binary with an orbital period of 0.14~yr and an inclination of 131.68$^\circ$. These values seem, however, incompatible with the absence of (CORAVEL) radial-velocity variations (33 measurements with a standard deviation of 0.413~\kms, to be compared with the average instrumental error of 0.380~\kms).

\begin{figure}
\includegraphics[width=8cm]{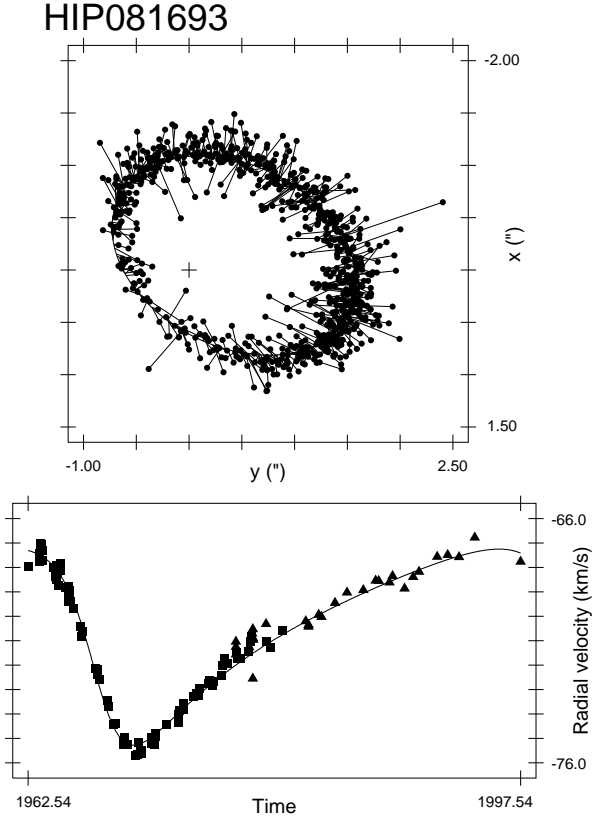}
\caption[]{\label{Fig:HIP81693}
Combined astrometric (upper panel) and spectroscopic (lower panel) orbits for WDS 16413+3136AB. The solid line in the lower panel corresponds to the predicted velocity curve for A. Individual measurements for the visual orbit were taken from  the WDS database at USNO.
}
\end{figure}

\noindent{\bf WDS 16413+3136AB.}
There are many visual orbits available for that system, the most recent ones by \citet{Baz1976} and \citet{Sod1999}. \citet{Scarfe1983} already computed an orbit combining  visual data and a collection of  Coud\'e radial velocities. Unpublished CORAVEL velocities given in Table~\ref{Tab:VrHD150680} nicely complement Scarfe's velocities, yielding a somewhat more accurate orbit as listed in Table~\ref{Tab:New_SB}.
A third component has been detected by IR speckle interferometry \citep{McCarthy1983}.

\begin{table}
\caption[]{
\label{Tab:VrHD150680}
New CORAVEL velocities for the components of WDS 16413+3136AB in complement of velocities published by \citet{Duquennoy1991}.  
}
\begin{tabular}{rrrrrr}
\hline\\
Date & JD - 2400000 &  \multicolumn{2}{c}{$Vr$} \\
        &          &   \multicolumn{2}{c}{(km/s)} \\
\hline\\
270490& 48009.500&  -68.18& 0.32\\
090891& 48478.459&  -67.56& 0.32\\
070592& 48750.578&  -67.49& 0.32\\
020393& 49049.692&  -67.57& 0.31\\
140494& 49457.592&  -66.77& 0.33\\
140797& 50644.431&  -67.76& 0.33\\
\hline\\
\end{tabular}
\end{table}

\noindent{\bf WDS 16555-0820AB.}  A triple system composed of M dwarfs, in which all three 
stars are visible in the spectra. Spectroscopic orbits are available for the inner (Ba-b) and outer pairs (A-Bab), both as SB2  \citep{Segransan2000}. The two orbits  are probably coplanar \citep{Mazeh2001}.  
The whole system is probably septuple.

\begin{figure}
\includegraphics[width=8cm]{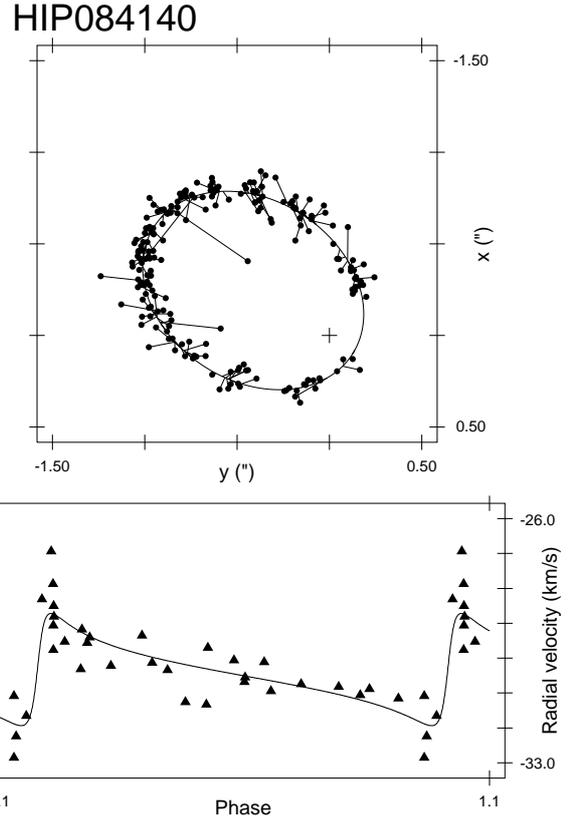}
\caption[]{\label{Fig:HIP84140}
Combined astrometric (upper panel) and spectroscopic (lower panel) orbit for WDS 17121+4540AB. The solid line in the lower panel corresponds to the predicted velocity curve for A. Individual measurements for the visual orbit were taken  from the WDS database at USNO.
}
\end{figure}

\noindent{\bf WDS 17121+4540AB.} A visual orbit for that system was obtained by \citet{Hrt1996a}. The available CORAVEL velocities shown in Table~\ref{Tab:VrHD155876} make it possible to derive a combined orbit for the first time. Visual observations are from the CHARA database. The new combined orbit is presented in Fig.~\ref{Fig:HIP84140}.

\begin{table}
\caption[]{
\label{Tab:VrHD155876}
New CORAVEL velocities for the components of WDS 17121+4540AB.  
}
\begin{tabular}{rrrrrr}
\hline\\
Date & JD - 2400000 &  \multicolumn{2}{c}{$Vr$} \\
        &          &   \multicolumn{2}{c}{(km/s)} \\
\hline\\
200477& 43254.623&  -83.29& 0.42\\
220477& 43256.617&  -31.08& 0.71\\
220477& 43256.633&  -32.85& 0.88\\
130378& 43581.648&  -28.31& 0.66\\
130378& 43581.662&    8.83& 0.93\\
270678& 43687.480&  -26.94& 1.10\\
180778& 43708.394&  -27.87& 1.11\\
210778& 43711.380&  -29.76& 0.73\\
220778& 43712.391&  -29.06& 0.68\\
240778& 43714.428&  -28.50& 0.68\\
140679& 44039.489&  -29.18& 0.93\\
140879& 44100.355&  -29.56& 0.68\\
010981& 44849.362&  -30.13& 0.84\\
250282& 45026.692&  -30.34& 0.76\\
160583& 45471.566&  -31.33& 0.57\\
300384& 45790.675&  -30.06& 0.73\\
280784& 45910.384&  -30.68& 0.52\\
130385& 46138.686&  -30.11& 0.77\\
110586& 46562.568&  -30.74& 0.72\\
200787& 46997.453&  -30.82& 0.63\\
230388& 47244.665&  -31.06& 0.61\\
070788& 47350.452&  -30.88& 0.52\\
060689& 47684.589&  -31.15& 0.46\\
270490& 48009.538&  -32.24& 0.71\\
190890& 48123.457&  -31.65& 0.71\\
070791& 48445.443&  -28.81& 0.68\\
051191& 48566.246&  -29.53& 0.64\\
080592& 48751.570&  -30.31& 0.66\\
190892& 48854.345&  -29.40& 0.60\\
210493& 49099.595&  -30.22& 0.53\\
140494& 49457.583&  -29.35& 0.51\\
260895& 49956.360&  -31.26& 0.65\\
110596& 50215.561&  -29.70& 0.50\\
140797& 50644.441&  -30.55& 0.47\\
070598& 50941.589&  -30.94& 0.46\\
\hline\\
\end{tabular}
\end{table}

\noindent{\bf WDS 17153-2636AB.} \citet{Irwin1996} derived precise B-A radial velocities (except for an acceleration component that they were unable to explain). Velocity measurements of $Vr$(A) and $Vr$(B) were obtained in 2012 by the HERMES spectrograph (Table~\ref{Tab:VrHERMES}).

\noindent{\bf WDS 17191-4638AB.} Indeterminate visual orbit.

\noindent{\bf WDS  17304-0104AB.} A combined astrometric/spectroscopic solution has been computed by \citet{Pourbaix2000} and has been used to compute the position of the orbital pole. 

\noindent{\bf WDS  17349+1234.}  Spectral types from \citet{Gatewood2005}. \citet{Kamper1989} performed a combined spectroscopic/astrometric analysis of this system, and their orbit has been recently updated by \citet{Hinkley2011}, from whom we adopted the orbital elements without requiring any change to comply with our conventions.

\noindent{\bf WDS 17364+6820Aa,Ab.} Velocity measurements of $Vr$(A) and $Vr$(B) obtained in 2012 by the HERMES spectrograph are listed in Table~\ref{Tab:VrHERMES}.

\noindent{\bf WDS 17465+2743BC.} HD 161797 ($\mu$ Her) is a quadruple system \citep{Raghavan2010}. The components of the wide pair STF 2220 AB ($35\arcsec$) have common proper motion.
The component A is an astrometric binary (period of 65 years, \citet{Hei1994a}). 
The components B and C form the visual binary AC 7 for which the orbital pole ambiguity has been lifted from unpublished CORAVEL velocities, yielding a drift of $Vr$(B) over 1975 -- 1995. 
\begin{figure}
\includegraphics[width=8cm]{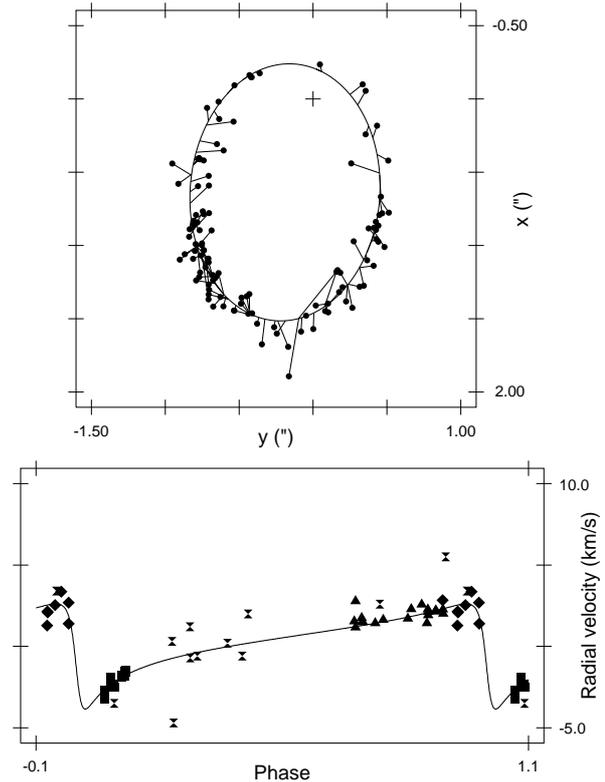}
\caption[]{\label{Fig:HIP88745}
Combined astrometric (upper panel) and spectroscopic (lower panel) orbit for WDS 18070+3034AB. The solid line in the lower panel corresponds to the predicted velocity curve for A. Individual measurements for the visual orbit were taken  from the WDS database at USNO. In the lower panel, triangles correspond to CORAVEL data, squares to velocities from \citet{Abt2006}, diamonds to \citet{Tokovinin2002}, and double triangles to the old radial velocities collected by \citet{Kamper1986}.
}
\end{figure}

\noindent{\bf WDS 18070+3034AB.} Based on available CORAVEL velocities (unpublished measurements are given in Table~\ref{Tab:VrHD165908}), a combined astrometric/spectroscopic solution (Table~\ref{Tab:New_SB} and Fig.~\ref{Fig:HIP88745}) improves the spectroscopic orbital elements provided by \citet{Abt2006} for that system.  AC 15 AB is a triple system, with the separation  Aa-Ab amounting to 0.228$\arcsec$ and that of AB to 0.851$\arcsec$ in 2005, according to \citet{Scardia2008}. 

\begin{table}
\caption[]{
\label{Tab:VrHD165908}
New CORAVEL velocities for the components of WDS 18070+3034AB in complement of velocities published by \citet{Duquennoy1991}.  
}
\begin{tabular}{rrrrrr}
\hline\\
Date & JD - 2400000 &  \multicolumn{2}{c}{$Vr$} \\
        &          &   \multicolumn{2}{c}{(km/s)} \\
\hline\\
190790& 48092.429&    2.15& 0.32\\
180890& 48122.388&    2.21& 0.33\\
020891& 48471.433&    2.27& 0.33\\
210891& 48490.384&    2.02& 0.32\\
\hline\\
\end{tabular}
\end{table}

\noindent{\bf WDS 18428+5938AB.} Velocity measurements (4 for $Vr$(A) and one for $Vr$(B)) obtained by the HERMES spectrograph in 2012  are listed in Table~\ref{Tab:VrHERMES}.

\noindent{\bf WDS 18570+3254AB.} A combined astrometric/spectroscopic solution confirms the spectroscopic orbital elements provided by 
\citet{Abt2006} for that system. BU 648 AB is a binary with an exoplanet \citep{Muterspaugh2010planet} in coplanar orbits.

\noindent{\bf WDS 19121+0254AB.} Masses are from \citet{AST2001}.

\begin{figure}
\includegraphics[width=8cm]{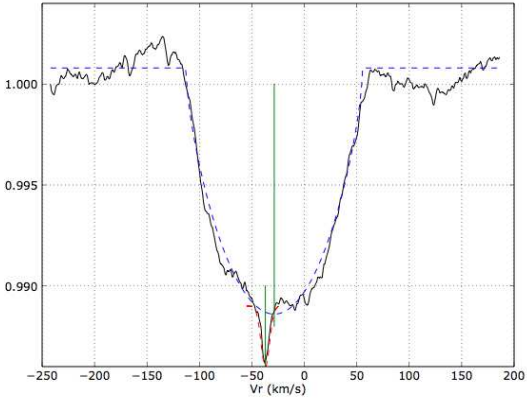}
\caption[]{\label{Fig:CCF_HIP95501}
Cross-correlation function of the HERMES spectrum of WDS 19255+0307Aa,Ab (obtained at JD~2455657.742) with a F0V template. 
Component Aa shows a rotationally broadened profile with $V_r \sin i = 84$~\kms, and component Ab is visible as a weak Gaussian 
[at $V_r$(Ab) = -37.70~\kms] superimposed on the rotationally broadened profile [centred on $V_r$(Aa) = -26.68~\kms].
Similar profiles have been observed over several consecutive nights.
}
\end{figure}

\noindent{\bf WDS 19255+0307Aa,Ab.} \citet{Kamper1989} performed a combined spectroscopic/astrometric analysis of the Aa-Ab pair. 
We adopted the orbital elements from their Table~V, except that 180$^\circ$ was added to their value of $\omega$, to conform with our convention that $\omega$ corresponds to the visual orbit of Ab around Aa, rather than to the spectroscopic orbit of Aa around the centre of mass. One HERMES measurement was obtained (Table~\ref{Tab:VrHERMES}) that reveals for the first time the companion in the cross-correlation function as a narrow peak on top of the broad peak ($V_r \sin i = 84$~\kms) due to the FV primary (Fig.~\ref{Fig:CCF_HIP95501}). 
The companion does not rotate fast. From the orbit described in \citet{Kamper1989}, we estimate that the HERMES data point [$V_r$(Aa) = -26.68~\kms] has been taken at orbital phase $0.54\pm0.14$, when the velocity is  $1.66\pm \mycom{0.15}{0.70}$~\kms\ higher than the centre of mass (CoM) velocity. The CoM velocity is thus estimated to be $-28.34\pm\mycom{0.15}{0.70}$~\kms. Combined with $V_r$(Ab) = -37.70~\kms, this yields a surprisingly high mass ratio $M_{Aa}/M_{Ab}$ of 5.6, or a mass of 0.29~\Msun\ for the companion if the primary is a F0V star of mass 1.6~\Msun.
The corresponding spectral type would then be M4V for the Ab component, 8.6 magnitudes fainter than the primary component.   
The close pair WDS 19255+0307Aa,Ab, seems to form a triple system with the wide pair (WDS 19255+0307AB = BUP 190, separation 96 to 133$\arcsec$), but the physical nature of this system is yet to be confirmed.   

\noindent{\bf WDS 21000+4004AB.} \citet{Fekel1978} found that the visual component A of KUI 103 is itself a spectroscopic binary with a period of 3.27~d, 
but no information is available about coplanarity \citep{Tokovinin2008}.
\citet{Pourbaix2000} has made a combined analysis of the visual and spectroscopic data. 
However, the authors note that this orbital solution is somewhat uncertain, given the large error bars and inconsistencies in the derived values for stellar masses and orbital parallax.
Thus, we decided to base our determination of the orbital pole of this system on the analysis of the published radial velocity measurements \citep{Pourbaix2000} and the orbital elements of the new visual orbit proposed by \citet{Doc2010d}.

\noindent{\bf WDS 21069+3845AB.} The mass ratio is taken from \citet{Pk2006b}. 
The node assignment is based on the difference $V_r$(B) - $V_r$(A) $\sim +1$~\kms\ $>0$, as provided by Table~1 of \citet{Pk2006b} and confirmed by unpublished CORAVEL measurements.

\noindent{\bf WDS 22234+3228AB.} The total mass seems too high for the spectral type.

\noindent{\bf WDS 22280+5742AB.} Masses are from \citet{Delfosse2000}. Among the many radial velocities available in the literature for the components of this system, we give the  highest weight to the measurement [$V_r$(A) = -32.7~\kms\ and $V_r$(B) = -33.2~\kms] by \citet{Gizis2002} because (i) that measurement is precisely dated  (JD~2450005.6) and on the same night for A and B, and (ii) the 0.5~\kms\ velocity difference is consistent with the prediction from the visual  orbit. The measurement [$V_r$(A) = -34.0~\kms\ and $V_r$(B) = -35.0~\kms] by \citet{Delfosse1998} is equally useful  (its epoch is not as precisely known as that report by Gizis, but  it was obtained between September 1995 and March 1997, a short time span with respect to the 44.7~yr orbital period);  the 1~\kms\ velocity difference agrees better with the prediction from the visual  orbit than Gizis' 0.5~\kms. 
The determination  of the {\it sign} of  $V_r$(B) - $V_r$(A) = -1~\kms is important and it is opposite to the prediction from the visual orbital elements from the 6th COVBS. It is thus necessary to add 180$^\circ$ to the values of $\Omega$ and $\omega$ listed in the  6th COVBS.
It is noteworthy that the velocity values listed in the Wilson - Evans - Batten catalogue \citep[][$V_r$(A) = -24.0~\kms\ and $V_r$(B) = -28.0~\kms]{Duflot1995} or in \citet{Reid1995} [$V_r$(A) = -16.0~\kms\ and $V_r$(B) = -29.0~\kms], although the differences are much larger than anticipated, always have $V_r$(B)~-~$V_r$(A)$<0$.

\noindent{\bf WDS 23317+1956AB.} Indeterminate visual orbit.

\noindent{\bf WDS 23524+7533AB.} Indeterminate visual orbit.
A triple system with A being SB2  with a period of 7.75~d \citep{Christie1934}.           

\section{Determining the orbital pole of a double star}
\label{Sect:pole}
\subsection{Orbital pole of a visual double star}

The true relative orbit (B/A) is related to the classical visual orbit of B relative to A. 
The pole of this orbit is defined by the unit vector perpendicular to the orbital plane and directed in such a way that an observer placed at its end sees that the B component has a direct motion (counter-clockwise).

The spatial motion of the binary components is described using the reference frame (A, $x$, $y$, $z$) centred on the A component with two axes in the plane tangent to the celestial sphere (which is denoted plane of the sky): 
A$x$ points north (position angle = 0$^\circ$), A$y$ points east (position angle = 90$^\circ$), and the third axis A$z$ is along the line of sight, pointing in the direction of increasing radial velocities (i.e. positive radial velocity). 
This reference frame is thus retrograde (i.e., viewed from the positive side of the A$z$-axis, a rotation of the A$x$-axis onto the A$y$-axis is carried 
out clockwise).

\begin{figure}
\includegraphics[width=9cm]{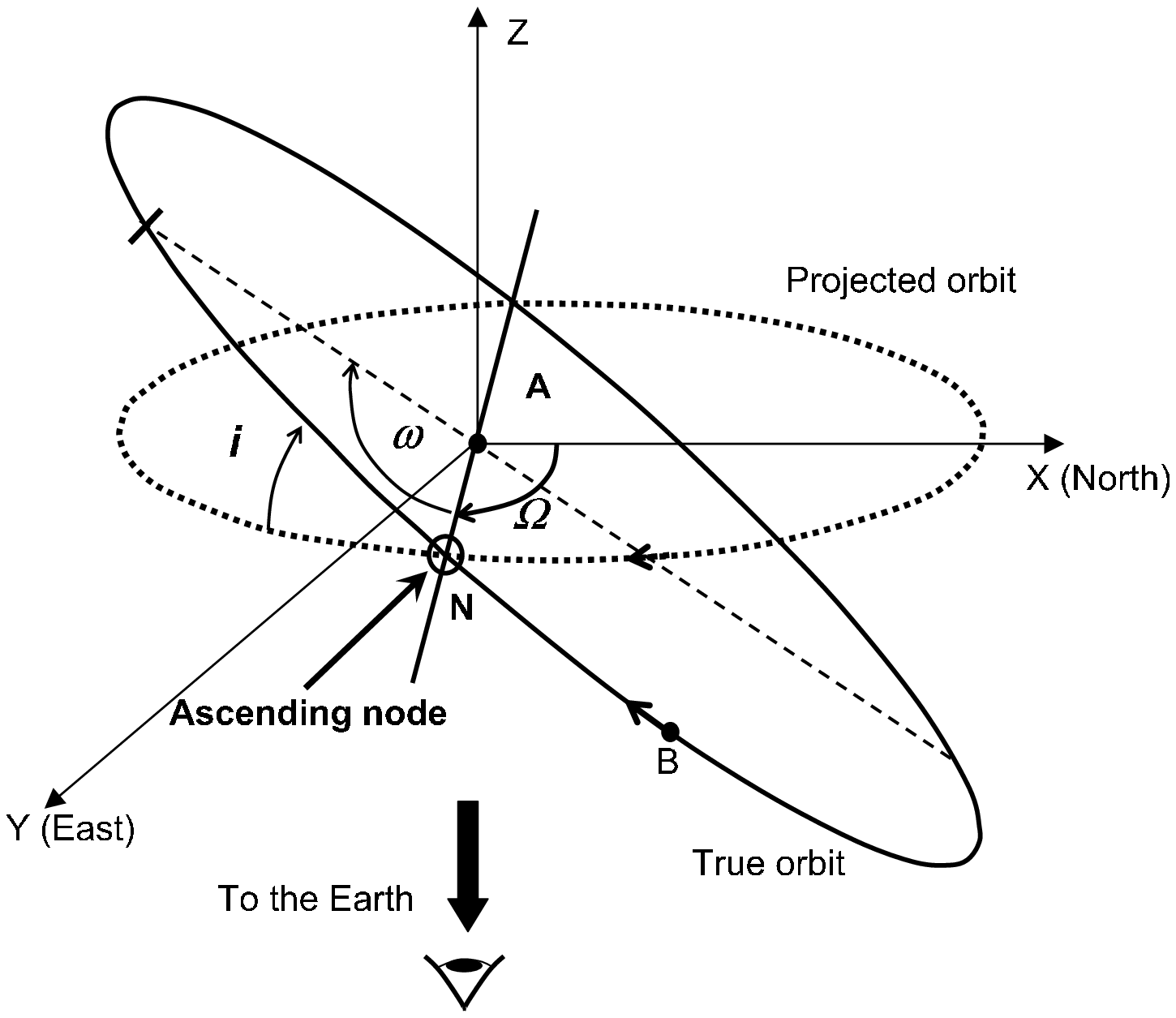}
\caption[]{\label{Fig:orbit}
True and the projected relative orbit (B/A) of a visual binary and its geometrical elements. The ascending node of the orbit is located at point N.
}
\end{figure}

The relative (B/A)  orbit is described by means of seven orbital elements:

(i) Four so-called dynamic elements specifying the properties of the Keplerian motion in the true orbit ($T$, the epoch of passage through periastron; $P$, the revolution period; $a$, the semi-major axis and $e$, the eccentricity).

(ii) Three so-called geometric elements that define the orientation of the true and apparent orbits, the latter 
being the projection of the true orbit on the plane of the sky
(see Fig.~\ref{Fig:orbit}):
\begin{itemize}
\item $\Omega$ is the position angle of the line of intersection between the true orbital plane and the plane  
of the sky. There are two nodes whose position angles differ by 180$^\circ$, but by convention, 
$\Omega$ is the position angle of the ascending node where the orbital motion is directed away from the Sun. 
If radial-velocity measurements are not available to identify the ascending node, a temporary value ranging between 0$^\circ$ and 
180$^\circ$ is adopted \citep{Finsen1934}. 

\item $i$, the inclination, is the angle between the planes of the projected orbit and of the true orbit, taken at the ascending node. The motion is direct if $0^\circ < i < 90^\circ$ and retrograde if $90^\circ < i < 180^\circ$.

\item $\omega$ is the argument of the periastron in the true orbit plane, measured in the direction of the orbital motion from the ascending node with a value ranging from 0$^\circ$ to 360$^\circ$.
\end{itemize}

Two other orbits may  be considered: the (A/AB) orbit of A around the centre of mass  AB and the (B/AB) orbit of B around the centre of mass.

For resolved binaries, astrometric observations give then access to the (A/AB) and (B/AB) orbits projected on the  plane of the sky.
For an unresolved binary, the astrometric measurements capture the position of the photocentre F located between A and AB. 
The astrometric orbit (F/AB) is similar to that of (A/AB), but with a smaller semi-major axis $a_{\mathrm{F}}$.

For a given projected orbit as derived from astrometric observations, two possible true orbits correspond that are symmetric with respect to the plane of the sky.

Therefore, even though the dynamic elements and the inclination $i$ are determined unambiguously, it is impossible to select between the two possible true orbits (and thus between the two possible ascending nodes $\Omega$) without a radial-velocity measurement.
The argument of periastron $\omega$ being measured from the ascending node, the ambiguity on  $\Omega$ thus propagates onto $\omega$.
One special case is $i$ = 0$^\circ$  or 180$^\circ$, the true orbit being then in the plane of the sky, and there is no ambiguity on the  ascending node (since there is no such thing as an ascending node in that case!). 
In all other cases, radial-velocity measurements for at least one component  are needed to lift the ambiguity on the value of $\Omega$, before the orientation of the orbital pole can be fixed , as was done by \citet{Dommanget2005} and \citet{Dommanget2006}.

\subsection{Radial velocities}

If $V_{\mathrm{r}}$(A), $V_{\mathrm{r}}$(B) and $V_{\mathrm{r,sys}}$ are the heliocentric radial velocities of components A and B and of the centre of mass  of the binary AB, and denoting $V_{\mathrm{A/AB}}$ and $V_{\mathrm{B/AB}}$ the radial components of the orbital velocity of A and B around the centre of mass, then

\begin{equation}
\label{radvel1}
V_{\mathrm{r}} \mathrm{(A,B)}= V_{\mathrm{(A,B)/AB}}+V_{\mathrm{r,sys}}.
\end{equation}

The orbital radial velocity of each component relates to the orbital elements by 

\begin{equation}
\label{radvel2}
V_{\mathrm{(A,B)/AB}} = 2 \pi \frac{a_{\mathrm{(A,B)}} \sin i}{P \sqrt{1-e^2}}[e \cos\omega_{\mathrm{(A,B)}}+\cos(\omega_{\mathrm{(A,B)}}+v)],
\end{equation}

where $v$ is the true anomaly, $a_{\mathrm{A,B}}$ are the semi-major axes, $P$ the orbital period, and $\omega_{\mathrm{(A,B)}}$ the argument of the periastron of the (A/AB) and (B/AB) orbits. 

For double-lined spectroscopic binaries (SB2), the measurements of $V_{\mathrm{A/AB}}$ and $V_{\mathrm{B/AB}}$ are possible; for single-lined spectroscopic binaries (SB1), only $V_{\mathrm{A/AB}}$ measurements are possible.

The relative radial velocity between the components, $V_{\mathrm{r}}~=~V_{\mathrm{r}}\mathrm{(B)}~-~V_{\mathrm{r}}\mathrm{(A)}$, can be calculated from the orbital elements of the (B/A) orbit using Eq.~\ref{radvel2}.

For an astrometric-binary orbit, the radial velocity of the photocenter $V_{\mathrm{r}}$(F) can be computed with the same formula (Eq.~\ref{radvel2}) where the semi-major axis 
$a_{\mathrm{F}}$ and the argument of the periastron $\omega_{\mathrm{F}}$  of the photocentric orbit have been inserted instead.

It is easy to show that

\begin{equation}
\label{radvel4}
\omega_{\mathrm{A}} = \omega_{\mathrm{F}} = \omega_{\mathrm{B}} + 180^\circ.
\end{equation}

The variation of $V_{\mathrm{r}}$ and $V_{\mathrm{r}}$(B) are the same and are opposite to that of $V_{\mathrm{r}}$(A).
By definition, the maximum of the $V_{\mathrm{r}}$(A) and $V_{\mathrm{r}}$(B) curves occurs at the passage of the ascending node in the true orbits (A/AB) and (B/AB), respectively, whereas the maximum of the $V_{\mathrm{r}}$ curve corresponds to the passage at the ascending node in the relative orbit (B/A). 

\subsection{Choosing  the correct ascending node}
\label{Thenode}
Generally, the ascending node $\Omega$ of the relative orbit (B/A) is determined by comparing the variations of the measured radial velocities $V_{\mathrm{r}}$ with the shape of the relative radial-velocity curve $\tilde{V}_{\mathrm{r}}$ [or $\tilde{V}_{\mathrm{r}}$(F)] computed from the visual orbital elements.
However, finding the ascending node of the relative orbit (B/A) in practice depends on the type of binary system and on the type of  radial-velocity measurements available, as described in Table~\ref{Tab:Nodechoose}.

\begin{table*}
\centering
\caption[]{
\label{Tab:Nodechoose}
The different ways to determine the ascending node of the relative orbit (B/A). In the table, the $\tilde{V}_{\mathrm{r}}$ quantities denote the values derived from the visual 
or astrometric elements through Eq.~\ref{radvel2}, the values without a tilde are the observed values.
}
\begin{tabular}{cccccc}
  \hline 
  System type          & Available orbit   & Available velocities  & Ascending node determination  & Case\\
  \hline 
	\\
Spectroscopic binary  & spectro-visual    &     from orbit       & $i$, $\Omega$, $\omega$ result from (B/A) orbit computation & 1\\
  \cline{3-4} 
	\\
resolved as           & visual and SB2      &    $V_{\mathrm{r}}$(B), $V_{\mathrm{r}}$(A) from orbit    & $\tilde{V}_{\mathrm{r}}$(B/A) compared to $V_{\mathrm{r}}$(B)-$V_{\mathrm{r}}$(A) & 2\\
  \cline{3-4} 
	\\
visual or astrometric & visual and SB1      &    $V_{\mathrm{r}}$(A) from orbit   & $\tilde{V}_{\mathrm{AB/A}}$ compared to $V_{\mathrm{r}}$(A) & 3\\
  \cline{3-4} 
	\\
binary                & astrometric and SB1 &    $V_{\mathrm{r}}$(A) from orbit & $\tilde{V}_{\mathrm{r}}$(F) compared to $V_{\mathrm{r}}$(A) & 4\\
  \hline
	\\
Visual binary         & visual            & $V_{\mathrm{r}}$(A) and $V_{\mathrm{r}}$(B)& $\tilde{V}_{\mathrm{r}}$ compared to $V_{\mathrm{r}}$(B)-$V_{\mathrm{r}}$(A) drift & 5\\
  \cline{3-4}
	\\
                      &                  & $V_{\mathrm{r}}$(A)        & $\tilde{V}_{\mathrm{r}}$ compared to $V_{\mathrm{r}}$(A) drift & 6\\
  \hline
	\\
Astrometric binary    & astrometric       & $V_{\mathrm{r}}$(A)        & $\tilde{V}_{\mathrm{r}}$(F) compared to $V_{\mathrm{r}}$(A) drift & 7\\
  \hline 
\end{tabular}
\end{table*}

\begin{figure}
\includegraphics[width=9cm]{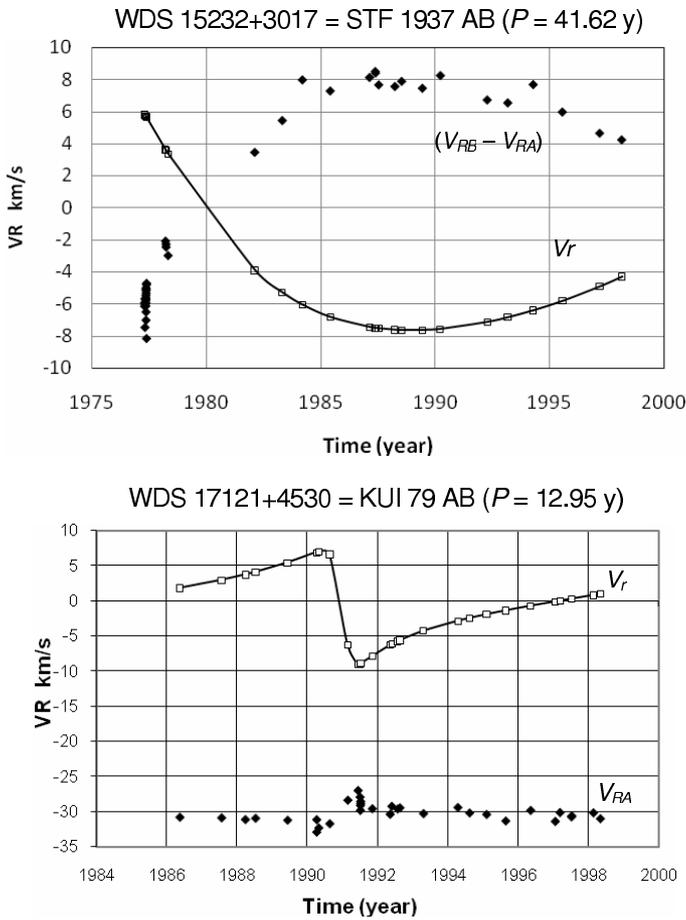}
\caption[]{\label{Fig:visbinVr}
Examples of lifting the ambiguity on the ascending node for a visual binary.
Top (cases~2, 5 of Table~\ref{Tab:Nodechoose}): the slope of the computed $\tilde{V}_{\mathrm{r}}$ curve is opposite to that of the $V_{\mathrm{r}}\mathrm{(B)} - V_{\mathrm{r}}$(A) measurements: it is then necessary to add 
$180^\circ$ to the values of $\Omega$ and $\omega$ given by the 6th Catalog of Orbits of Visual Binary Stars (6th COVBS) \citep{hartkopf2001}.
Bottom (cases~3, 6 of Table~\ref{Tab:Nodechoose}): the slope of the computed $\tilde{V}_{\mathrm{r}}$ curve is opposite to that of $V_{\mathrm{r}}$(A) measurements; there is no need to change the values of $\Omega$ and $\omega$ listed in the 6th COVBS.  
}
\end{figure}
For spectroscopic systems with a combined spectro-visual orbit available (case~1 of Table~\ref{Tab:Nodechoose}), the ascending node of the relative orbit is unambiguously determined \citep[e.g.,][]{Pourbaix2000}. The values of the orbital elements $i$, $\Omega$ and $\omega$ can then be directly used to compute the orbital pole orientation.

For visual binaries with available radial-velocity measurements, if the slope of the computed $\tilde{V}_{\mathrm{r}}$ curve is the same as that of the $V_{\mathrm{r}}$(B) - $V_{\mathrm{r}}$(A) measurements (cases~2 or 5 of Table~\ref{Tab:Nodechoose}), or opposite to that of the $V_{\mathrm{r}}$(A) curve (cases~3 or 6 of Table~\ref{Tab:Nodechoose}), there is no need to change the values of the position angle of the ascending node $\Omega$ and of the argument of the periastron $\omega$, from the values adopted for computing $\tilde{V}_{\mathrm{r}}$ in Eq.~\ref{radvel2}. 
In the opposite situation, it is necessary to add $180^\circ$ to the values of $\Omega$ and $\omega$.
Examples are given in Fig.~\ref{Fig:visbinVr}.

For astrometric binaries (cases~4 or 7 of Table~\ref{Tab:Nodechoose}), if the drift of the computed $\tilde{V}_{\mathrm{r}}$(F) curve is opposite that of the $V_{\mathrm{r}}$(A) measurements, there will be no need to change the values of the position angle of the ascending node $\Omega_{\mathrm{F}}$ or of the argument of the periastron $\omega_{\mathrm{F}}$. 
In the opposite case, it is necessary to add $180^\circ$ to these values.
Thus the ascending node of the (B/A) orbit will be $\Omega = \Omega_{\mathrm{F}} + 180^\circ$ and $\omega = \omega_{\mathrm{F}} + 180^\circ$. 

It is important to note that the ambiguity-free values listed in Table~\ref{Tab:poles} are correspond to the relative orbit (B/A). 
Hence, in Tables~\ref{Tab:New_SB} and \ref{Tab:SB2}, which provide newly determined orbital elements of spectro-visual orbits, we list the values of $\omega_B$, contrary to usage with spectroscopic orbital elements, to ensure consistency with Table~\ref{Tab:poles}.

\subsection{Determining the direction of the pole of the true relative orbit}
\label{Thepole}

This section describes the determination of the direction of the pole of the true relative orbit (B/A), including the case of astrometric binaries for which the B component is invisible.

In the reference frame (A, $x'$, $y'$, $z'$) attached to the (B/A) orbit, the direction of the orbital pole is given by the unit vector ($x'$, $y'$, $z'$) = (0, 0, 1).

\begin{figure}
\includegraphics[width=9 cm]{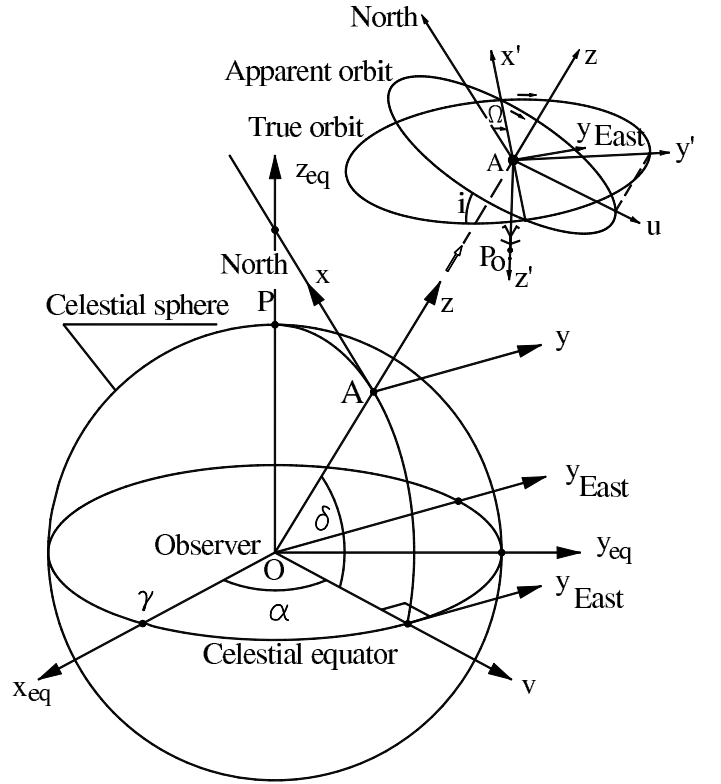}
\caption[]{\label{Fig:coordpole1}
Orientation of the true and apparent relative orbits (B/A) with respect to the celestial sphere and the corresponding reference frames. $(\alpha, \delta)$ are the equatorial coordinates of the binary system.
}
\end{figure}

The ambiguity on the position angle of the ascending node was solved along the method described in Sect.~\ref{Thenode}, the geometric orbital elements $i$, $\Omega$ and $\omega$ are known, so that the equatorial coordinates of the pole ($x_{\mathrm{eq}}$, $y_{\mathrm{eq}}$, $z_{\mathrm{eq}}$) can be computed from the equatorial coordinates of the binary (Right Ascension $\alpha$, Declination $\delta$) by a succession of reference-frame changes as shown in Figure~\ref{Fig:coordpole1}:
\begin{itemize} 
\item From (A, $x'$, $y'$, $z'$) to (A, $x'$, $u$, $z$) by a rotation of angle $i$ around the A$x'$ axis, 
\item to (A, $x$, $y$, $z$)  by a rotation of angle $\Omega$ around the A$z$ axis,
\item to (O, $v$, $y$, $z_{\mathrm{eq}}$) by a rotation of angle $\delta$ around the A$y$ axis,
\item and finally to (O, $x_{\mathrm{eq}}$, $y_{\mathrm{eq}}$, $z_{\mathrm{eq}}$) by a rotation of angle $\alpha$ around the A$z_{\mathrm{eq}}$ axis.
\end{itemize} 

It is then easy to show that in the equatorial reference frame (O, $x_{\mathrm{eq}}$, $y_{\mathrm{eq}}$, $z_{\mathrm{eq}}$), the direction of the orbital pole is given by the unit vector with the coordinates  

\begin{equation}
\label{EquCoord}
\begin{array}{l}
x_{\mathrm{eq}} = \cos\alpha\sin\delta \sin\Omega \sin i-\cos\alpha\cos\delta\cos i-\sin\alpha\cos\Omega\sin i\\
y_{\mathrm{eq}} = \sin\alpha\sin\delta\sin\Omega\sin i-\sin\alpha\cos\delta\cos i+\cos\alpha\cos\Omega\sin i\\
z_{\mathrm{eq}} = -\cos\delta\sin\Omega\sin i-\sin\delta\cos i\\
\end{array}
\end{equation}

We stress that our convention, which leads to a polar direction opposite to that of the system when $i = 0^\circ$, is opposite to that of \citet{Batten1967}.  In the terminology of \citet{Batten1967}, our poles are thus those that would be derived by applying the right-hand rule  to the orbital motion.

The Galactic coordinate system  
is the most appropriate for studying the distribution of the orbital poles in a Galactic framework. Therefore, the coordinates ($x_{\mathrm{eq}}$, $y_{\mathrm{eq}}$, $z_{\mathrm{eq}}$) of the unit vector giving the direction of the pole are converted into the coordinates  ($x_{\mathrm{g}}$, $y_{\mathrm{g}}$, $z_{\mathrm{g}}$) in the Galactic frame using the relation \citep[for details see][]{Johnson1987}

\begin{equation}
\label{Galcoord1}
\left(
	\begin{array}{c}
	x_{\mathrm{g}}\\
	y_{\mathrm{g}}\\
	z_{\mathrm{g}}\\
	\end{array}
\right) = [T]
\left(
	\begin{array}{c}
	x_{\mathrm{eq}}\\
	y_{\mathrm{eq}}\\
	z_{\mathrm{eq}}\\
	\end{array}
\right).
\end{equation}

The terms of the transformation matrix $[T]$ are computed from the J2000 equatorial coordinates of the Galactic centre and pole given by the {\it Institut de M\'ecanique C\'eleste et de Calcul des Eph\'em\'erides} (IMCCE):
\begin{itemize}
\item North Galactic pole: 
{\newline$\alpha_{\mathrm{0}}= 12^h 51^{m} 26.28^s$ and $\delta_{\mathrm{0}}= 27^\circ 7^\prime 41.7^{\prime\prime}$}
\item Galactic center:
{\newline$\alpha_{\mathrm{G}}= 17^h 45^{m} 37.20^s$ and $\delta_{\mathrm{G}}= -28^\circ 56^\prime 10.2^{\prime\prime}$}
\end{itemize}

This leads to the transformation matrix:

\begin{equation}
\label{Amatrix}
[T]~=~
\left(
\begin{array}{rrr}
-0.054875 & -0.873437 & -0.483835 \\
0.494110 & -0.444829 & 0.746982 \\
-0.867666 & -0.198077 & 0.455984 \\
\end{array}
\right).
\end{equation}

After the coordinates ($x_{\mathrm{g}}$, $y_{\mathrm{g}}$, $z_{\mathrm{g}}$) are computed, the Galactic longitude $l$ and latitude $b$ are deduced using the relations

\begin{equation}
\label{Galcoord2}
	\begin{array}{lcc}
	\cos l \cos b = x_{\mathrm{g}} \\
	\sin l \cos b = y_{\mathrm{g}} & \textrm{~with~} &  0^\circ~\le~l~<~360^\circ \\
	\sin b = z_{\mathrm{g}} & \textrm{~with~} &  -90^\circ~\le~b~\le~90^\circ .\\
	\end{array} 
\end{equation}

\end{document}